\documentclass{article}
\usepackage[english]{babel}

\usepackage[letterpaper,top=2cm,bottom=2cm,left=3cm,right=3cm,marginparwidth=1.75cm]{geometry}

\usepackage{titling}
\usepackage{amsmath}
\usepackage{graphicx}
\usepackage[colorlinks=true, allcolors=blue]{hyperref}
\usepackage{placeins}
\usepackage{multirow}
\usepackage{authblk} 
\usepackage{lineno}
\usepackage{xcolor}
\usepackage{textcomp}
\usepackage{xspace}
\usepackage{url}
\usepackage{nicefrac}
\usepackage{subcaption}
\usepackage{comment}
\usepackage{microtype}
\usepackage{amssymb}

\newcommand{\um}{{\textmu}m\xspace}
\newcommand{\umsq}{{\textmu}m\textsuperscript{2}\xspace}
\newcommand{\us}{{\textmu}s\xspace}
\newcommand{\gevc}{GeV/${c}$\xspace}
\newcommand{\SR}{synchrotron radiation\xspace}
\newcommand{\BNLReportNumber}{BNL Formal Report\\BNL-229253-2025-FORE}

\pretitle{%
  \begin{flushright}\small \BNLReportNumber\end{flushright}%
  \vspace{5em}%
  \center\LARGE\bfseries
}
\posttitle{\par\vspace{0.5em}}

\preauthor{\center\large}
\postauthor{\par}

\predate{\center\large}
\postdate{\par}

\title{Closeout Report of BNL LDRD 23-050\\ \vspace{0.4cm}A Second EIC Detector: Physics Case and Conceptual Design\vspace{2cm}}

\author[1]{Jihee~Kim}
\author[1]{Cheuk-Ping~Wong}
\author[1]{Thomas~Ullrich}
\author[1]{Zhoudunmin~Tu}
\author[1]{Brian~Page}
\author[1]{Elke~Aschenauer}
\author[1]{Alexander~Jentsch}
\author[1]{Alexander~Bazilevsky}
\author[1]{Alexander~Kiselev}
\author[1]{Oleg~Kjeld~Eyser}
\author[1]{Xiaoxuan~Chu}
\author[1]{Zhengqiao~Zhang}
\author[1]{Evgeny~Shulga}
\author[1]{Akio~Ogawa}

\author[2]{Barak~Schmookler}
\author[3]{Ciprian~Gal}
\author[4]{Grzegorz~Kalicy}
\author[4]{Tanja~Horn}
\author[5]{Anselm~G.~Vossen}
\author[6]{Charles~Hyde}
\author[7]{Zuhal~Seyma~Demiroglu}

\affil[1]{Brookhaven National Laboratory, Upton, New York, USA}
\affil[2]{University of Houston, Houston, Texas, USA}
\affil[3]{Thomas Jefferson National Accelerator Facility, Newport News, Virginia, USA}
\affil[4]{The Catholic University of America, Washington, D.C., USA}
\affil[5]{Duke University, Durham, North Carolina, USA}
\affil[6]{Old Dominion University, Norfolk, Virginia, USA}
\affil[7]{Stony Brook University, Stony Brook, New York, USA}

\date{\today}

\begin{document}
\maketitle

\newpage
\tableofcontents
\setlength{\parskip}{6pt}
\newpage
\section{Introduction} 

\subsection{LDRD 23-050} 

This document is the closeout report for LDRD 23-050, a type-A LDRD project awarded in FY2022 under the title \emph{``A Second EIC Detector: Physics Case and Conceptual
Design''}. The project was motivated by the strong interest within the EIC community in a second general-purpose detector and interaction region, and by the recognition that such a detector is essential to fully exploit the scientific potential of the EIC over its multi-decade lifetime.

The key goals of the LDRD were to (i) strengthen the case for a second EIC detector,
building on the arguments already articulated in the community Yellow Report
\cite{bib:YR}, (ii) provide a realistic detector concept that is complementary to the current project detector, ePIC, in terms of physics reach, precision, and control of systematics, and
(iii) broaden the overall physics program of the EIC facility. Since a possible second detector is expected to be realized with a delay of several years relative to the first detector, the project explicitly aimed at identifying technologies that are not yet sufficiently mature for ePIC but could be deployed on the later timescale of a second detector, thereby providing genuine complementarity and room for innovation.

LDRD 23-050 provided support for two postdoctoral researchers and partial support for
several BNL staff members, and the effort ramped up in mid-2023. Work was carried out in close collaboration with colleagues from the EIC project, the Cold QCD and STAR groups at BNL, and with substantial input from members of the EIC User Group and the ePIC Collaboration as reflected in the author list. Over the period from October 2022 through September 2025, this team developed and refined a physics program tailored to a second detector, translated the resulting physics requirements into detector-level performance targets, and explored
multiple conceptual detector layouts capable of meeting these targets within the constraints of the second interaction region.

The expected outcome, as stated in the original proposal, was a document detailing the physics potential and requirements of a second EIC detector, accompanied by a
comprehensive conceptual detector design and an outline of the remaining R\&D needed to bring the relevant technologies to maturity. The present report summarizes the
progress that has been achieved toward these goals. It consolidates the physics studies, detector concepts, and technology assessments developed under this LDRD and places them in the broader context of worldwide detector R\&D. While the evolving priorities of the EIC project and the substantial community effort required to design and construct ePIC naturally limited the scope of EIC-wide activities toward a second detector during this period, the work documented here is intended to provide a foundation and a point of reference for future efforts. Whatever form a second EIC detector may ultimately take, we
hope that this report will serve as a useful guide for colleagues who continue to advance this program in the near- and mid-term future.

\subsection{Background and Motivation} 

The Electron-Ion Collider (EIC) was designated the highest-priority new construction project in the 2015 U.S. Nuclear Physics Long Range Plan~\cite{osti_1296778}, and its scientific case was strongly endorsed by a 2018 National Academy of Sciences review~\cite{NAP25171}. The EIC physics program was outlined in a community-led White Paper~\cite{Accardi:1498519}, while detailed detector requirements and potential technologies were presented in the comprehensive Yellow Report~\cite{ABDULKHALEK2022122447}. The resulting general-purpose detector, ePIC, is designed to support a broad physics program and is currently planned for one interaction point at IP-6. Although the EIC can accommodate two interaction regions (IR-6 and IR-8) and two interaction points (IP-6 and IP-8), only one detector is included in the current project scope. To realize a second EIC detector, a compelling scientific case must be developed to support national and international funding efforts.

The broader EIC community strongly supports the construction of a second detector, as outlined in a dedicated chapter of the Yellow Report~\cite{ABDULKHALEK2022122447}. The primary arguments are as follows:

\begin{description}
\item[Cross-Checking:] 
Independent measurements from two detectors help prevent erroneous conclusions caused by analysis errors, instrumental malfunctions, or statistical fluctuations—particularly when probing new phenomena. Nearly all modern colliders have employed at least two detectors for this reason. Experience from Fermilab and  CERN programs provides multiple examples where independent experiments rapidly cross-checked one another, uncovering issues and helping to avoid premature or incorrect claims (see Ref.~\cite{Grannis:2023vxf} for illustrative case studies).

\item[Cross-Calibration:] Complementary detector designs enable cross-calibration, reducing systematic uncertainties. Gains in precision arise not only from increased data statistics, but also from independent measurements that help identify and control detector-specific biases.
\item[Primary Physics Focus:] A second detector can be optimized for a different set of physics goals, expanding the overall scientific reach of the EIC. Complementary programs and unique scientific motivations strengthen the case for funding a second interaction region and detector.
\item[Technology Redundancy:] No single detector design can be guaranteed to perform flawlessly, and pre-operation environmental modeling has inherent limitations. Employing different technologies across two detectors mitigates risk and enhances the overall robustness of the EIC program.
\end{description}

This LDRD report outlines the physics potential and detector requirements for a second EIC detector. It presents a comprehensive conceptual design that meets these requirements and describes the proposed technologies. The report is accompanied by a detailed overview of relevant R\&D efforts, including technologies that may be pursued within the timeline of the second EIC detector project.
\section{New Physics Opportunities}
\label{sec:new-physics-opportunities}
As previously mentioned, having two general-purpose collider detectors to support the full EIC science program provides significant advantages and enhances complementarity. This setup allows the use of complementary detector technologies targeting similar physics goals, thereby enhancing scientific coverage, promoting innovation in detector design, and increasing robustness across measurements. Furthermore, variations in interaction-region designs and primary physics emphases between the detectors can help optimize sensitivity across the full range of the EIC physics program. The two-detector configuration also creates opportunities to explore potential facility upgrades and the
new scientific capabilities they may unlock.

Examples of such facility upgrades and their corresponding physics potential include:

\begin{description}
\item[Double-polarized eD collisions:] Transversely polarized electron--deuteron
  collisions provide a promising avenue to probe transversity distributions. Such measurements
  could offer new insights into the role of gluons in nuclear binding.

\item[Positron beams:] Incorporating positron beams alongside electron beams enables
  precise comparisons of electron--proton and positron--proton electroweak interactions. This
  allows fundamental studies of quark axial and vector couplings and significantly enhances the physics reach of exclusive measurements.

\item[Real photon beams:] Real photon beams produced via Compton scattering can be
  polarized and used for unique spectroscopy studies. These capabilities are particularly valuable for exploring the formation of new hadronic states, such as charmed mesons, and provide a complementary approach to investigations performed at LHCb and Belle~II.
\end{description}

Similarly, variations in interaction-region (IR) designs and primary physics focuses between the two detectors open up additional opportunities:

\begin{description}
\item[Secondary focus integrated into the IR:] Incorporating a secondary focus into the
  interaction region enables detection of particles with transverse momentum down to
  $p_T \sim 0$. This is particularly valuable for studying nuclear structure, as it provides access to low-momentum particles emitted from nuclei. Implementing this feature, however, requires extended auxiliary detector coverage and capabilities.

\item[Fixed targets integrated into the EIC:] The inclusion of fixed-target capabilities
  within the EIC framework offers complementary physics opportunities to those pursued by
  experiments such as STAR, LHCb, and ALICE. It provides access to high-$x$ physics in both
  $ep$ and $pp$ collisions. Careful consideration must be given to detector acceptance and
  kinematics for fixed-target running.
\end{description}

Given these possibilities, the second EIC detector should be designed not only to complement the first detector but also to provide unique physics opportunities that enhance the overall EIC science program. The following subsections highlight a selection of these opportunities in greater detail.

\subsection{Opportunities Through Alternative Technologies}
As discussed above, complementarity between the two EIC detectors can be achieved through the use of different detector technologies. Below we highlight a few illustrative examples, covering only a subset of the 36 subsystems included in the ePIC detector design:

\begin{description}
\item[Magnet:] A second detector could adopt an alternative solenoid design optimized for
  different physics priorities. For example, a stronger magnetic field and a larger inner bore radius than in ePIC (1.7~T solenoid) would improve momentum resolution, especially at high $p_T$, and provide additional flexibility for integrating large-volume tracking or calorimetry. Such a design would improve tracking precision, allow more space for detector services, and enable greater detector depth. However, this approach also introduces technical challenges and risks, including the complexity of manufacturing a higher-field, larger-radius solenoid.

\item[Tracker:] A complementary tracking system could incorporate alternative technologies,
  such as a gaseous detector (e.g.\ a time-projection chamber or drift chamber) combined with
  outer layers of silicon-based precision trackers. This configuration could enhance pattern
  recognition, improve tracking efficiency, and enable particle identification for low-momentum particles via dE/dx measurements.

\item[PID:] Particle identification (PID) systems could be optimized for different momentum regimes compared to ePIC. For example, the second detector might prioritize high-momentum PID in the forward region using a simplified RICH (Ring Imaging Cherenkov) detector. Additionally, a stronger investment in time-of-flight (TOF) technology—potentially targeting a time resolution of $\sim$10~ps—could significantly enhance PID performance in the low-intermediate momentum range.

\item[Barrel EMCAL:] While ePIC employs a hybrid imaging electromagnetic calorimeter combining silicon sensor layers with a Pb/SciFi section, the second detector could explore alternatives such as homogeneous crystal calorimeters—e.g., an improved version of the current SciGlass technology. However, such calorimeters generally require more longitudinal space to achieve good energy resolution for electromagnetic showers, particularly for electrons.

\item[Barrel HCAL:] Instead of a conventional hadronic calorimeter, the second detector could consider integrating a dedicated muon detection system in the barrel region. This approach would enhance muon identification capabilities and provide additional leverage in physics channels involving heavy flavor or electroweak final states.
\end{description}

This complementarity in detector technologies—despite the detectors serving similar experimental goals—plays a crucial role in cross-calibration. It helps reduce systematic uncertainties associated with a single detector configuration when combining data from both the first and second EIC detectors for the same measurement. 

\subsection{Physics Enabled by IR Design}
Specifically, the pre-conceptual design of IR-8 improves the acceptance of low transverse momentum protons and light nuclei in exclusive reactions at very low $t$, as well as nuclear breakup products in incoherent processes. This enhancement increases the potential for diffractive physics studies and tagging measurements at the EIC. This section presents a few examples of these opportunities.

\subsubsection{Interaction Region overview}
For the EIC physics program, the detector must provide broad acceptance for both charged and neutral particles, not only in the central region but also in the far-forward and far-backward regions. To enable measurements across a wide range of physics processes, specialized detectors are integrated into the interaction region lattice. The IR incorporates multiple tracking and calorimetry systems along the beamline to measure particle momenta and energies. 

\begin{figure*}[!hbt]
    \centering
    \includegraphics[width=\linewidth, height=3cm]{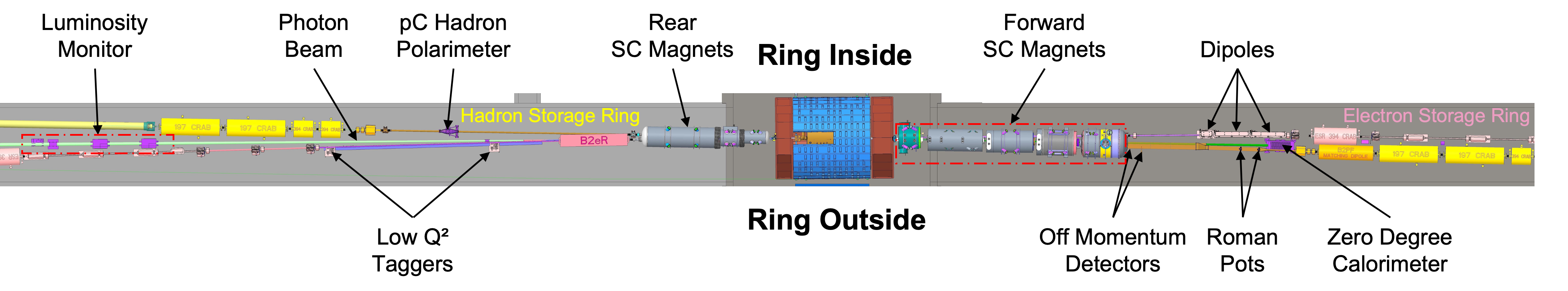}
    \caption{Full layout of the first interaction region (IR-6) featuring a 25~mrad crossing angle and the associated beamline instrumentation. Taken from~\cite{PhysRevD.111.072013}.}
    \label{fig:ir6}
\end{figure*}

The IR-6 and IR-8 interaction regions share luminosity between their respective detectors at the same center-of-mass energy and present similar advantages and challenges from the perspective of accelerator design. However, their configurations include distinct design features. IR-6, shown in Figure~\ref{fig:ir6}, with a 25~mrad crossing angle, enables transverse momentum ($p_T$) measurements in the range of 0.2--1.3~GeV and provides neutron acceptance up to a polar angle of approximately $\theta \sim 4.5^\circ$.

\begin{figure}[!hbt]
    \centering
    \includegraphics[width=1\linewidth, height=10cm]{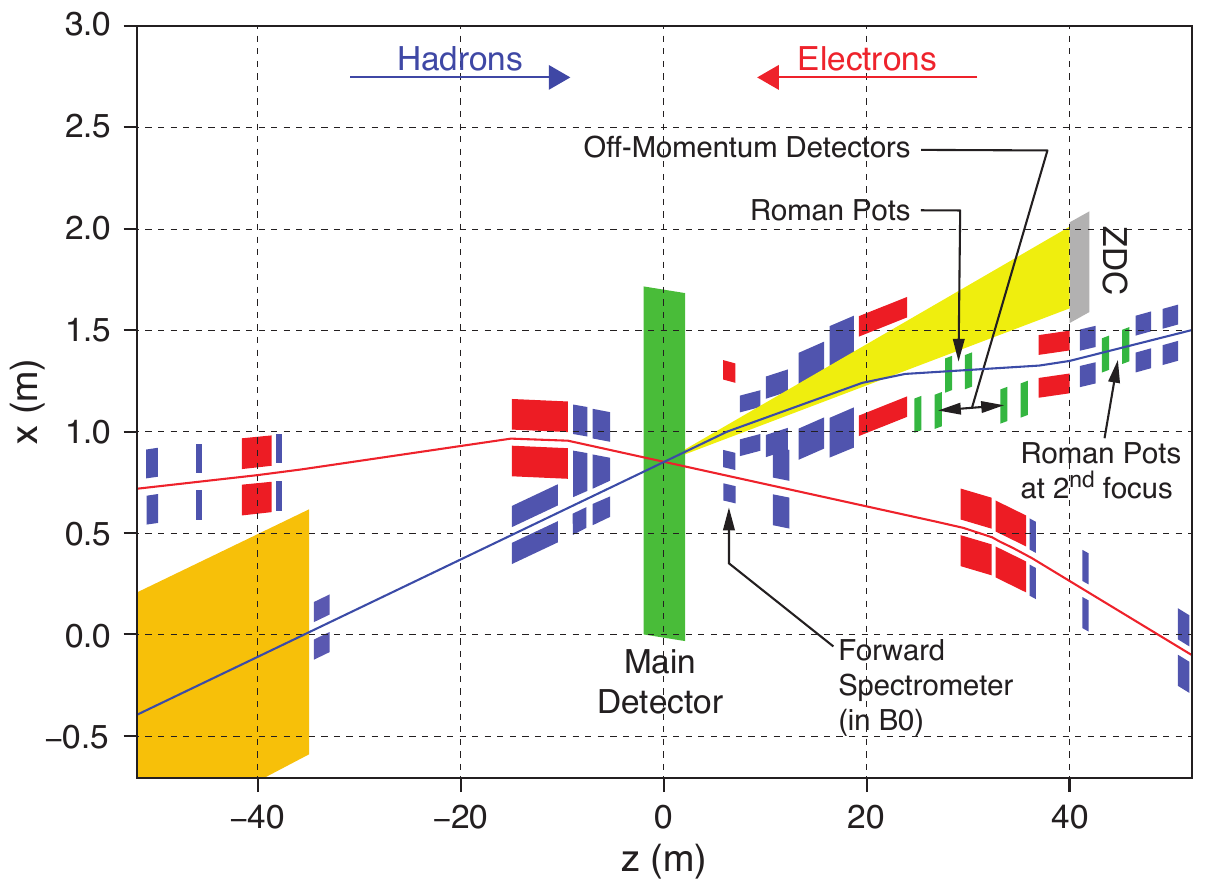}
    \caption{Schematic layout of the second interaction region (IR-8) with a 35~mrad crossing angle, illustrating the lattice design, potential detector locations, and the position of the secondary beam focus. Taken from~\cite{PhysRevD.111.072013}.}
    \label{fig:ir8}
\end{figure}

In contrast, IR-8, illustrated in Figure~\ref{fig:ir8}, is constrained by the geometry of the existing experimental hall and tunnel, leading to a larger optimal crossing angle of 35~mrad. This larger angle introduces different blind spots in pseudo-rapidity and makes it more challenging to achieve acceptance at high pseudo-rapidity values (e.g., $\eta \sim 3.5$) in the central detector. As a compensating feature, the proposed hadron beamline for IR-8 incorporates an optical configuration with a secondary focus approximately 45~m downstream of IP-8, achieved by placing additional dipole and quadrupole magnets in the lattice. This design enhances forward detection acceptance despite the larger crossing angle.

\subsubsection{Secondary Focus}
The concept of a secondary focus is to create a narrow beam profile in the transverse plane—similar to the focus at the interaction point—allowing the detection of particles scattered at very small angles (near $\sim 0$~mrad) and with minimal changes in magnetic rigidity. This feature benefits the second EIC detector by providing complementary capabilities to the first detector, particularly for exclusive, tagging, and diffractive physics programs.
 
Placing silicon detectors around the secondary focus enables them to approach closer to the beam core, significantly enhancing forward acceptance for scattered protons, ions from diffractive interactions, and nuclear fragments that would otherwise remain undetected due to their proximity to, or overlap with, the beam envelope. This configuration increases sensitivity to particles with low transverse momentum ($p_T <$ 200~MeV) and those exhibiting small longitudinal momentum loss (high $x_{L} \sim \frac{p_{\textrm{proton}}}{p_{\textrm{beam}}}$), which correspond to minimal magnetic rigidity loss and typically travel very close to the nominal beam orbit.

Crucially, improved detection of nuclear fragments enables better separation of coherent and incoherent diffractive events, a key requirement for the 3D imaging of nuclei. It may also allow for partial or full momentum reconstruction of the detected fragments, further advancing the EIC nuclear physics program. 

\subsubsection{Tagging Efficiency for the Incoherent Diffractive Subprocess}

\begin{figure}[t]
    \centering
    \includegraphics[width=0.8\linewidth, height=10cm]{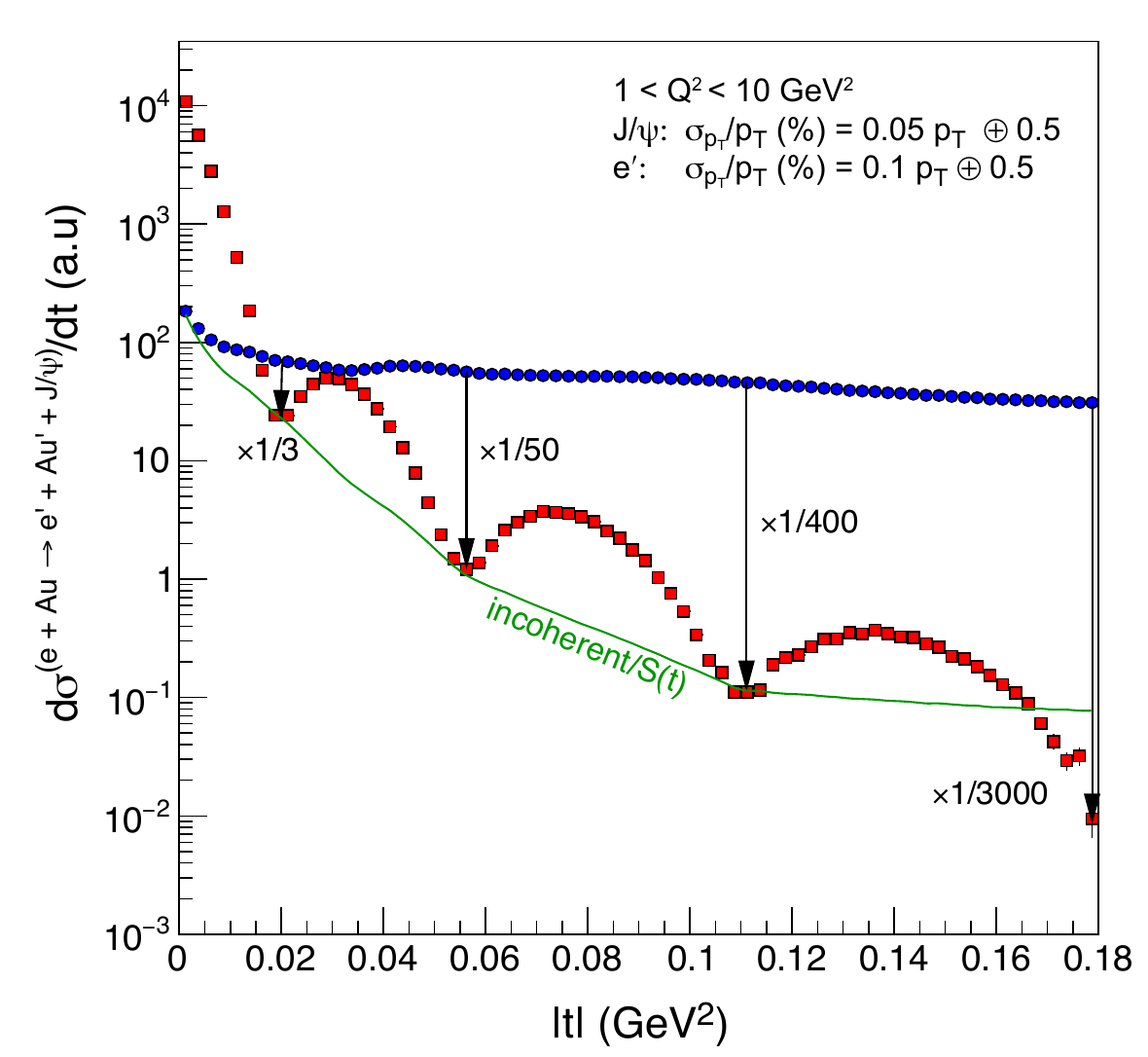}
    \caption{Coherent (red) and incoherent (blue) cross section for diffractive $J/\Psi$ vector meson production in e + Au collisions, assuming nominal resolution in the range  $1 < Q^{2} < 10$~GeV$^2$. Adapted from~\cite{ABDULKHALEK2022122447}.}
    \label{fig:eicyr_coherent_incoherent_cross_section}
\end{figure}

In the exclusive, tagging, and diffractive physics program at the Electron-Ion Collider (EIC), a key measurement is coherent diffractive vector meson production~\cite{Accardi:1498519}, which provides direct access to the spatial distribution of partons within nuclei. The vector meson is primarily produced at mid-rapidity, within the acceptance of the central detector, making its detection experimentally clean and accessible. In contrast, identifying the recoiling nucleus or its breakup products requires dedicated far-forward detectors positioned up to approximately 30~meters downstream along the hadron beamline.

Precise identification of the coherent subprocess component is essential because the observed diffractive vector meson production includes contributions from both coherent (nucleus remains intact) and incoherent (nucleus breaks up) processes. The incoherent process constitutes a dominant background, particularly at higher momentum transfers. For example, as shown in Fig.~\ref{fig:eicyr_coherent_incoherent_cross_section}, the incoherent contribution dominates for $|t|\geq0.015$~GeV$^2$, obscuring the diffractive pattern of the coherent signal.

However, tagging a coherent nuclear recoil is experimentally challenging, especially for heavier nuclei. A key strategy for isolating coherent events—up to the third diffractive minimum—is to tag and veto incoherent events with extremely high efficiency (greater than 99~\%). This requires detecting nuclear breakup fragments scattered at small angles near the beamline. Far-forward detectors are thus crucial for identifying these fragments, including both charged hadrons and neutral particles, to effectively separate coherent from incoherent diffractive processes.

In this study, we evaluate the potential of the current design of the EIC’s second interaction region (IR-8), with a particular focus on the role of the secondary focus in suppressing incoherent diffractive contributions. The simulations incorporate the necessary far-forward detectors adapted to the IR-8 hadron beamline layout and its current magnetic field configuration. Using the BeAGLE event generator~\cite{PhysRevD.106.012007}, we assess the capability to separate coherent and incoherent events, and quantify the improvement in separation power provided by tagging nuclear fragments at the secondary focus.

To study the veto efficiency for incoherent diffractive events, it is sufficient to detect activity from at least one nuclear breakup fragment in any single far-forward detector. The detailed veto conditions applied in the analysis are as follows:

\begin{itemize}
    \item Veto 1: Any registered hits in the ZDC hadronic calorimeter (HCAL)
    \item Veto 2: Hits in the Roman Pot (RP) station closest to the secondary focus, located outside a 10$\sigma$ safe-distance cut
    \item Veto 3: Hits in at least two layers of the Off-Momentum Detector (OMD)
    \item Veto 4: Hits in at least two out of the four layers of the B0 tracker
    \item Veto 5: Energy deposits exceeding 100~MeV in the B0 electromagnetic calorimeter (EMCal)
    \item Veto 6: Energy deposits exceeding 100~MeV in the ZDC EMCal
\end{itemize}

\begin{figure*}[]
    \centering
    \includegraphics[width=\linewidth]{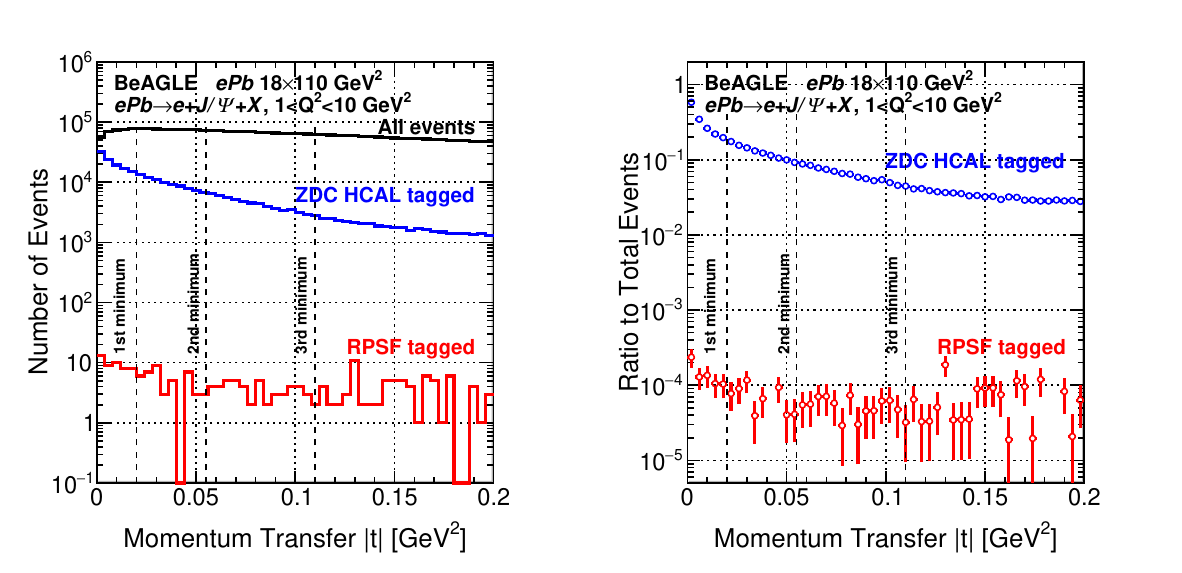}
    \caption{Left: Number of non-vetoed incoherent diffractive events in ePb collisions as a function of momentum transfer $t$. The black curve shows all incoherent events; the blue curve shows events surviving after ZDC tagging and vetoing; the red curve shows events surviving after both ZDC and Roman Pot (RP) tagging and vetoing. Right: Vetoing power as a function of $t$, where each curve represents the inefficiency histogram of a given veto selection normalized to the total incoherent sample. Only selections with significant impact are shown.  The term ``RPSF" refers to the Roman Pot located at the Secondary Focus. Adapted from~\cite{PhysRevD.111.072013}.}
    \label{fig:vetoing_power}
\end{figure*}

The left panel of Fig.\ref{fig:vetoing_power} shows the number of non-vetoed incoherent diffractive events in ePb collisions as a function of momentum transfer $t$, while the right panel shows the ratio of non-vetoed to total incoherent events. These plots highlight the tagging performance of the Zero Degree Calorimeter (ZDC) and the Roman Pot (RP) at the Secondary Focus (RPSF) as functions of $t$. As seen, a significant improvement in veto efficiency is achieved when combining ZDC and RPSF tagging. Unlike in the IR-6 design\cite{PhysRevD.104.114030}, where RP tagging of heavy nuclear fragments is limited, the IR-8 design allows far more effective tagging due to the smaller beam size at the secondary focus—comparable to that at the interaction point—enabling the RP to be positioned closer to the beam core. 

This enhanced tagging capability has a substantial impact on incoherent event suppression. According to the EIC Yellow Report~\cite{ABDULKHALEK2022122447}, a strong rejection factor is required to resolve the diffractive minima (“dips”) in coherent vector meson production, especially for $J/\Psi$. At the third diffractive minimum, the incoherent cross section exceeds the coherent one by approximately a factor of 400. Thus, a suppression factor greater than 400:1 is necessary to distinguish the coherent signal.

In the IR-6 configuration~\cite{PhysRevD.104.114030}, this level of suppression is sufficient only to resolve the first diffractive minimum. In contrast, the IR-8 design enables suppression of incoherent contributions at all three minima, achieving a vetoing power exceeding $10^{3}$ for the current geometry integrated in the simulation without implementing a beam pipe as well as no background and detector noise. Figure~\ref{fig:vetoing_power} demonstrates that tagging at the secondary focus yields superior veto performance across the full $t$-range, with rejection factors up to $10^{3}$. 

After all veto conditions are applied, the remaining non-vetoed events primarily consist of low-multiplicity nuclear fragments with high atomic mass (e.g., A = 208 and 207 for Pb). These fragments carry momenta very close to that of the beam and remain within the beam envelope, rendering them undetectable regardless of the IR design.

Additionally, we evaluated the impact of the beam pipe on the vetoing efficiency by implementing a simplified model based on the current IR-6 design, extending up to the neutron exit window of the ZDC. Although a reduction in neutron acceptance is observed, the overall effect on vetoing efficiency is minimal. This is primarily because the dominant contribution to the veto power originates from the tagging of heavy nuclear fragments at the secondary focus. Moreover, each incoherent event typically produces multiple neutrons and nuclear fragments, many of which can still be intercepted by the Roman Pot at the secondary focus. Consequently, the overall vetoing performance remains largely unaffected. Further details are provided in Ref.~\cite{PhysRevD.111.072013}.
    
\subsubsection{Isotopes}
    
Recent studies~\cite{Bertulani:2024mqe,CORE:2022rso} suggest that electron-nucleus collisions at the EIC will produce a large variety of nuclear fragments. These fragments will be produced from the disintegration of the excited, residual nucleus that remains following the initial electron-nucleus interaction and intranuclear cascade. Direct detection and identification of the nuclear fragments is possible in the far-forward region of IR2, provided that tracking and Cherenkov detectors are optimally placed relative to the location of the proposed secondary focus.

The physics motivations for detecting these nuclear fragments and associated gamma decays are as follows:

\begin{itemize}

\item Simultaneous measurements of the production cross sections for a large number of nuclear fragments using different ion beams and varying event kinematics will improve and test production models. This will lead to the improvement of the fast Abrasion-Fission model as well as a better understanding of the underlying reaction mechanism.
\item The high energy of the produced nuclear fragments (100~GeV/nucleon) will allow for the direct detection of fragments with lifetimes $>1$~ns. There is potential for the discovery of new neutron-deficient isotopes in the $Z = 89 - 94$ range. In this range, the EIC will have advantages over Rare Isotope Beam facilities due to differences in flight time and possibly higher production cross sections.
\item Many of the decay gammas will be Lorentz up-shifted to energies much larger than the background photons present in the detector area. This will allow for clean detection and identification of these gamma rays, which can be used to study the level-structure of the nuclear fragments.
\item Detection and identification of the nuclear fragments alongside a determination of the total neutron energy in the Zero-Degree Calorimeter (ZDC) will provide an event-by-event reconstruction of the atomic number and mass number of the excited, residual nucleus. This will constrain the parameters used in intranuclear cascade models (e.g. formation time, mass model)~\cite{Ferrari:1995cq} and provide an observable that correlates with event centrality.

\end{itemize}

Simulations of electron-heavy nucleus collisions at EIC energies using the \textit{BeAGLE} event generator~\cite{Chang:2022hkt} show that, to first approximation, the produced nuclear fragments will have the same momentum per nucleon ($p/A$) as the incoming ion beam and no angle relative to the beam. Under this approximation, the rigidity ($R = p/Z$) of the isotope is directly related to the ratio of its mass and atomic numbers ($A/Z$) as

\begin{equation}
R_{Rel} = (R - R_{beam})/R_{beam} = \left(\frac{A}{Z}\right) / \left(\frac{A_{beam}}{Z_{beam}}\right) - 1\text{ ,}
\end{equation} 
where $R_{Rel}$ is the relative rigidity of the outgoing nuclear fragment with respect to the incoming ion beam.

\begin{figure}[h!]	
	\centering
	\begin{subfigure}[b]{0.49\textwidth}
	    \centering
         \includegraphics[keepaspectratio=true,width=2.9in,page=1]{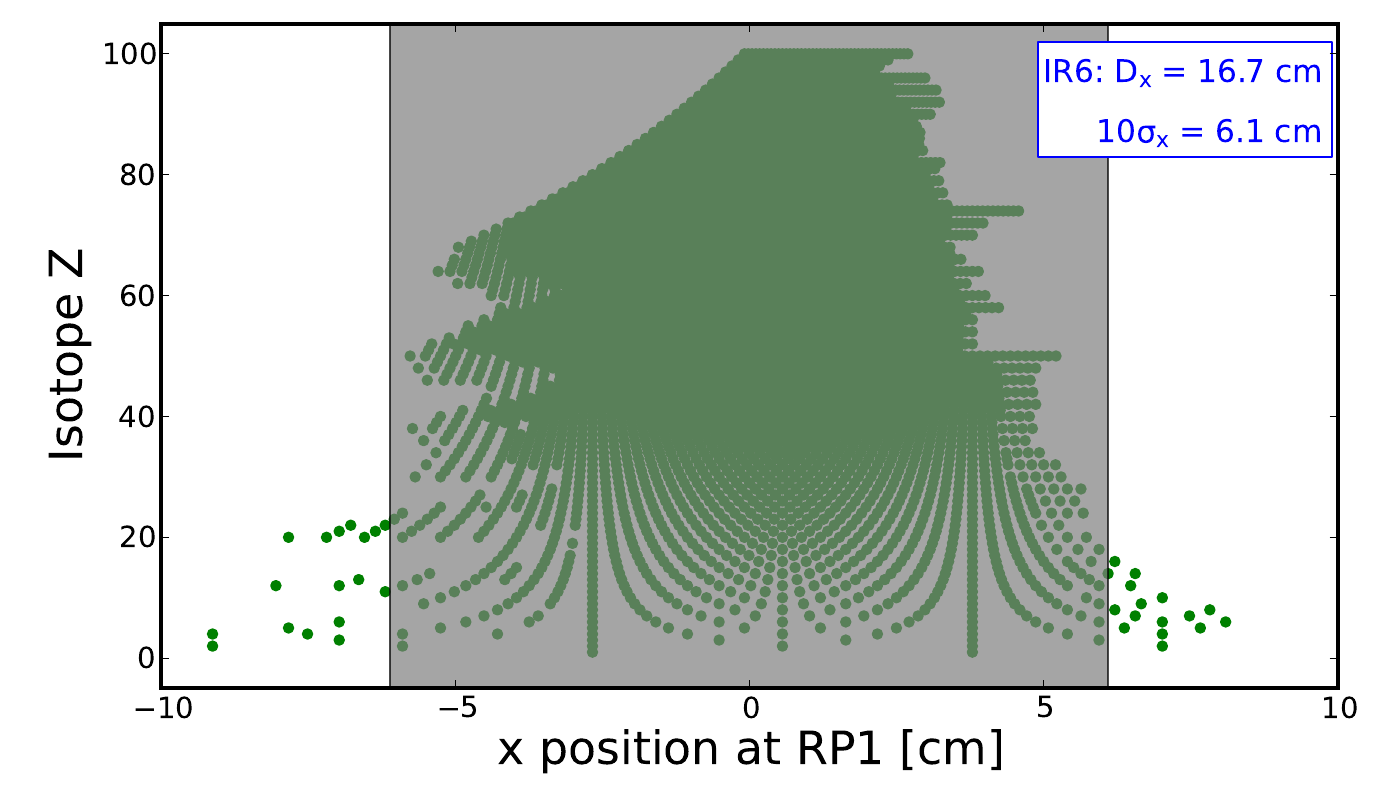}		
	\end{subfigure}
	\hfill
	\begin{subfigure}[b]{0.49\textwidth}
	    \centering
		\includegraphics[keepaspectratio=true,width=2.9in,page=1]{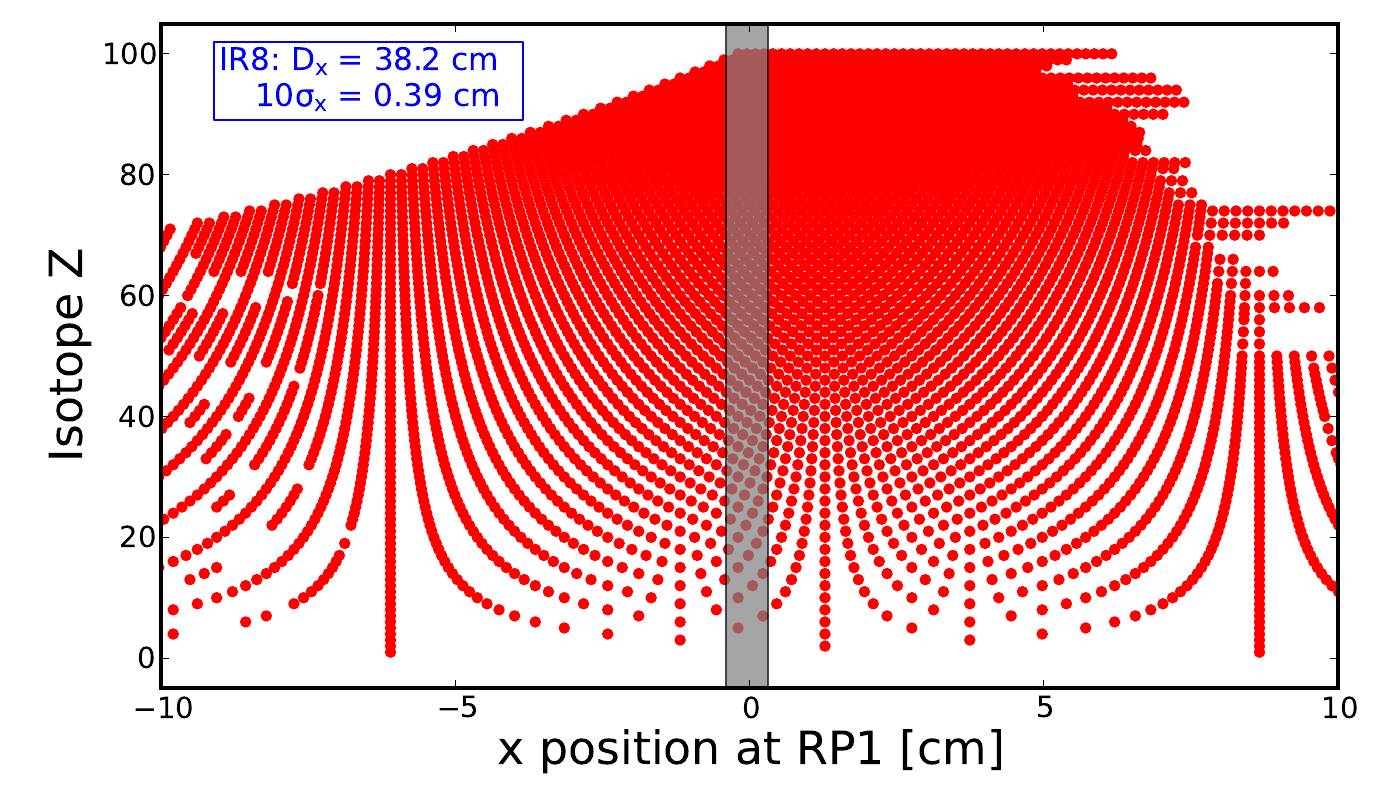}
	\end{subfigure}
	\caption{Left: Fragment $Z$ vs. hit position in the first RP for IR1. Right: Fragment $Z$ vs. hit position for a RP located near the secondary focus for IR2. The gray box on each plot shows the 10$\sigma$ beam exclusion area, using the beam parameters in table~3.5 of the 2021 EIC CDR~\cite{osti_1765663}. The plots are made assuming a $^{238}$U beam, with both known and predicted nuclear isotopes included as data points.}
	\label{fig:rigidity}
\end{figure}

Within this approximation, a nuclear fragment's hit location along the dispersive direction at a Roman Pot (RP) tracking detector provides a measurement of $A/Z$. Figure~\ref{fig:rigidity} shows the expected hit positions for both known and predicted nuclear isotopes at a RP detector in IR1 and at a RP detector located near the secondary focus of IR2, assuming a $^{238}$U beam. Nuclear fragments with the same $Z$ and different $A$ values are shown at the same vertical position in the plots. In addition, using the beam parameters taken from the 2021 EIC Conceptual Design Report (CDR)~\cite{osti_1765663} for a heavy-ion beam at 110~GeV/A colliding with an electron beam at 18~GeV, the 10$\sigma$ beam exclusion regions that set the acceptance limits for the RPs are shown by the gray boxes. As can be clearly seen, IR2 has the potential to detect fragments with rigidity values differing by only 1\% from that of the beam.  At the RP in IR2, fragments with the same $Z$ that differ by a single neutron are expected to be separated by 1.5~mm for $Z = 100$ and 5~mm for $Z = 25$.

In order to uniquely determine the atomic and mass numbers of the detected nuclear fragment, a direct measurement of $Z$ is needed. The simplest way to do this is by placing a Cherenkov detector behind the RPs near the secondary focus. The number of Cherenkov photons produced by the fragment will be proportional to $Z^2$.

\begin{figure}[h!]	
	\centering
    \includegraphics[keepaspectratio=true,width=0.7\textwidth]{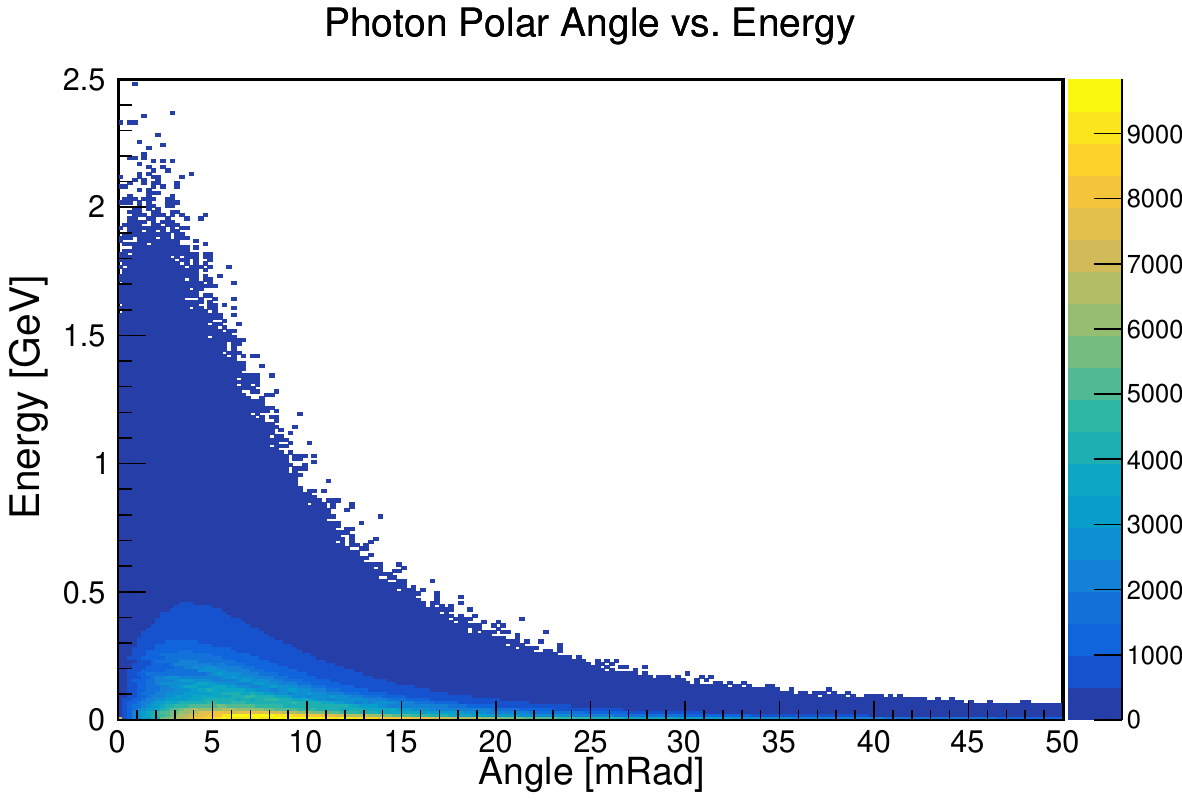}
	\caption{Energy vs. polar angle in the lab frame for photons produced using a \textit{BeAGLE} + \textit{FLUKA} simulation of a 110~GeV/A $^{238}$U beam on a 18~GeV electron beam. According to the simulation, the vast majority of the photons produced in the far-forward region originate from gamma decays of the produced nuclear fragments.}
	\label{fig:photon}
\end{figure}

Measuring decay gammas in coincidence with the detection of a nuclear fragment is also important as the level transitions reveal the structure of the nuclear fragment. The decay gammas are produced isotropically in the fragment's rest frame but can be Lorentz upshifted significantly in the lab frame. The gamma decay spectra in the lab frame is shown in figure~\ref{fig:photon} using a \textit{BeAGLE} + \textit{FLUKA} simulation. The shift of the decay gammas to higher energies in the lab frame, as well as the requirement that these photons be detected in coincidence with a nuclear fragment, means that photon background will be small. LYSO crystals that do not require cryogenics can therefore be used for this measurement. In addition, while spectroscopy would benefit from a good photon acceptance, it is not a critical requirement.

\subsection{Beyond the Standard Model}
\label{subsec:bsm}

The versatility and extraordinary luminosity reach of the EIC opens the door to studies of physics beyond the Standard Model (BSM). Recent theoretical analyses have shown that, with appropriate experimental capabilities, the EIC can provide constraints on BSM theories that are complementary to those from other planned facilities around the world. This is particularly important in light of the absence of new-particle discoveries at the LHC, which suggests that the new physics required to address outstanding issues within the Standard Model may lie at energy scales significantly higher than the electroweak scale.

In scenarios where the scale of new physics is much larger than the weak scale direct observation of new particles would be impossible. However, the effect of new heavy particles on SM observables would be possible with high precision measurements. This scenario is best analyzed within the Standard Model Effective Field Theory (SMEFT) framework. The extension to the SM is done through the inclusion of a Lagrangian of the form 
\begin{equation}
  \mathcal{L}_\mathrm{SMEFT} = \sum_\alpha c_\alpha \mathcal{O}_\alpha,
\end{equation}
where the operators $O_\alpha$ contain only SM fields and are typically of dimension greater than 4, so the couplings $c_\alpha$ (Wilson coefficients which encode the BSM contribution) are suppressed. Multiple groups have contributed to the development of a global SMEFT analysis of the existing world data~\cite{Ellis:2020unq,Ethier:2021bye}. From these analyses it has become clear that measurements at a single facility would not be able to adequately constrain the large parameter phase space leading to so called ``flat directions". This ambiguity can be only be disentangled through different types of measurements. The EIC with its ability to polarized both electrons and protons allow it to uniquely constrain certain Wilson coefficients (as has been shown in~\cite{Boughezal:2022pmb} for $C_{eu}$ and $C_{lu}$).

The SMEFT framework has been used further to analyze the potential impact of the EIC on Charged Lepton Flavor Violation (CLFV). CLFV is a consequence in many models that explain neutrino masses and while $\mu \leftrightarrow e$ have constraints on the level of $10^{-13}$, $e \leftrightarrow \tau$ conversions have weaker constraints on the level of $10^{-8}$. The work in~\cite{Cirigliano:2021img} has shown that the EIC is as constraining as the LHC, and together complementary to low energy measurements planned at Belle II. Experimentally the challenge lies in $\tau$ tagging. Initial studies have shown that both the 3-pion final state~\cite{Zhang:2022zuz} and the single muon~\cite{Banerjee:2022xuw} final state of the $\tau$ decay have the potential to provide a relatively low background measurement. Moreover, the EIC can constrain lepton number violation models by unambiguously measuring the charge of the~tau. 

Another set of models that can provide CLFV relies on the existence of flavor violating axion like particles (ALPs). The study in \cite{Davoudiasl:2021mjy} has shown that the EIC is most competitive in the region of phase space with heavy BSM particles (light particles are better constrained by the fixed target measurements) and which have distinctive experimental signatures with very low backgrounds. Additionally, the study relies on eA scattering to take advantage of the $Z^2$ enhancement in coherent scattering imposing a limit on the scale of the interaction $Q^2\leq (100{\rm\ MeV})^2$. Here again, experimentally the challenge is the detection of the resulting $\tau$ lepton. However, more interestingly is the need to detect it in pseudorapidity regions that are relatively background free, such as the far backward region.

Strengthening the case for detection in the far backward regions ($-6<\eta<-4$), studies~\cite{Davoudiasl:2025rpn} have shown that the EIC can have a significant impact on searches of dark vector bosons. Similarly to the ALP studies discussed above the production is enhanced through the detection in coherent $eA$ collisions. The key experimental capability to be able to make the measurement feasible would be tracking in the far backward region far away from the interaction region ($z\approx 5$~m) in order to distinguish displaced vertexes.  

The EIC will deliver measurements of nucleon and nuclear structure functions across a wide ($x$,~$Q^2$) range using both neutral current and charged current DIS. Inclusive observables $F_2$, $F_L$, $F_3$ (and polarized $g_1$, $g_5$) test QCD and separate electroweak/flavor combinations. The scaling violations and longitudinal structure $F_L$ probe small-$x$ dynamics and, using multiple $\sqrt{s}$ measurements the EIC can address H1-ZEUS tension in $F_L$~\cite{Jimenez-Lopez:2024hpj}. The inclusive program leverages the EIC's high luminosity, polarization, and nuclear beams, while acknowledging limitations such as smaller center of mass energy and the absence of positrons. 

These measurements then feed global fits where EIC data are combined with existing hadron, lepton, and neutrino results using modern PDF tools that handle correlated systematics, nuclear corrections and heavy quark schemes in uniform way. Within this program, charged current DIS with charm tagging gives direct sensitivity to strange PDF and helps separate $s$ from $\bar s$; together with world data it reduces sea and gluon PDF uncertainties, improves $\alpha_s$ and clarify light-flavor asymmetries $(\bar d - \bar u;s - \bar s)$. On nuclei, $eA$ measurements map modifications from shadowing to the EMC region, separate coherent and incoherent channels~\cite{Olness:CFNSwkshop}. Combining proton, light-ion, and heavy-ion data over a wide ($x$,~$Q^2$)  range reduce the nuclear correction uncertainties that currently limit precision predictions for neutrino and photon initiated processes. The main path has a clear flow: precise structure function measurements $\to$ global fits with consistent theory $\to$ robust flavor decomposition and $\alpha_s$ $\to$ better inputs for electroweak and BSM studies.

All the proposed measurements and studies have a consistent theme that is complementary to the ePIC detector. Detector capabilities such as muon detection, $\tau$ tagging, far backward detection, open windows to new tests of BSM theories. For these searches to be efficient we need to look in areas where the SM signatures are minimized, necessitating a different detector design than the base EIC detector.

\section{Lessons Learned from the ePIC Detector}
\subsection{Silicon-Focused tracking}
\label{sec:ePICTracking}

Precision tracking is essential in EIC experiments. Beyond charged particle tracking, it also provides vertexing for event characterization and heavy-flavor particle reconstruction, as well as angular measurements for PID detectors. Tracking at the EIC presents unique challenges. First, the tracking system must be robust against backgrounds, particularly those induced by beam-related processes, such as synchrotron radiation. Second, the EIC imposes stringent requirements on transverse momentum ($p_T$) resolution. This is especially important in the backward region ($\eta<-3$), where the excellent $p_T$ resolution is critical for accurate scattered electron measurements. 

This subsection analyzes the ePIC tracking system design based on a series of simulation studies, focusing on its limitations and potential improvements to meet the EIC’s demanding tracking requirements.

\paragraph{The ePIC Tracking System Configuration}

\begin{figure}[h]
    \centering
    \includegraphics[width=0.52\textwidth]{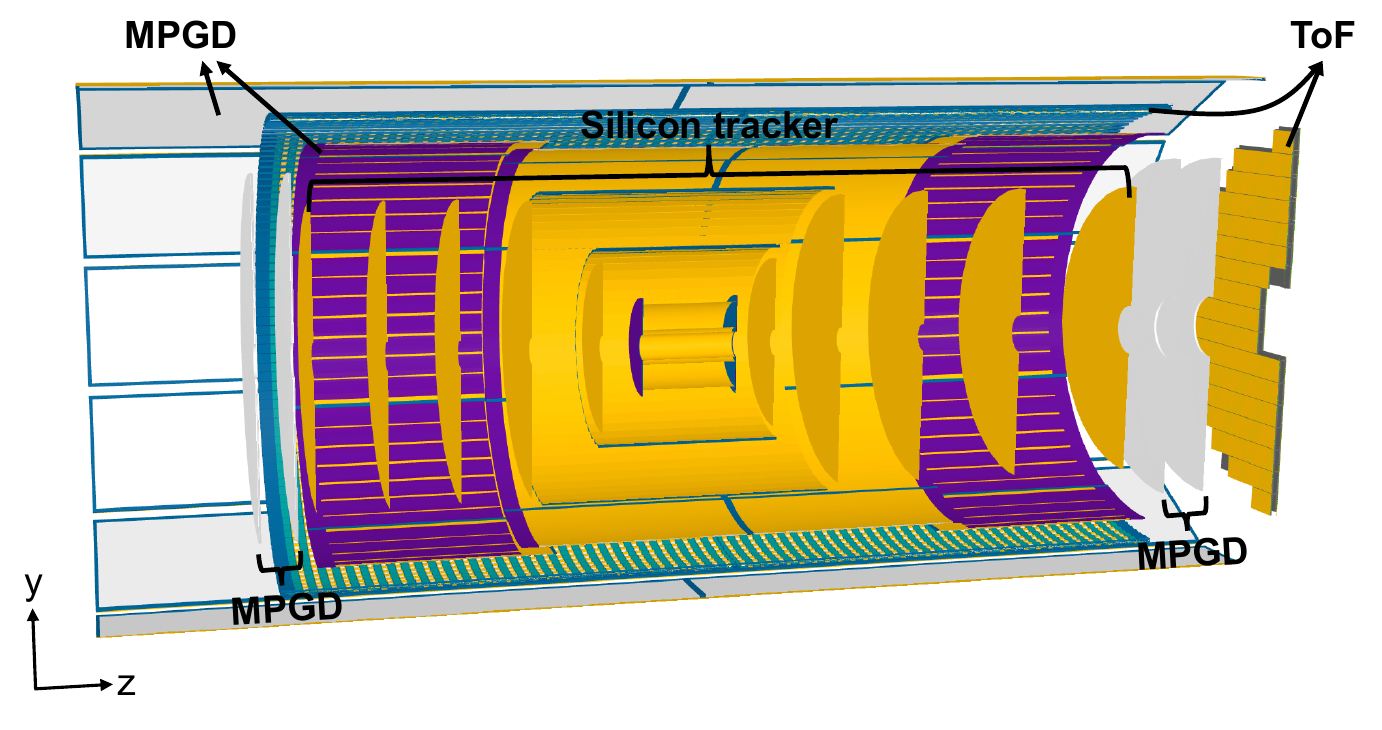}
    \caption{The ePIC tracking system. The backward direction points to the left.}
    \label{fig:ePICtracking}
\end{figure}

\begin{figure}[h]
    \centering
    \includegraphics[width=0.5\textwidth]{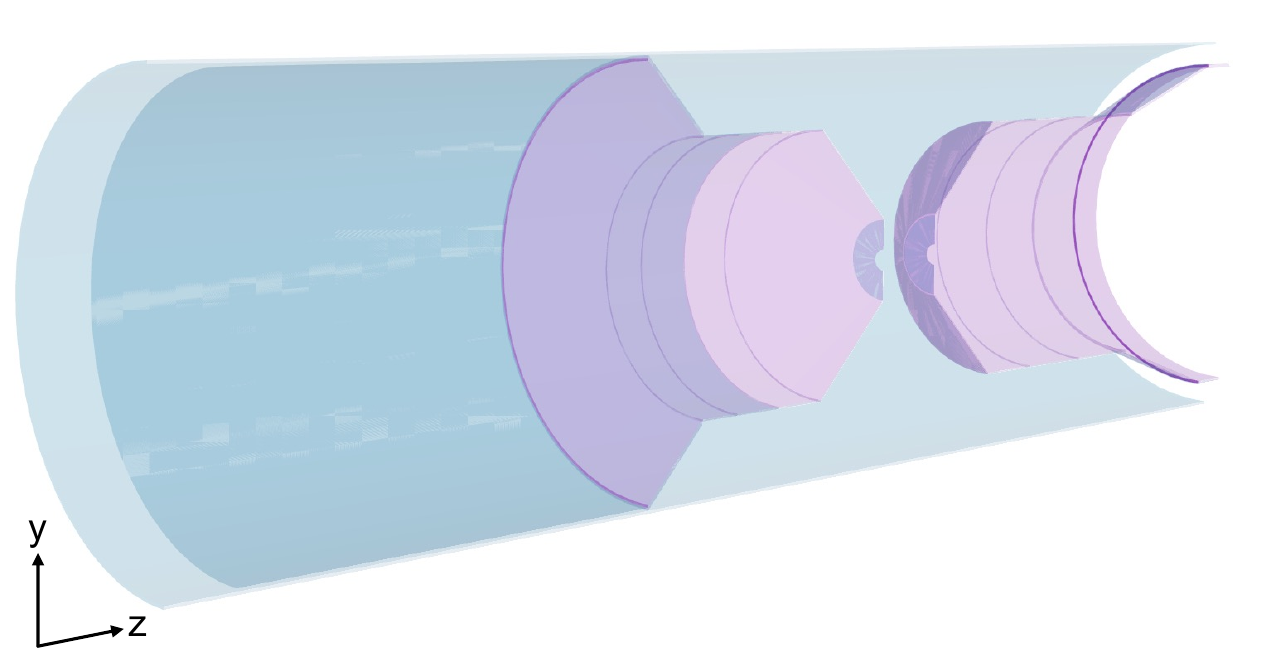}
    \caption{Support structure of the ePIC tracking system.}
    \label{fig:ePICtracking_supportCone}
\end{figure}

The ePIC (version 2023.10.0~\cite{bib:ePIC_github}) tracking system and its support structure, depicted in Fig~\ref{fig:ePICtracking} and \ref{fig:ePICtracking_supportCone} respectively, is immersed in a $1.7$~T magnetic field and consists of an inner silicon tracker, Micro-Pattern Gaseous Detectors (MPGD), and AC-coupled LGAD time-of-flight detectors (TOF) in the outer region.

The silicon tracker is composed of three parts: five silicon barrel layers and five silicon disks in both the backward and forward directions. Each silicon layer or disk includes a $40$~\um-thick silicon wafer with a pixel size of $20\times20$~\umsq. The barrel layers span radii from $3.6$~cm to $42$~cm, while the disks have outer radii between $24$~cm and $42$~cm. The backward disks are position between $z=-105$~cm and $z=-25$~cm, and the forward disks between $z=25$~cm and $z=135$~cm. 

The innermost three layers of the silicon are designed for vertexing and utilize the ITS3 curved wafer-scale sensor with a material budget of $0.05$\%. The outer two layers and all the disks employ the ITS3-based large-area sensors with a stave design, which requires a $150$~\um aluminum foil and a $120$~\um carbon fiber support structure. The material budget is $0.55$\% for each of the outer two layers and $0.24$\% for each disks.

Located in the outer barrel are the AC-coupled LGAD TOF and two layers of MPGDs utilizing Micromegas and hybrid GEM-$\mu$RWELL technologies. The hybrid GEM-$\mu$RWELL MPGD is also employed in the backward and forward regions. Positioned behind the forward GEM-$\mu$RWELL MPGD is the AC-coupled LGAD TOF. Together, these outer tracking subsystems extend the ePIC tracking system to an outer radius of approximately $70$~cm, and longitudinal coverage form $z=-120$~cm to $z=188$~cm. The spatial resolutions of the MPGD and TOF are $150$~\um and $30$\um, respectively. 

\paragraph{Limited number of hits for tracking}
The ePIC tracking design emphasizes the silicon tracker, which occupies the majority of the tracking volume. The silicon-focused design enables high precisions in tracking and vertexing due to the fine spatial resolution of the silicon sensors. However, the high cost of silicon sensor limits the number of silicon layers and disk, resulting in a relatively low number of tracking hits per track. The ePIC tracking system provides five to nine tracking hits per track. While five or more hits are generally sufficient for pattern recognition in track reconstruction, certain measurements, such as $\Lambda^0\rightarrow p+\pi^{-}$, which have a decay length of several tens of centimeters in the central pseudorapidity region, may lack hits in the innermost layers of the silicon tracker~\cite{bib:YR,bib:EIC_lambda}. This can degrade the tracking performance for the decay products ($p$ and $\pi^{-}$), and hinder the reconstruction of $\Lambda^0$. Moreover, the tracking performance can be significantly impacted if any section of the tracking system becomes non-operational.


\paragraph{Susceptible to Background}
While the bunch crossing at the EIC is expected every $10$--$12$~ns, the integration window of the silicon tracker of ePIC is much longer at $2$~\us. Although possibility of event pile up is low at the EIC, the long integration window makes the tracker susceptible to continuous synchrotron radiation background. This background consists of X-ray photons in keV range, produced as a results of the bending of the electron beam by the beam magnets. Synchrotron radiation photons interact with the detector primarily via photoelectric effect and Compton scattering, generating electrons that contribute to background noise. The intensity of these photons depends on both the beam current and power. For $ep$ collisions at $18\times275$~GeV with an electron beam current of $0.277$~A, synchrotron radiation photons are produced at a rate on the order of $10^{18}$ photons per second. While a gold coating on the central beam pipe can significantly reduce the number of photons reaching the detector, it is estimated that more than $6500$ photons per $2$~\us integration window will still survive and reach the tracking detector. Therefore, a shorter integration window is particularly important for reducing synchrotron radiation background.


\paragraph{Challenge in backward tracking}
Another challenge for ePIC tracking design is meeting the stringent transverse momentum resolution requirements in the backward region for scattered electron measurements in exclusive events. The Yellow Report~\cite{bib:YR} (Fig.~8.85) defines the requirement for the transverse momentum resolution of scattered electrons in diffractive measurements to be
\begin{align}
    \sigma\left(\frac{\Delta p_{T}}{p_{T,MC}}\right)=\sigma\left(\frac{p_{T,RE}-p_{T,MC}}{p_{T,MC}}\right)\leq0.1p_{T,MC}\oplus0.5 \text{ }(\%) \text{ ,}
    \label{eq:res}
\end{align}
where $p_{T,MC}$ and $p_{T,RE}$ are the generated and reconstructed transverse momenta of the scattered electron in GeV/$c$, respectively. The symbol $\sigma$ represents the root-mean-square (RMS) of the relative difference between the reconstructed and generated transverse momenta.

To evaluate potential improvements to the ePIC tracking design for meeting the Yellow report tracking requirements in the backward region, six modifications to the detector setup were tested through simulations studies. These changes were applied cumulatively in the following order:
\begin{enumerate}
    \item Placing the detector in vacuum instead of air,
    \item Increasing the magnetic field from $1.7$~T to $2$~T,
    \item Removing the support cone of the tracking detectors,
    \item Removing the service parts (carbon fiber and aluminum layers) behind each silicon disk to mimic the ALICE ITS3 sensor~\cite{ALICEITS3},
    \item Adding a sixth silicon disk at $z=-2$~m. The backward MPGD, pfRICH and electromagnetic calorimeter are removed in this setup to give room for the additional silicon disk,
    \item Removing the beam pipe.
\end{enumerate}

\begin{figure}[h]
    \centering
    \includegraphics[width=\textwidth]{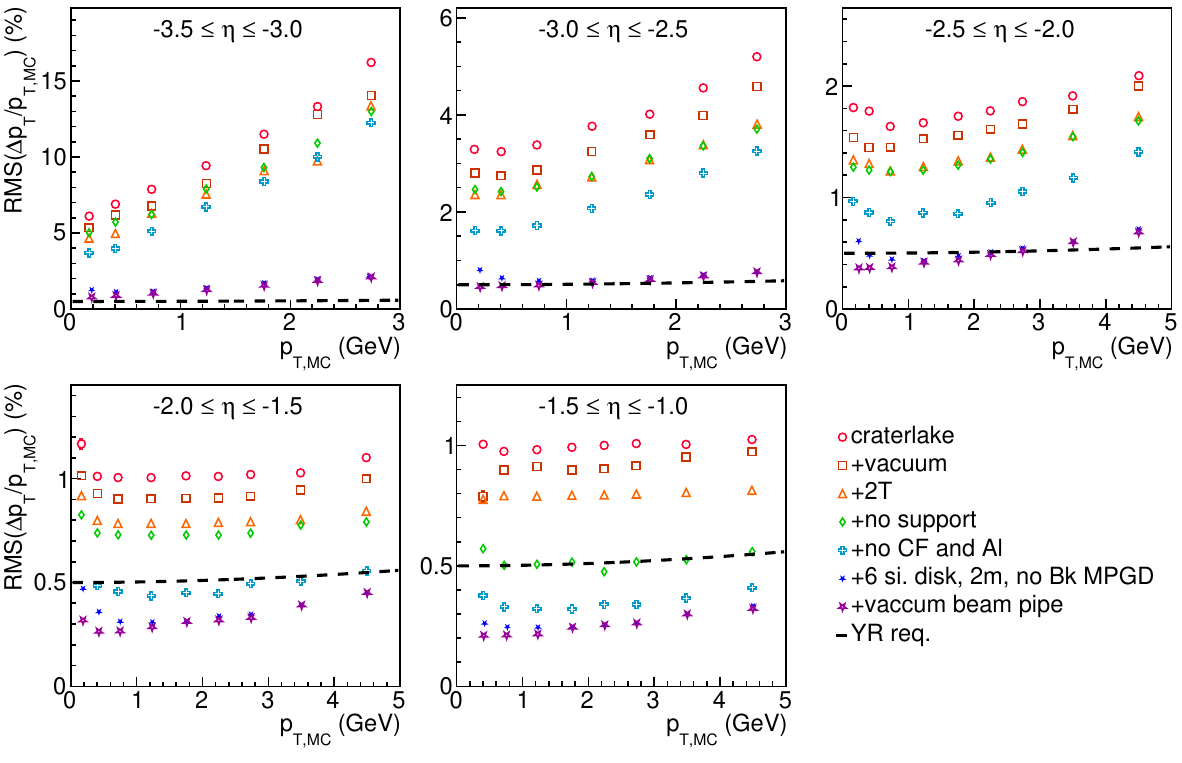}
    \caption{Transverse momentum resolution of electrons in the backward region for various detector configurations. The dash line indicates the Yellow Report requirement for diffractive measurements in exclusive events.}
    \label{fig:ptRes}
\end{figure}

Figure~\ref{fig:ptRes} shows the improvement in transverse momentum resolution for single electron under various detector setup modifications. Introducing a vacuum environment reduces the material budget effect, as longer traversal paths between disks result in thicker layers of air. Scaling the magnetic field from $1.7$~T to $2$~T improves the resolutions by a factor of approximately $1.176$, consistent with the inverse relationship between magnetic field strength momentum resolution. A significant improvement is observed in the range $-1.5<\eta<-1$ when the support cone is removed. Removing the carbon fiber and aluminum layers of each silicon disk further improves the resolutions, particularly at $\eta>-3$, where the electron traverse a longer path inside the layers. Adding the sixth silicon disk at $z=-2$~m improves the resolution across a broad pseudorapidity range. with even more pronounced gains in the more backward region. Finally, reducing the beam pipe material by setting it to vacuum leads to better transverse momentum resolution at low $p_T$ over a wide pseudorapidity range.

As demonstrated in these simulations, meeting the Yellow Report tracking requirement with an silicon-focused tracker remains challenging at the pseudorapidity  range ${-3.5<\eta<-3}$, even when employing the ALICE ITS3-type silicon sensors with stitching technology and assuming idealized vacuum beam pipe. Potential approaches to further improve the resolution of an silicon-focused tracker design include extending tracker length beyond $-2$~m and reducing pixel size. However, increasing the tracker length may compromise the available space for other sub-detectors. The ALICE ITS3 sensor is expected achieve a pixel pitch as small as ${10\times10}$~\umsq which could improve the resolution by up to a factor of four compared to the current ePIC silicon tracker design.

As an alternative to the all-silicon tracking design, a mixed-technologies approach, such as an inner silicon tracker combined with an outer gaseous detector, typically results in a thicker material budget. However, this approach offers more hits per track, which enhance tracking performance through increased hit redundancy. This approach also complements the ePIC tracking system by using different tracking technologies.

\FloatBarrier
\subsection{Complexity of the dRICH}
\label{ePIC-dRICH}

A dual-radiator RICH (dRICH) concept used in ePIC, is an elegant solution providing a continuous Cherenkov PID coverage in a wide momentum range, with a $\pi$/K separation on at least a 3$\sigma$ level up to $\sim$50~GeV/c momenta. However, it comes at a cost, in particular because one has to use the same photon detector matrix for both radiators. Since aerogel is in general not transparent below $\sim$300~nm, one cannot use photon detectors sensitive in a deep UV range, where a majority of Cherenkov photons from a gas radiator is produced. Therefore, e.g. use of a robust, low material budget and cheap MPGD technology with a CsI photocathode is excluded. A photon detector matrix needs to be taken out of acceptance, into the area with a strong solenoid fringe field, oriented at a large angle to the photosensor plane. Therefore, not only a conventional dynode PMT, but also a Micro-Channel Plate PMT (MCP-PMT) usage is excluded, and Silicon Photomultipliers (SiPMs) become the only feasible option. SiPMs have a high dark count rate (DCR), dozens of kHz per cm$^2$ at best. They need to be cooled down to below -20~C in order to suppress the DCR to an acceptable level, but also to decrease aging. Their use in ePIC will require regular in-situ annealing process, yet the performance is expected to degrade over time, also increasing a burden to the Data Acquisition System (DAQ) since noise hits are indistinguishable from single Cherenkov photons by any other means but timing and cannot be easily rejected on an FPGA level in real time. Besides this, placing a photodetector plane outside of acceptance results in a so-called off-axis optical configuration, which causes substantial spherical aberrations for Cherenkov photons produced in a gas radiator. This effect ultimately limits a Cherenkov angle resolution, and as a consequence limits an achievable momentum range.

An alternative PID solution for a hadron-going endcap can be a use of two independent RICH detectors in the same acceptance. One of them would be an aerogel-based high-performance proximity focusing RICH (hpRICH) with High-Rate Picosecond Photodetectors (HRPPDs) as photosensors, 
see section \ref{hpRICH}. Such a proximity focusing RICH would then be complemented by a single-radiator compact gaseous RICH detector with an MPGD/CsI based UV-sensitive readout placed in the same acceptance. This gaseous RICH would then have optimal on-axis focusing configuration, potentially higher Cherenkov photon count per track, and therefore either extend a 3$\sigma$ $\pi$/K separation  momentum reach beyond 50~GeV/c or increase a hadron sample purity at the same momentum or both. 

Using a pair of such proximity focusing  hpRICH detectors in both electron- and hadron-going endcaps would also allow one to improve time of flight performance of the EIC detector overall, since a majority of tracks detected in the endcaps will have timing measurement accuracy order of $\sim$10~ps due to the use of HRPPD photosensors. 

\subsection{A magnetic spectrometer for acceptance between the main detector and the Roman pots} \label{sec:B0}

ePIC and IP6 have been designed in part to ensure strong capabilities for exclusive physics. When probing at moderately high-$x$ ($> 0.1$) with lower EIC beam energy configurations many far-forward final states have scattering angles large enough to be outside the standard acceptance region for the Roman pots and Zero-Degree calorimeter while still staying outside the central detector fiducial acceptance. The solution to this gap between the central and far-forward regions in IP6 was to purpose the first hadron dipole magnet after the IP as both a steering magnet and a magnetic tracking spectrometer. The magnet in question in IP6 is called ``bending magnet 0" due to its being a dipole and the first magnet on the outgoing hadron side, and it is called ``B0" for short. In the ePIC design, AC-LGAD silicon technology is used to instrument the spectrometer, while $PbWO_{4}$ crystals are used to instrument an electromagnetic calorimeter. Pixilated AC-LGADs with 500~\um pitch can provide spatial resolution as good as $\sim20$~\um (using charge sharing to push performance beyond normal standards) and also provide timing resolution $\sim$30~ps, enabling 4D tracking. In ePIC, the spatial and timing resolution of the current ePIC AC-LGADs meet the requirements for the B0 tracking system.

In IP6 the B0 magnet itself proves to be a challenging component to build due to the necessary large aperture to allow full acceptance of the tracking system (up to 20~mrad in polar scattering angle) and the sharing of bore space between hadron and electron beams, while ensuring that the field is shaped in such a way to avoid the presence of a dipole field on the electron beam as it approaches the IP. This magnet has to also allow for all of the necessary services and support structures for the tracking and calorimetry subsystems.

In the case IP8, other options should be explored. One option would be to consider a tracking system with more layers and smaller pixels to further improve the transverse momentum resolution (limited to around 7\% in the IP6 B0 implementation). This option would require a different approach for photon detection likely using a silicon pre-shower, which could also double as an extension of the tracking system. Another option to consider would be a different machine element implementation and doing away with the spectrometer option all together. Instead of a large-bore dipole, a smaller bore dipole could be installed closer to the IP to avoid overlap with the electron quadrupole on the opposite side, and then either tracking layers or an imaging HCAL (ePIC SiPM-on-tile, silicon + Pb-absorber sampling and imaging calorimeter) could be placed where the tracker currently sits in the IP6 configuration, but without the acceptance limitations imposed by the magnet support structure. This, in addition to a different beam pipe size to change the angular acceptance gap between the Roman pots and the IP8 B0 could make for a nice complementarity in both fiducial acceptance and technology between IP6 and IP8. Significant work would be required by machine experts to come up with a feasible lattice solution.

In addition to a different magnet and layout as a possible option for the IP8 B0 detector, different silicon technology could be employed for the tracking sub-system. Technologies that would allow for some increase in capabilities would ones which provide the timing capabilities of present ePIC AC-LGADs, but with increased precision in space. This would imply either smaller pixels on present AC-LGADs with a potentially new, denser readout (high power consumption a potential challenge), or advancement of monolithic AC-LGAD concepts. All of this requires R\&D resources to advance and represent a challenge for deployment, but is perhaps feasible on the second detector timescale.

\subsection{Needs for faster timing}

Complementing traditional detector systems by a high-resolution timing capability on a level of dozens of ps (4D tracking, 5D calorimetry) is a noticeable modern trend in High Energy and Nuclear Physics experiments, in addition to installing dedicated high resolution timing layers. EIC second detector can pursue both options. High resolution timing capability allows one to reduce adverse effects of out-of-bunch backgrounds in the event reconstruction, extend the range of low momenta particle identification using time-of-flight technique, as well as (similar to ePIC detector) improve a $t_0$ reference estimate in the experiment on event-per-event level. Several options can be considered on the technology side: LGAD sensors (used in ATLAS HGTD and CMS ETL upgrades), LYSO:Ce crystals (CMS BTL upgrade) or MCP-PMTs (either as dedicated compact timing layers or as part of respective RICH detectors).

\subsection{Space for services}
Since the experimental halls for interaction points 6 (ePIC) and 8 (the second detector) are inherited from STAR and sPHENIX, respectively, the space constraints are predetermined. The hermetic requirements of an EIC detector—demanding tracking, particle identification, and calorimetry in all directions—further complicate the design. To proactively address these space constraints during detector development, the following strategies are recommended:
\begin{itemize}
    \item Involve engineers at the earliest stages.
    \item Define the detector envelope early.
    \item Incorporate electronics, services, and cooling into the detector design from the start.
    \item Decouple detector and electronics whenever possible.
    \item Pursue unified digitization and cooling solutions across subsystems where feasible.
\end{itemize}

\subsection{Software (flexibility, ACTS)}

The software framework for the EIC second detector is being developed with strong synergy to the ePIC software ecosystem. This approach allows reuse of many proven components—including the event data model, reconstruction framework, and geometry description tools—ensuring consistency across EIC efforts and reducing the overhead for new development. Leveraging the ePIC infrastructure provides an efficient starting point for detector concept studies while maintaining compatibility with existing analysis workflows. The current code repository for the EIC second detector can be found here, \url{https://github.com/eic/D2EIC}.

A central part of the framework is DD4hep (Detector Description for High Energy Physics), a unified toolkit for describing detector geometries and conditions across the full experiment lifecycle—from simulation and reconstruction to visualization and alignment. DD4hep acts as a high-level wrapper around GEANT4, which handles detailed particle transport and interaction physics. In this structure, DD4hep manages the materials, geometry hierarchy, and sensitive elements, providing a coherent interface for both simulation and reconstruction applications.

While DD4hep and GEANT4 together provide a robust environment for detailed detector studies, their complexity makes them less convenient for rapid prototyping and iterative optimization. The process of defining detailed geometries, compiling configurations, and running full GEANT4 simulations introduces significant turnaround time, which can slow down the early stages of detector concept development.

To complement this, the EIC second detector team has integrated fast simulation frameworks that enable quick evaluations of detector performance. These include: Delphes, a modular and widely used fast-simulation package that parameterizes detector responses (such as tracking, calorimetry, and particle identification) based on user-defined performance inputs. It allows for quick physics performance studies across a wide range of detector configurations.
ACTS-based tools, leveraging the A Common Tracking Software (ACTS) framework, which provides a modern, geometry-aware tracking toolkit. When coupled with simplified material maps or smearing models, ACTS enables fast tracking studies that maintain realistic kinematic and resolution behavior consistent with full simulations.

This hybrid approach—combining DD4hep/GEANT4 for detailed simulations and Delphes/ACTS for fast evaluations—allows flexible and efficient exploration of the detector design space. It ensures that early physics studies and subsystem optimization can proceed rapidly, while maintaining a clear migration path toward detailed simulations and reconstruction validation. Current Effort can be summarized as follows, 

\begin{itemize}
    \item Framework base: Built on ePIC software infrastructure to ensure compatibility and reuse of established components.
    \item Geometry description: Implemented using DD4hep as the unified interface to GEANT4 for full simulation studies.
    \item Fast simulation: Incorporates both Delphes and ACTS-based tools to enable rapid detector and physics performance studies.
    \item Integration: Ongoing work connects fast-simulation outputs with reconstruction and analysis workflows for consistent validation.
\end{itemize}

Here are a few recommendations for future development on the software of the second detector:
\begin{itemize}
    \item Dual-tier simulation strategy: Maintain both detailed (DD4hep-based) and simplified (fast-simulation) modes, ensuring a consistent geometry interface so that fast studies can easily transition to full simulation.
    \item Simplified geometry tools: Develop higher-level configuration layers or parameterized geometry interfaces that allow non-expert users to modify detector concepts without deep DD4hep knowledge.
    \item Unified data model: Ensure that data formats from fast and full simulations remain compatible to streamline validation and comparison studies.
    \item 
    Benchmarking framework: Establish a standard set of performance benchmarks (e.g., tracking efficiency, momentum resolution, PID performance) to evaluate detector concepts consistently across different software setups.
    \item Community engagement: Coordinate with the broader EIC software community to align developments, avoid duplication, and share tools that improve usability and documentation.
\end{itemize}

The EIC second detector software has advanced significantly by combining established full-simulation frameworks with flexible fast-simulation capabilities. DD4hep provides the backbone for detailed GEANT4-based detector modeling, while Delphes and ACTS-based tools enable rapid physics and design optimization studies. Moving forward, emphasis on modularity, user accessibility, and consistency across software layers will be critical to support both efficient R$\&$D and long-term integration within the broader EIC software ecosystem.

\section{Second Detector Concepts}
\subsection{Overview}

Our studies in the context of this LDRD focused only on \emph{general-purpose} detector concepts that can satisfy the full set of physics requirements across central and forward/backward regions, while remaining practical to integrate, operate, and maintain. The aim is balanced performance in tracking, particle identification (PID), calorimetry, and muon identification, rather than optimizing narrowly for a few single channels.

A primary boundary condition is the geometry of the IR and the associated integration envelope. The available radial and longitudinal space, the proximity of machine elements, and service routing (power, cooling, and data) constrain detector dimensions and technology choices. Within this envelope, the \emph{solenoidal magnet}, the practical only choice for a magnet not affecting the electron beam, is the dominant global trade: field strength, bore radius, and length must be co-optimized with tracking and PID. Higher fields improve momentum resolution and pattern recognition at high~\(p_T\), but they also increase track curling and effectively reduce acceptance for low-\(p_T\) particles—precisely where several PID techniques have leverage. Experience from ePIC further shows that an undersized magnet leaves too little room for the services of inner detectors, PID, and calorimetry; accordingly, the magnet envelope is being sized with explicit integration margin.

For the \emph{main tracker}, we have surveyed multiple concepts. Options include a gaseous core---a time projection chamber (TPC) or drift chambers---for high hit multiplicity, robust pattern recognition, and \(dE/dx\) information, potentially \emph{augmented by scintillating-fiber (SciFi) layers} to sharpen momentum resolution and provide timing handles. Hybrid concepts that pair an \emph{inner MAPS silicon} system with an \emph{outer TPC} can meet central-region resolution targets while maintaining resilience to background; reduced-dimension TPC variants remain viable if compactness is prioritized. The effects of background such as beam-gas and synchrotron radiation is of concern for large volume gaseous detectors. 

PID coverage is treated as a coupled optimization with the magnet and tracking envelope. We retain flexibility among \emph{time-of-flight} (e.g., AC-LGAD-class timing), \emph{DIRC/RICH} approaches, and \emph{TPC-based \(dE/dx\) and cluster counting}, mapping each technique to the momentum/rapidity bands it serves best and to the available radii. 

For \emph{muon identification}, \emph{instrumented return-yoke} concepts in the barrel and \emph{dedicated muon stations} in strategic regions need to be considered. Newer layouts that decouple muon timing and shielding from calorimetry are under study to improve purity without excessive mass.

Throughout the effort, we emphasized \emph{technology readiness} and synergy with \emph{ongoing R\&D} at neighboring experiments, and opportunities to reuse or adapt proven designs. This reduces risk and schedule pressure while preserving a clear \emph{upgrade path} as machine parameters and physics priorities sharpen. As detailed later, the current lack of dedicated generic R\&D program for the EIC is a serious issue.

\subsection{Magnet}
The magnet is essential for charged-particle tracking. The strength of the magnetic field is inversely proportional to the momentum resolution and sets the limit for $p_T$ coverage. From an engineering perspective, one must consider the operating temperature, current, and stored energy of a superconducting magnet, as well as its cooling system. This section focuses on the superconducting solenoid magnet, which is widely used in high-energy collider experiments.

\begin{figure}
    \centering
    \includegraphics[width=0.5\linewidth,angle=-90]{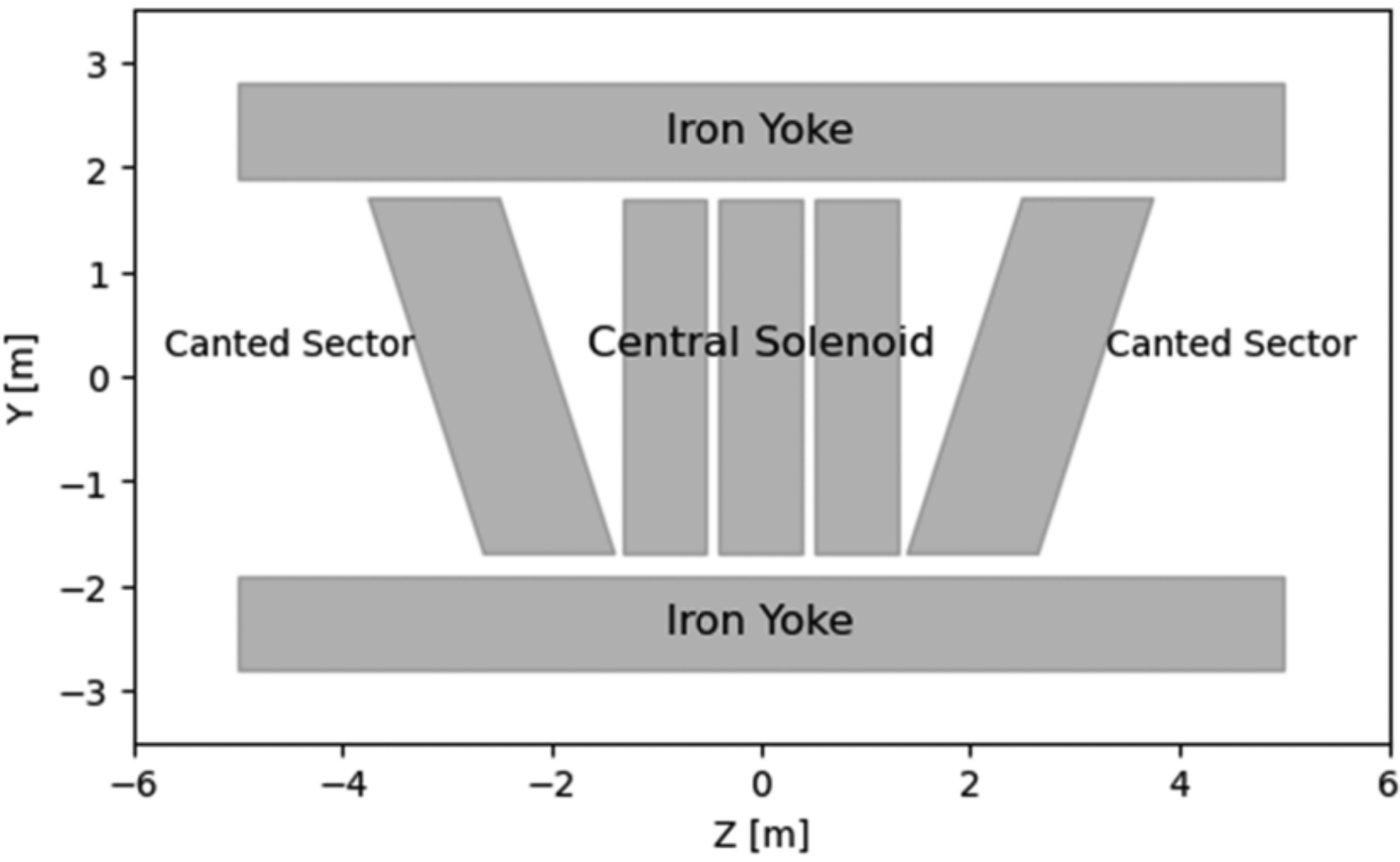}
    \caption{Concept of a canted magnet for the ALICE3 upgrade~\cite{bib:ALICE3magnet}.}
    \label{fig:ALICE3magnet}
\end{figure}

Superconducting solenoid magnets provide a relatively even longitudinal magnetic field, typically ranging from 0.25~T to 1.4~T~\cite{bib:STAR_magnet2,bib:STAR_magnet,bib:sPHENIX_TPOT,bib:ALICEdet2000}. While a stronger magnetic field improves tracking resolution at high $p_T$, it reduces the tracking efficiency for low $p_T$ particles and makes the system more susceptible to soft backgrounds spiraling inside the detector. For reference, the EIC detector design requires $p_T$ coverage down to $0.1$~\gevc~\cite{bib:YR}. This tradeoff can be addressed by the novel canted magnet design proposed for the ALICE3 upgrade~\cite{bib:ALICE3magnet}, illustrated in Fig.~\ref{fig:ALICE3magnet}. This design incorporates a standard solenoid coil at the center to provide a longitudinal magnetic field, along with canted coils at each end that superimpose dipole fields. The canted coils are tilted in opposite directions so that the dipole fields are in opposite directions, avoiding the need for beam correction. This hybrid longitudinal-dipole configuration reduces the longitudinal field in the central region for low $p_T$ coverage while enhancing tracking performance in the backward and forward regions.

The dimensions of the magnet must also be carefully considered. A larger the bore size (the magnet diameter) is generally preferred, as it alleviates spatial constraints and allows the magnet to be positioned at the outer layer of the detector. This placement minimizes passive material and radiation length in front of the inner subsystems. However, production cost increases with the magnet length, radius squared, and field strength squared. Conversely, a smaller radius reduces the stored magnetic energy quadratically, lowering operational costs.

To ease the production and operational costs of superconducting magnets, the IDEA magnet~\cite{bib:FCC_IDEA_1,bib:FCC_IDEA_2} design proposed a ``small and thin'' magnet using high temperature superconductors (HTS) that can operate at 20~K. This compact design allows the magnet to fit between the particle-identification detector and the calorimeters. It requires a high-yield-strength superconductor for self-support and a thin ($\sim25$~mm), highly thermally conductive aluminum layer to reduce passive material and overall material budget. Operating at 20~K rather than 4~K as in conventional superconducting magnets, reduces cryogenic cooling costs by a factor of five. The high operating temperature also allows higher current densities, reducing the amount of conductor and support material required. 

\begin{figure}[h]
    \centering
    \includegraphics[width=0.5\linewidth]{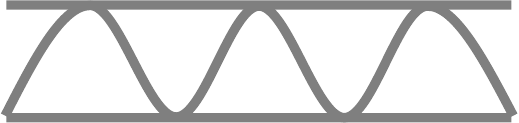}
    \caption{Illustration of the sandwich with corrugated insert cryostat design.}
    \label{fig:IDEAmagnet_corrugatedSandwich}
\end{figure}

The IDEA design also emphasizes a low-material-budget cryostat, employing a metallic-vacuum structure such as carbon-fiber honeycomb or a sandwich design with corrugated insert illustrated in Fig.~\ref{fig:IDEAmagnet_corrugatedSandwich}. Current IDEA magnet and cryostat designs are about $30$~cm thick, corresponding to a material budget of 74\% at normal incident. This can be reduced by an additional 14\% with a carbon-fiber honeycomb cryostat. Challenges for this design include the need to revitalize the commercial production of aluminum-stabilized Nb-Ti cables.

\subsection{Tracking}
\label{sec:PingTracking}
Section~\ref{sec:ePICTracking} discussed the advantages and disadvantages of a silicon-focused tracking design. While the silicon-focused approach offers excellent spatial precision, it lacks hit redundancy and operates with a long integration window, making it less robust against background. To address these limitations, this section explores alternative tracking technologies that offers complementarity capabilities to the ePIC tracking design.

\subsubsection{Gaseous Detector}
Two prominent high hit redundancy technologies are gaseous detectors: the Time Projection Chamber (TPC) and the Drift Chamber (DC). These detectors have been widely used in major collider experiments, including ALICE~\cite{bib:ALICE_TPC}, BELLE~II~\cite{bib:BELLEII_detReport}, STAR~\cite{bib:STAR_iTPC_upgrade}, and sPHENIX~\cite{bib:sPHENIX_ACTS_tracking}, and are also proposed for future experiments, such as the FCC~\cite{bib:FCC_IDEA_1} and the BELLE II upgrade~\cite{bib:BELLE_II_upgrades}. The tracking systems in these experiments typically combine inner silicon layers, which provide fine spatial resolution, with an outer gaseous detector that features a low material budget and provides a large number of hits on the order of $O(10)$ to $O(100)$. In addition to tracking, gaseous detectors also offer particle identification at low momentum below a few GeV/$c$. Section~\ref{sec:PID} will discuss this capability in more detail.

\paragraph{Drift chamber, carbon fiber wires}
The latest R\&D on DC, such as for the IDEA design at the FCC~\cite{bib:FCC_IDEA_1}, focuses on reducing material budget, enhancing uniformity of the equipotential surface, and advancing particle identification capabilities. The IDEA DC is a four-meter-long detector with inner and outer radii of $0.35$~m and $2$~m, respectively. It consists of $112$~layers and over $34$~thousands wires. Despite its large volume, the material budget is minimized to 1.6\% in the barrel region and $5$\% in the forward and backward regions to reduce multiple scattering, photon conversions and hadronic interactions. This low material budget is achieved through the advanced wiring techniques developed for MEG~II, as well as the use of a lightweight gas mixture of helium and isobutane (He:iC\textsubscript{4}H\textsubscript{10}) in a $9:1$ ratio, which provides a typical drift velocity of $2.2$~cm/\us. 

In addition to its long drift distance, the IDEA DC is designed to utilize fast readout electronics with a sampling rate on the order of GHz, enabling cluster timing resolutions below a few tens of nanoseconds. This allows precise determination of the impact parameter between a track and a sense wire by leveraging weighted information from all detected clusters, leading to improved spatial resolution~\cite{bib:clusterCounting}. With excellent cluster timing, The IDEA DC is expected to achieve a transverse spatial resolution of $100$~\um.

\begin{figure}[h]
    \centering
    \includegraphics[width=0.8\linewidth]{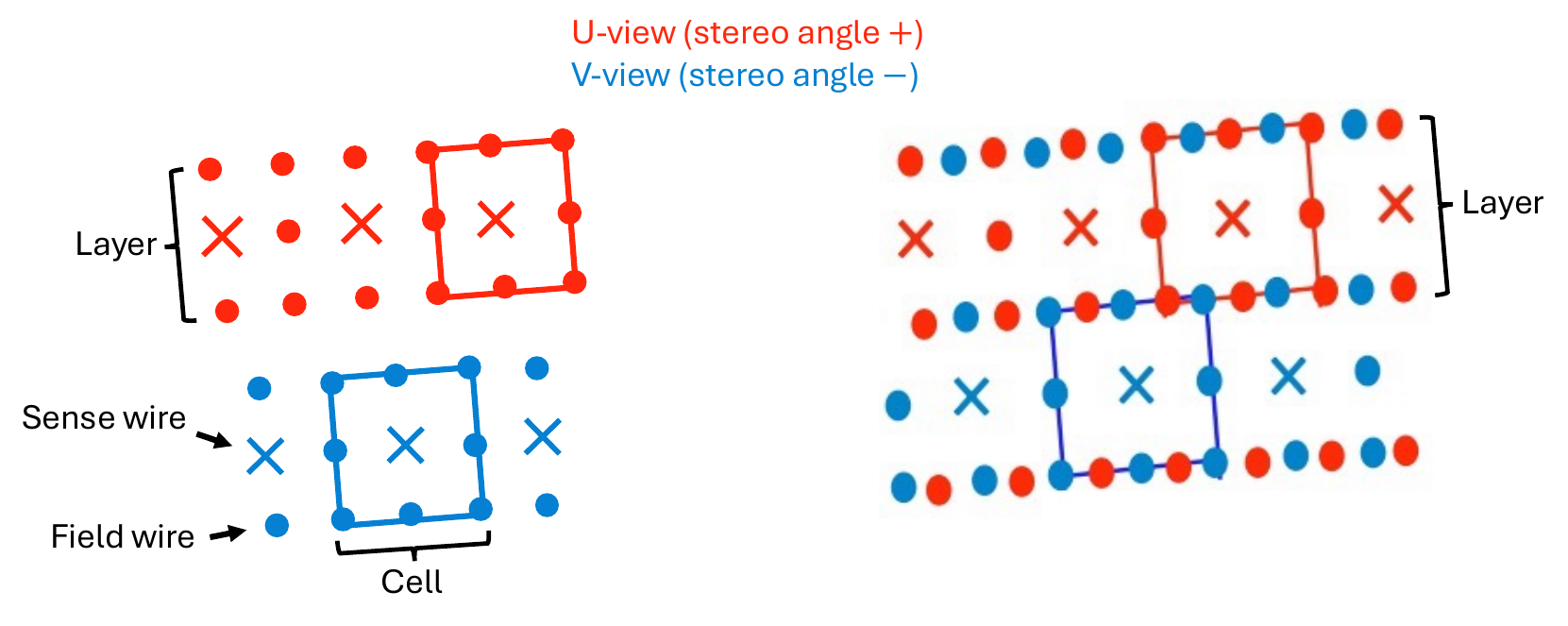}
    \caption{Left: Typical wire arrangement in a drift chamber, where field wires in the same layer share the same stereo angle. Right: Enmeshed wire configuration in the IDEA drift chamber, featuring alternating stereo angles for field wires and a higher number of field wires per cell~\cite{bib:FCC_IDEA_1}.}
    \label{fig:IDEA_DC_wireArrangement}
\end{figure}

A high field-to-sense wires ratio is achieved by alternating the wire orientations at positive and negative stereo angles, as illustrated in Fig.~\ref{fig:IDEA_DC_wireArrangement}. This configuration allows for the use of finer field wires, with a diameter of $40$--$50$~\um. In addition to reducing the material budget, this wire arrangement improves both spatial granularity and uniformity of the electric field. 

The cell size $w$ of the IDEA DC ranges from $12$ to $14.5$~mm, resulting in drift time of $400$~ns or less. The cell size, along with the choice of wire material, is constrained by the the wire tension $T$ which must lies between the limits imposed by electrostatic stability and mechanical elasticity. This relationship is expressed as
\begin{align}
\text{Electrostatic stability condition}&\leq T \leq\text{Elastic limit condition}\nonumber\\
\frac{C^2V^2}{4\pi\epsilon}\cdot\frac{L^2}{w^2}&\leq T \leq YTS\cdot\pi r^2\text{ ,}
\end{align}
where $C$ is the capacitance per unit length, $V$ is the voltage between anode and cathode wires, $L=4$~m is the length of the DC, $YTS$ is the yield strength of the wire material, and $r$ is the wire radius. To enable a fine cell size with thin wires while maintaining electrostatic stability, high-yield-strength materials are required.


Traditional materials, such as copper or aluminum, while excellent conductors, add substantial mass. Carbon fiber (CF), by contrast, offers an outstanding strength-to-weight ratio, excellent thermal stability, and very low radiation length, making it highly attractive for detector mechanical supports. However, its electrical conductivity is several orders of magnitude lower than metals, and most importantly, it easily creates discharge due to the rough surface with micro hair structure.

Currently, there is a separate but related R\&D for studying the CF wire. The goal of this R\&D is to combine the mechanical advantages of carbon fiber with the electrical and reflective properties of thin metallic coatings. By depositing a nanometer- to micrometer-scale metallic layer—such as silver, aluminum, or copper—onto CF filaments or wires, it may be possible to create multifunctional materials that serve as both mechanical support and electrical conductor, while keeping the total material budget minimal. Such developments could impact multiple detector technologies, including: low-mass support and grounding structures in silicon trackers, conductive wire supports for thin foils or gaseous detectors, lightweight biasing or signal paths integrated into composite components, and  radiation-tolerant alternatives for cables and connectors.

The R\&D effort, conducted at BNL and Stony Brook University, focuses on developing and characterizing metal-coated carbon fiber wires produced using a high-vacuum evaporation system. The current setup employs an electron-beam evaporator that allows controlled deposition of thin metallic layers under high vacuum, ensuring clean and uniform coatings.

\textbf{Substrate preparation.} Commercial carbon fiber wires or bundles (typically 5–10~\um filament diameter) are cleaned and mounted onto a copper plate with rotatable holders inside the evaporator chamber. Surface treatment (ion source) is often applied to improve adhesion between the carbon surface and the metallic film, as carbon’s graphitic surface tends to be chemically inert. However, in this project, the current vacuum quality cannot sustain both ion source and the electron source at the same time. A dedicated cryo pump has been prepared to be installed to improve this aspect. 

\textbf{Deposition Process} The coating material—initially silver (Ag) for its high conductivity and favorable deposition properties—is evaporated from a crucible or filament under vacuum conditions ($~10^-6$ torr). The carbon fiber is rotated or translated during deposition to ensure uniform coverage. Film thickness can be tuned by deposition rate and time, typically targeting 50–500~nm for initial tests. The process parameters of interest include: 1) Deposition rate (controlled via quartz-crystal microbalance (QCM) feedback). 2) Substrate temperature during coating and surface activation methods for improved adhesion with Ion Source(not available now but will be in the future). 3) Post-deposition thermal or mechanical stress testing.

\begin{figure}[h]
    \centering
    \includegraphics[width=0.99\linewidth]{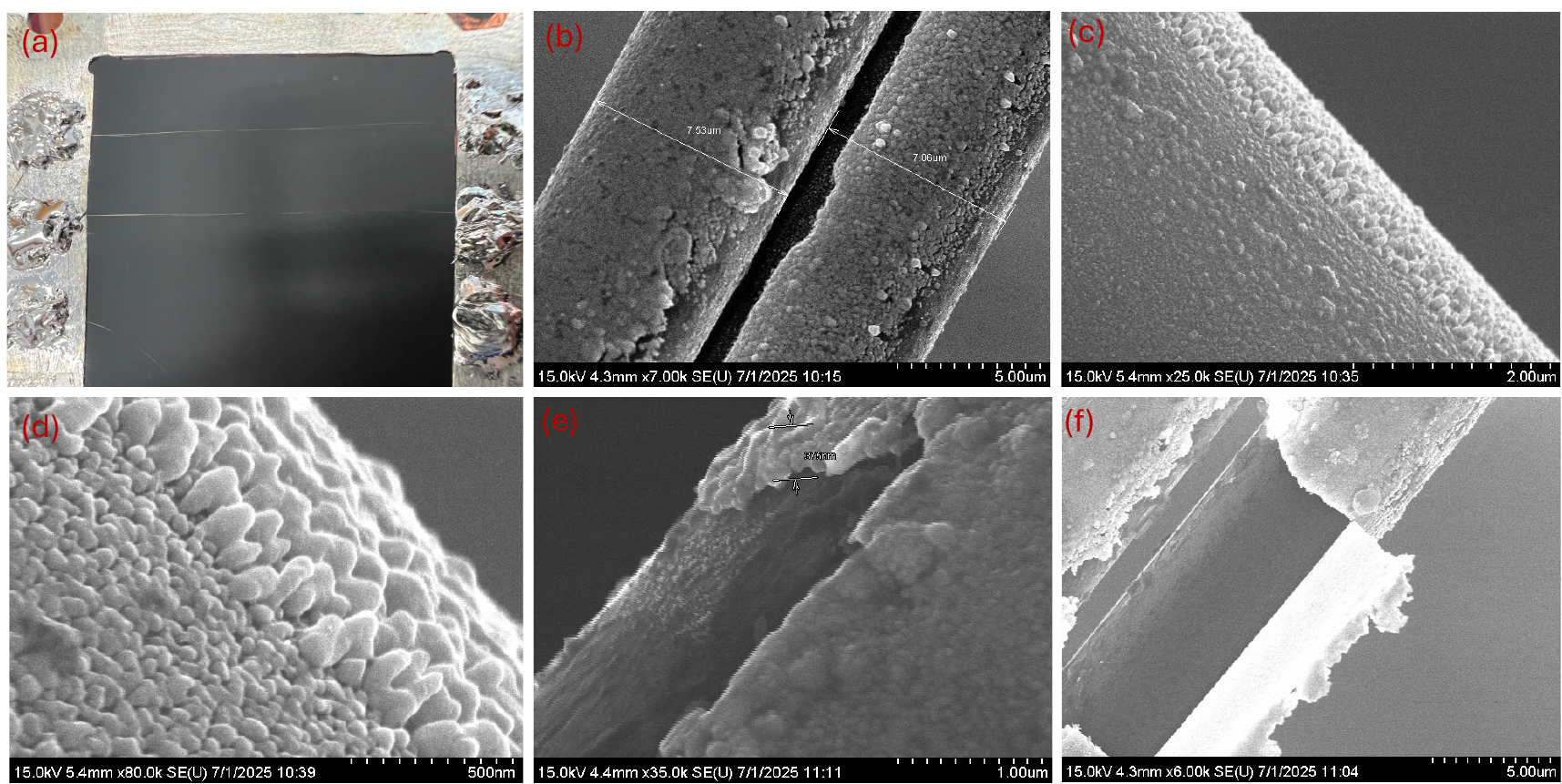}
    \caption{(a) The carbon fiber wire solder on a copper hollow plates. (b) The same wire under the Scanned Electron Microscope (SEM) after coating with a thin layer (a few hundreds of nanometers) of Silver. There are two wires about 7~\um thick which cannot be separated by hand. (c) The area where up and down side are met, where up side has coating and down side has none. (d) zoom in on (c). (e) and (f) delamination found on some spots. }
    \label{fig:CF-wire-coating}
\end{figure}

\textbf{Characterization} The coated fibers are examined using optical microscopy, scanning electron microscopy (SEM) to assess film morphology, uniformity, and composition. See Fig.~\ref{fig:CF-wire-coating}. Electrical resistivity will be measured using a four-point probe or dedicated micro-contact setup. Mechanical flexibility and adhesion strength are evaluated by bending and abrasion tests. The goal is to quantify both electrical conductivity improvement and mechanical robustness relative to bare CF. These studies are ongoing and have been planned, which is beyond this current project.

\textbf{Material Budget Reduction.} Replacing copper or aluminum wires with metal-coated CF could significantly reduce the overall radiation length ($X/X_0$) contribution of detector cabling and supports. Even a few 100~nm silver coating contributes negligible mass compared to conventional metallic wires while achieving surface conductivities sufficient for grounding, biasing, and smooth out the surface roughness.

\textbf{Technical Challenges.} Despite the promising outlook, several challenges remain in optimizing the coating process and ensuring the composite’s long-term stability. Adhesion and surface compatibility:
carbon fiber’s chemically inert surface can lead to poor adhesion of metallic films, particularly under thermal or mechanical stress. This can be seen from Fig.~\ref{fig:CF-wire-coating} (e) and (f). Surface activation, inter-layer materials (e.g., Chromium as adhesion promoters), or chemical functionalization may be necessary to improve film bonding. For the first coatings, no Cr was used. Uniformity of Coating: achieving consistent coverage around individual filaments within a CF bundle is difficult, especially for non-line-of-sight deposition techniques like thermal evaporation. Non-uniform coatings can result in variable conductivity or weak mechanical points. Mechanical Durability: Thin metal films can crack or delaminate during fiber bending or tensioning. Balancing film thickness for conductivity versus flexibility will be key. Future tests may explore composite layering or sputtered coatings with controlled grain structure for enhanced ductility. Electrical Contact and Integration: 
Reliable electrical connection between coated CF and standard metallic connectors must be established. The interface resistance, solderability, or compatibility with conductive adhesives requires systematic study. Scalability and Cost:
While small-scale evaporator-based coating is feasible for R$\&$D, scaling the process to meter-length wires or woven CF fabrics will require adaptation to continuous or roll-to-roll coating systems.

\textbf{Future Directions.} The ongoing and planned R\&D steps include: optimization of silver coating parameters: systematic variation of deposition rate, substrate temperature, and film thickness to maximize conductivity and adhesion, followed by quantitative electrical and mechanical testing. Exploration of Alternative Metals: investigate copper, aluminum, or nickel coatings for different conductivity and adhesion profiles. Nickel, for example, may provide improved durability and oxidation resistance, while aluminum offers extremely low mass. Adhesion Layer Development:
Introduce a thin interlayer (e.g., Ti, Cr) between carbon and metal to enhance bonding. The optimal interlayer thickness and deposition sequence will be studied.

In short, the metal-coated carbon fiber wire R\&D at BNL and Stony Brook represents an innovative step toward ultra-lightweight, multifunctional materials for next-generation collider detectors. By merging the mechanical and thermal advantages of carbon fiber with the high conductivity of thin metallic coatings, this approach offers a promising route to minimize the material budget in future tracking and vertex systems at the EIC, FCC, and other high-luminosity facilities.

Early results indicate that thin-film silver coatings can be deposited successfully on CF using standard evaporator techniques. The next phase will focus on optimizing adhesion, testing mechanical resilience, and extending the approach to alternative metals and composite structures. With continued development, such materials could play a critical role in realizing low-mass, high-performance detector systems, supporting the precision physics goals of future colliders.

\paragraph{GridPix}

\begin{figure}[h]
    \centering
    \includegraphics[width=0.48\linewidth]{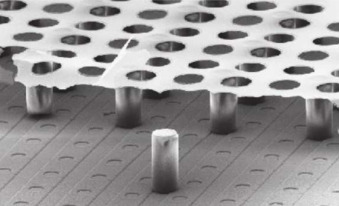}
    \includegraphics[width=0.48\linewidth]{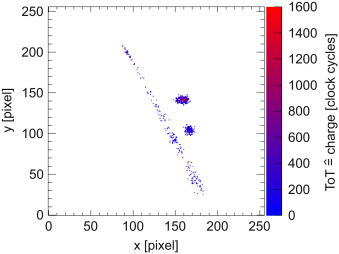}
    \caption{Left: SEM picture of a GridPix detector. The aluminum grid is partially removed for better visibility. Right: Event picture of a track accompanied with two transition radiation photons~\cite{KAMINSKI2017233}.}
    \label{fig:gridpix_slice_track}
\end{figure}

As was already mentioned silicon tracking has become the most wanted technology for particle detectors in recent years.
In the typical configuration, charged particles traverse the silicon itself releasing enough ionization to
be detected with no or low avalanche. GridPix is a CMOS pixel readout chip with an integrated amplication grid added by MEMS postprocessing techniques~\cite{Ligtenberg:2020ofy,KAMINSKI2017233, VANBEUZEKOM2025170397}. GridPix is instead designed to collect the much less dense ionization trail coming from gas ionization. In this case the measurements are driven by few electrons per pixel, the signal is too small for room temperature electronics. Thus, an avalanche process is required to boost the signal size above the detection threshold.
In the GridPix technology the pixels themselves are located below an etched aluminum grid (mesh).
This grid yields suffcient gain to provide electronics enough signal from a single electron to be measured with efficiency $> 90\%$. Below the grid is an array of $55\times55$~\umsq pixels, providing high-resolution measurements.
Using typical Argon-based mixtures, diffusion is most often sufficient that each electron from a cluster lands in a different hole. In this way, the GridPix detector technology promises a final realization of the long-sought dream of cluster counting. Additionally, when coupled to a low diffusion gas, the GridPix technology will act as a high resolution tracker competitive with all silicon systems. Example of an electron track in the Xe dominated gas along with two transition radiation photons is shown in Fig.~\ref{fig:gridpix_slice_track} from~\cite{KAMINSKI2017233}. 

There is a long-known limitation for detectors with large gas volumes. Ions created during the avalanche stage are slowly drifting through the gas volume. This is known as the positive ion backflow (IBF) problem. For high-intensity measurements, IBF is a limiting factor as it creates distortions of charged particle tracks. There are several ways to metigate this limiting factor. One of them is to couple GridPix with a passive bi-polar wire grid or GEM, additional electrode, which will limit the IBF. In the presence of the magnetic field, the grid more effectively attenuates the ion current than the electron current going through it. 

The GridPix TPC has been supported as a generic R\&D for EIC~\cite{GridPixSoW2022}. 


\subsubsection{Simulation of a mixed-technology tracking system}
\label{sec:det2_tpc_sim}
A series of simulations using EicRoot~\cite{bib:eicroot} were conducted to study a mixed-technology tracking system consisting of an inner silicon tracker and an outer TPC. These simulations evaluate the systems's tracking performance, resilience to background and dimensional requirements for the TPC.

\begin{figure}[h]
    \centering
    \includegraphics[width=0.6\linewidth]{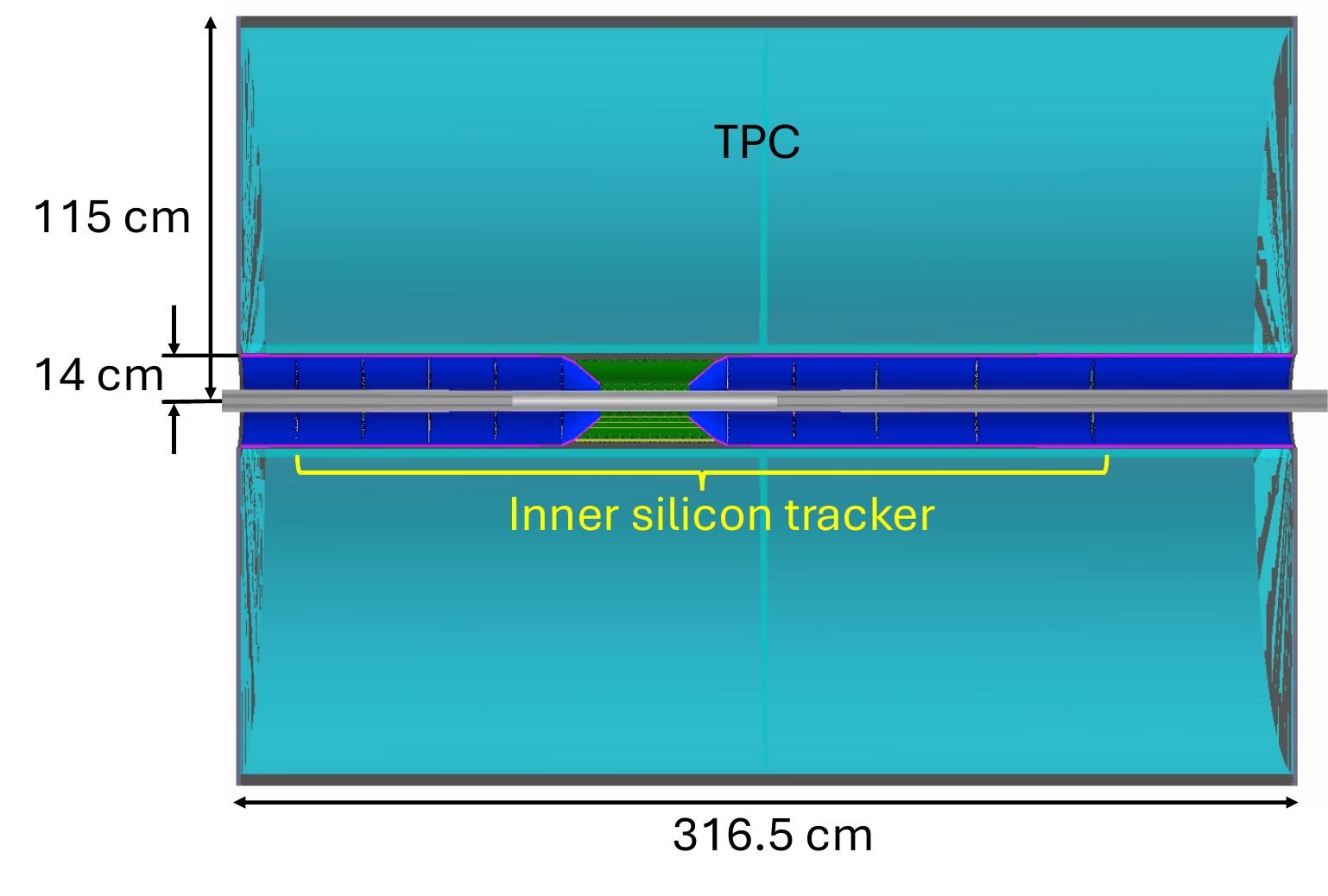}
    \caption{Layout of the mixed-technology tracking system used in the simulation. The gray area represents the beam pipe. The dark blue and magenta regions correspond to the support structures made of carbon fiber and aluminum, respectively.}
    \label{fig:det2_TPC_layout}
\end{figure}

\paragraph{Simulation Setup}
The simulated tracking system, placed in a magnetic field of $2$~T, consists of an inner silicon tracker and an outer TPC, as shown in Fig~\ref{fig:det2_TPC_layout}. The beam pipe in the simulation comprises a $40$~cm-long center section and two $410$~cm-long extensions on either side. The center beam pipe is constructed from $800$~\um-thick beryllium with an outer radius of $3.23$~cm, and its inner surface is coated with a $5$~\um-thick gold layer. Each extension is made of $1$~mm-thick aluminum, with the radius gradually increasing from $3.23$ to $3.45$~cm.

The inner silicon tracker adopts the mechanical design of the ALICE ITS2 MAPS stave, but features a pixel size of $20\times20$~\umsq, matching that of the ALICE ITS3 sensors. It consists of three subsystems: a three-barrel-layer vertexing silicon tracker (VST), a five-disk backward silicon tracker (BST), and five-disk forward silicon tracker (FST). The VST is positioned at radii between $3.6$~cm and $12$~cm. The BST spans the range $-105<z<-25$~cm, and the FST spans $25<z<135$~cm. All the disks share the same outer radius of $13$~cm. The support and service structure for the inner silicon tracker are implemented in cone or tube shapes, consisting of a $2$~mm-thick carbon fiber and a $1$--$2$~mm-thick aluminum foil. 

The TPC is $316.5$~cm in length, with inner and outer radii of $14$~cm and $115$~cm, respectively. It is centered at $z=36.75$~cm. These dimensions are maximized within the available space at the interaction region and accommodate the ePIC particle identification detectors. The inner field cage of the TPC composes of three layers: a $2.62$~mm-thick carbon fiber layer, a $150$~\um-thick kapton layer, and a $50$~\um-thick aluminum layer. The outer field cage has the same composition, except the carbon fiber is thicker at $9.28$~mm. The endcaps are single aluminum sheets with a thickness of $13.4$~mm, and the central membrane is a $100$~\um-thick kapton sheet. The TPC is filled with a gas mixture of Argon, CF\textsubscript{4} and iC\textsubscript{4}H\textsubscript{10} in a $95:3:2$ ratio. Electron cloud clustering is not implemented in the simulation; instead, hit smearing is applied post-simulation. The diffusion uncertainty and the effective number of electrons are set to $28$~\um/$\sqrt{\text{cm}}$ and 20, respectively~\cite{bib:sPHENIC_TPC_database}, resulting in a transverse dispersion of $6.258$~\um$/\sqrt{\text{cm}}$. A Space charged-induced uncertainty of $50$~um~\cite{bib:sPHENIX_TPC_spaceCharge} is assumed along with one pad row per $1$~cm. The transverse intrinsic resolution is set to $55$~\um~\cite{bib:GridPix_Kaminski2016}, based on the GridPux sensor resolution. 

\begin{figure}[h]
    \centering
        \includegraphics[width=0.5\linewidth]{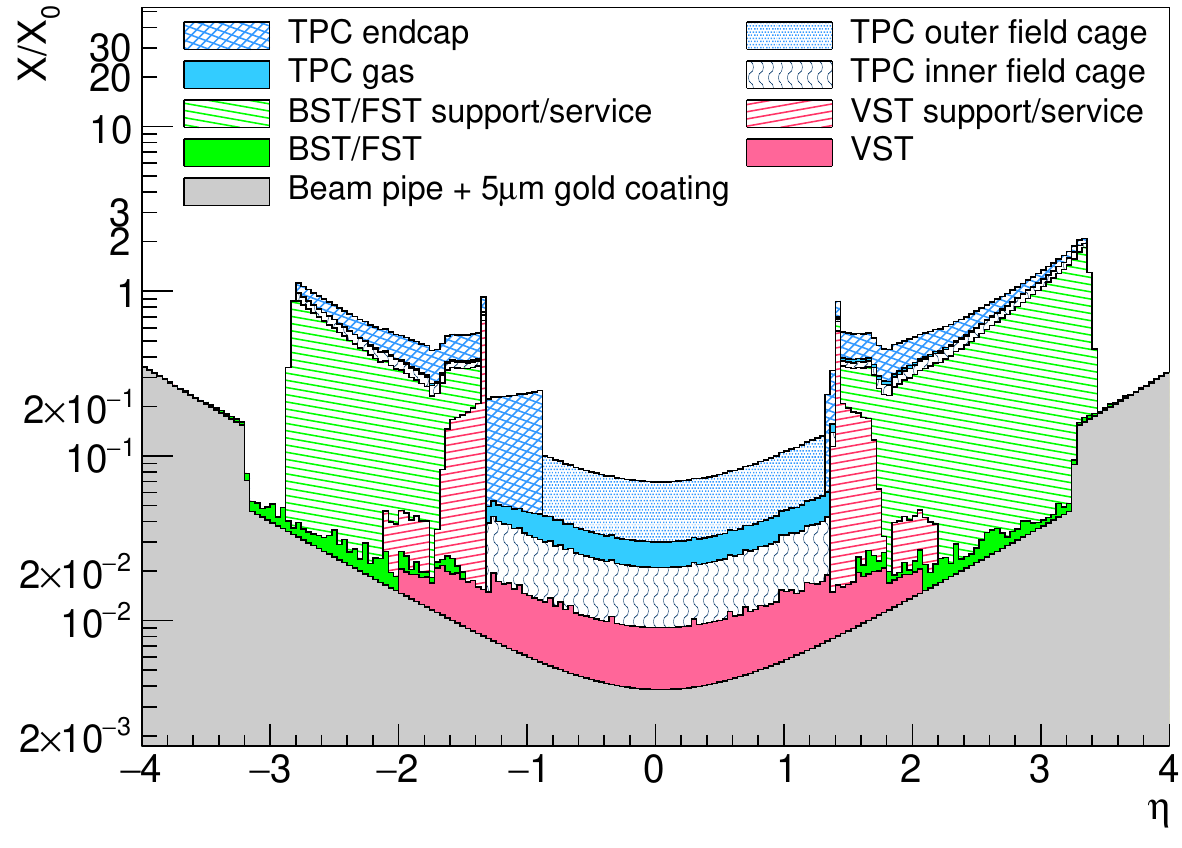}
    \caption{Material budget of the mixed-technology tracking design as a function of pseudorapidity, shown on a logarithmic scale.}
    \label{fig:det2_radlenLog}
\end{figure}

\begin{figure}[h]
    \centering
        \includegraphics[width=0.48\linewidth]{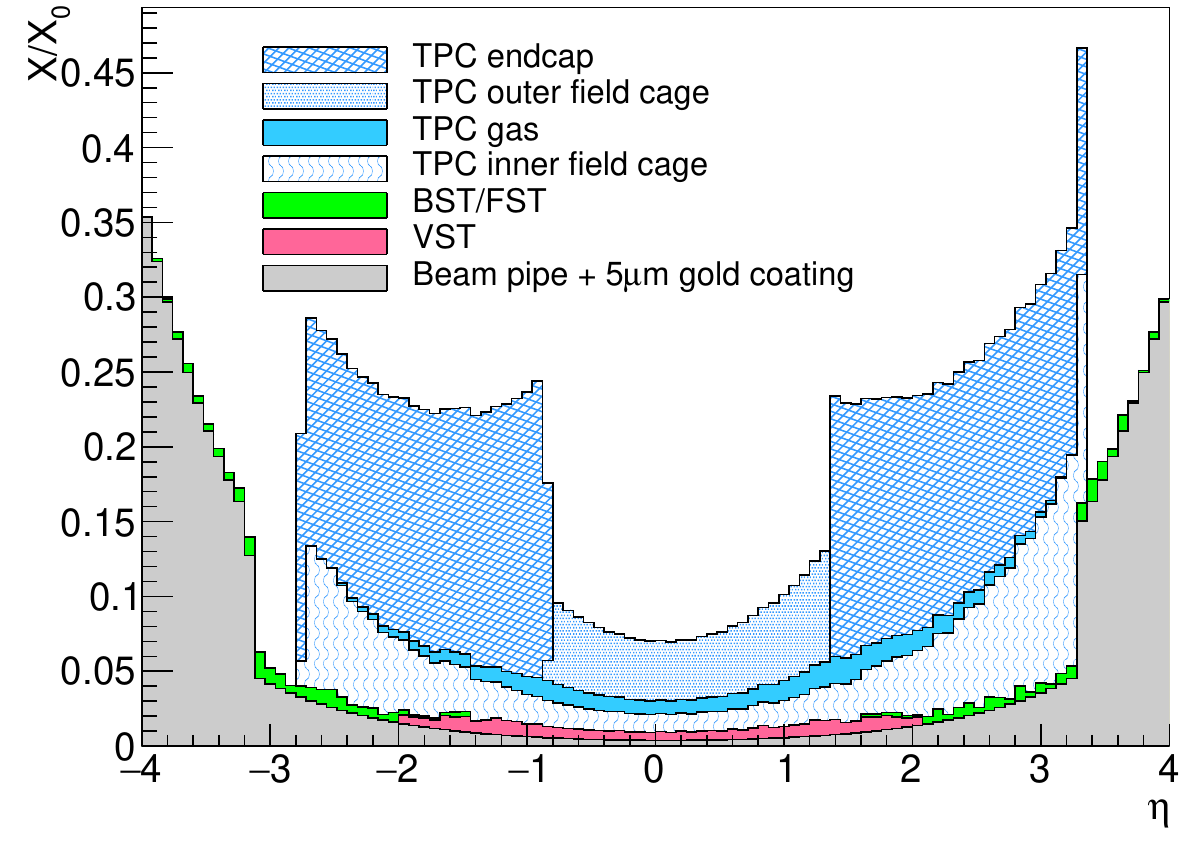}
        \includegraphics[width=0.48\linewidth]{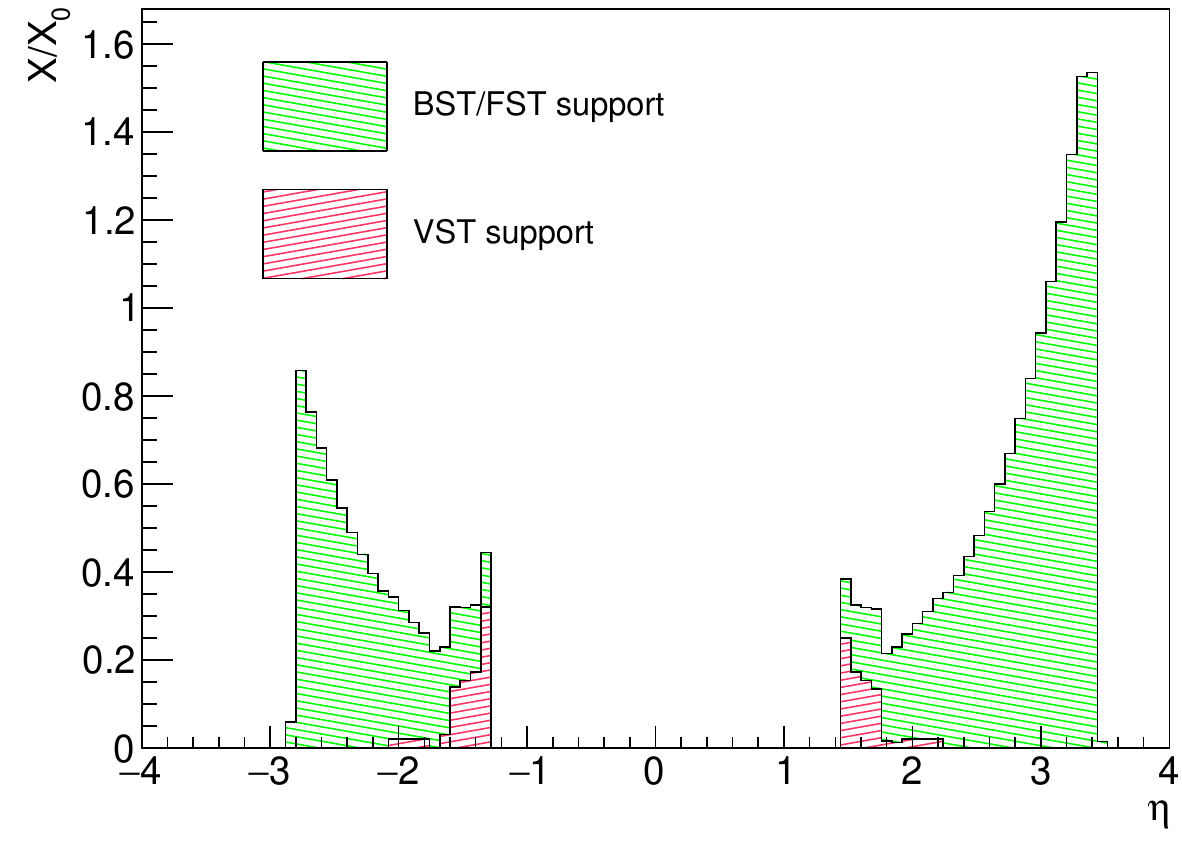}
    \caption{Left: Material budget of the mixed-technology tracking design excluding the inner silicon support structures. Right: Material budget of the inner silicon support structures.}
    \label{fig:det2_radlen}
\end{figure}

Figure~\ref{fig:det2_radlenLog} and \ref{fig:det2_radlen} show the material budget of the mixed-technology tracking design. The effective material budget is approximately $3$\% at $\eta=0$. The VST provide acceptance in the pseudorapidity $-2<\eta<2$, while the BST and FST cover the backward and forward regions with $1.7<|\eta|<3.8$. It should be noted that the acceptance of BST and FST depends on the specific beam pipe design. The active area of the TPC, defined by the gas volume, covers the range $-2.5<\eta<3$. 


\begin{figure}[h]
    \centering
    \includegraphics[width=\linewidth]{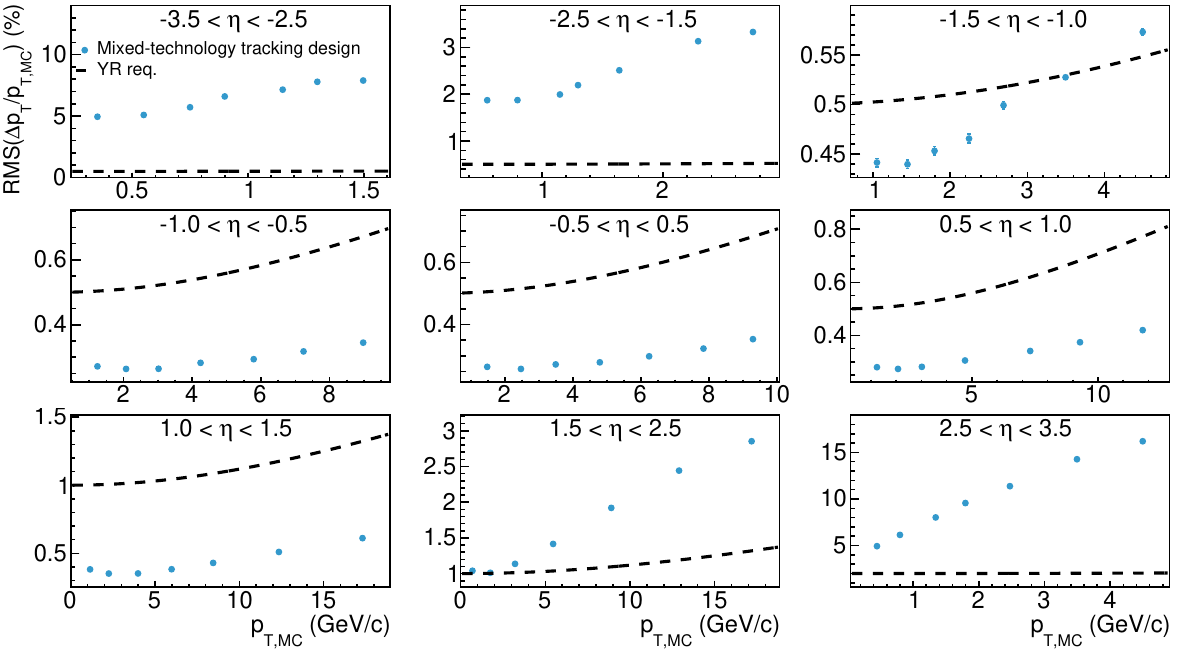}
    \caption{Transverse momentum resolution of the mixed-technology tracking design compared to the Yellow Report requirement.}
    \label{fig:det2_TPC_comparePtResYRreq}
\end{figure}

\paragraph{Tracking Performance}
Figure~\ref{fig:det2_TPC_comparePtResYRreq} shows the $p_T$ resolution of the mixed-technology tracking design for single charged pions, compared to the Yellow Report requirements. The mixed-technology design meets the Yellow Report requirements within the central region, $-1.5<\eta<1.5$. At $|\eta|>1.5$, near the edge of the TPC acceptance, the number of TPC hits per track decreases, and the increased material budget leads to degradation in $p_T$ resolution. It is important to emphasize that, similar to the ePIC tracking system, additional outer tracking layers in conjunction with the inner silicon disks, are necessary in the $|\eta|>2$ region to improve the tracking performance.

\paragraph{Syncrotron Radiation Background}
Two detector configurations are simulated to study the amount of synchrotron radiation background inside a TPC. In the first setup, an 5~\um-thick \SR absorber gold coating is absent from the inner surface of the beam pipe. Additionally, the service and support structures for the inner silicon tracker are not included. In the second setup, both the gold coating and the service and support structures are implemented. The \SR photons, produced by an 18~GeV electron beams with a beam current of $0.277$~A, are integrated over a $5$~\us window, corresponding to the estimated timing window of the inner silicon tracker in the ePIC detector at the time of this simulation study.

\begin{figure}[h]
    \centering
    \includegraphics[width=\linewidth]{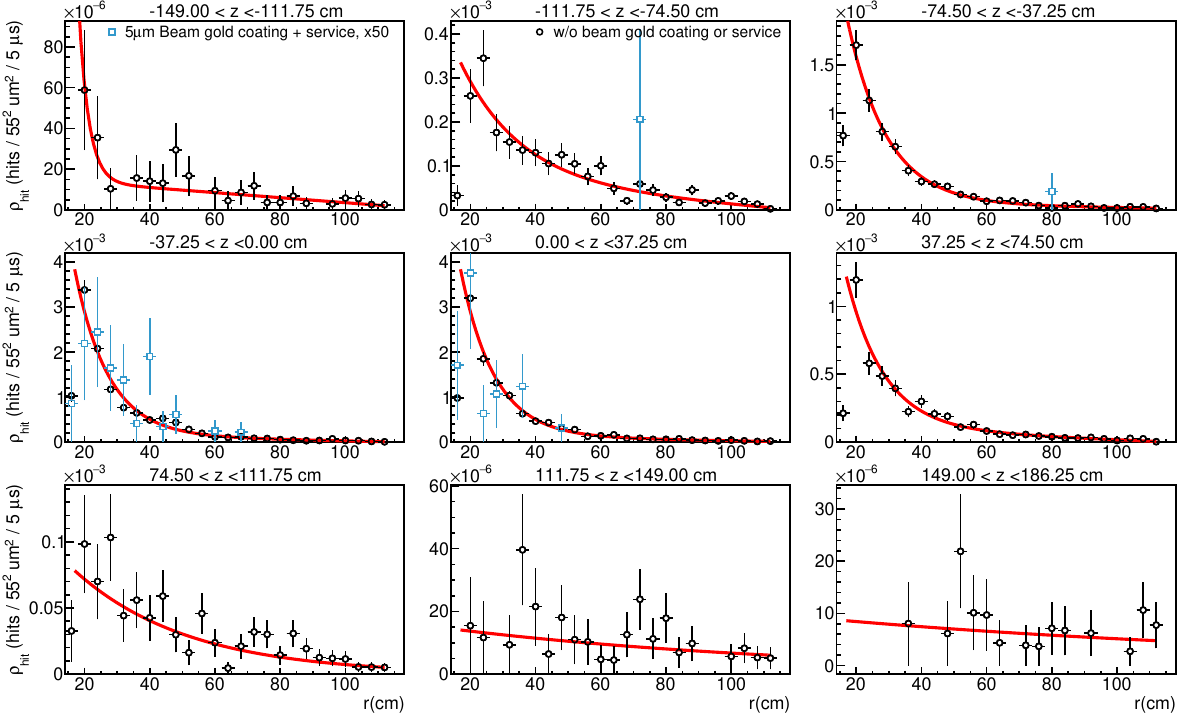}
    \caption{Synchrotron radiation hit density inside the TPC as a function of radius in different $z$ ranges. Red lines indicate exponential fits to the simulated data, shown as black markers.}
    \label{fig:det2_TPC_SR_HitDensityRes55pdf}
\end{figure}

Figure~\ref{fig:det2_TPC_SR_HitDensityRes55pdf} shows the \SR photon hit density as a function of radius in the TPC across various $z$-ranges. Each $z$-range span $37.25$~cm, corresponds to a $5$~\us drift time, assuming a drift velocity of $74.5$~\um/ns~\cite{bib:sPHENIC_TPC_database}. The hit density is integrated from the given $z$-range to one endcap of the TPC and normalized to a pixel size of $55^2$~\umsq, corresponding to the expected pixel size of the GridPix sensor. The hit density falls exponentially with increasing radius. The highest \SR photon hit density occurs near $z=0$~cm and decreases with increasing $|z|$. The maximum hit density is approximately $3\times10^{-3}$~hits/$55^{2}$~\umsq/$5$~\us in the configuration without the gold coating, service and support structures, and about $50$~times lower when both are included. The \SR photon hit density is higher in the backward region ($z<0$~cm) than in the forward region ($z>0$~cm).

\begin{figure}[h]
    \centering
    \includegraphics[width=\linewidth]{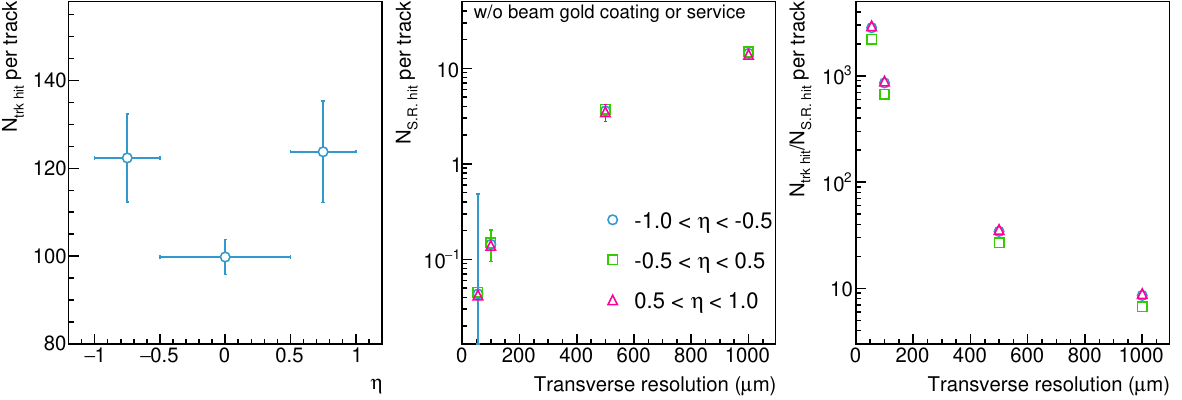}
    \caption{Left: Number of track hits per track. Center: Number of synchrotron radiation photon hits located near the track. Right: Ratio of track hits to nearby synchrotron radiation photon hits.}
    \label{fig:det2_TPC_SR_HitPerTrk}
\end{figure}

To estimate the number of \SR photon hits that could be misidentified as track hits, the hit density from the first setup (without the gold coating, service and support structures) is used for its higher statistics. This hit density is further scaled for three additional pixel sizes, $100^2$~\umsq, $500^2$~\umsq, and $1000^2$~\umsq. The number of hits per track at $|\eta|<1$, estimated from single charged pion events without a magnetic field, ranges from $100$ to $125$, as shown on the left of Fig.~\ref{fig:det2_TPC_SR_HitPerTrk}. The central panel of Figure~\ref{fig:det2_TPC_SR_HitPerTrk} shows that fewer than one \SR photon hit lies near a track if the TPC sensor resolution is finer than $100^2$~\umsq. This number rises above $20$ for resolution of $1000^2$~\umsq. The right panel of Fig.~\ref{fig:det2_TPC_SR_HitPerTrk} shows that the ratio of the track hits to \SR photon hits exceeds $1000$ for a $55^2$~\umsq resolution, but drops below $10$ when the resolution reaches $1000^2$~\umsq. It is important to note that the total number of synchrotron radiation hits is expected to decrease by approximately a factor of $50$ when the beam pipe gold coating and the silicon support and service structures are included in the detector configuration.

\begin{figure}
    \centering
    \includegraphics[width=\linewidth]{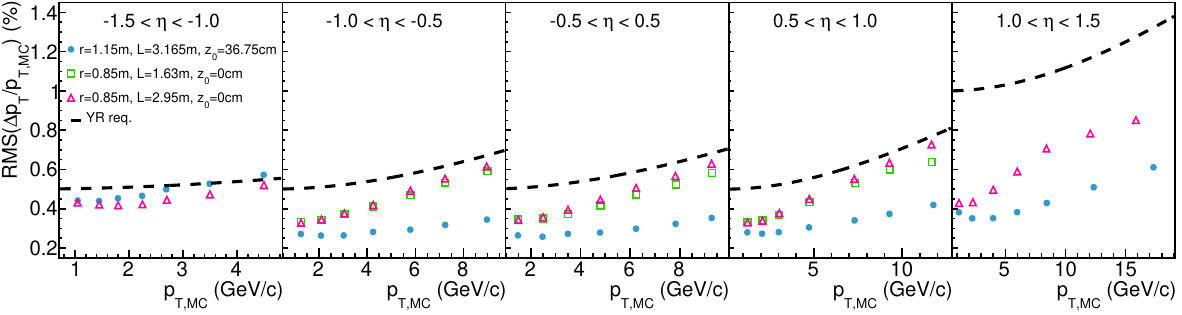}
    \caption{Transverse momentum resolutions with various TPC dimensions.}
    \label{fig:det2_TPC_ptRes_compareDimension}
\end{figure}

\paragraph{TPC Dimension}
While a large TPC offers better tracking performance, it also necessitates a greater number of sensor channels, leading to increased complexity in cooling and cable management. Therefore, it is beneficial to identify the minimum TPC dimensions that still deliver adequate tracking performance. To investigate this, two reduced-dimension TPC configurations are tested in the simulation. Both share a smaller outer radius of $85$~cm and are centered at $z=0$~cm, but differ in length, $1.63$~m and $2.95$, corresponding to pseudorapidity acceptances of $|\eta|<\pm1$ and $|\eta|<\pm1.5$, respectively. Figure~\ref{fig:det2_TPC_ptRes_compareDimension} shows the $p_T$ resolutions for these configurations. The results indicate that a reduced outer radius of $85$~cm can still provide the sufficient tracking performance in the central pseudorapidity regions.

\FloatBarrier

\subsubsection{Scintillating Fiber (SciFi)}
Scintillating Fiber trackers (SciFi) had seen a resurgence in recent years, notably in the LHCb and Mu3E experiments. 

\begin{figure}[h]
    \centering
    \includegraphics[width=0.48\linewidth]{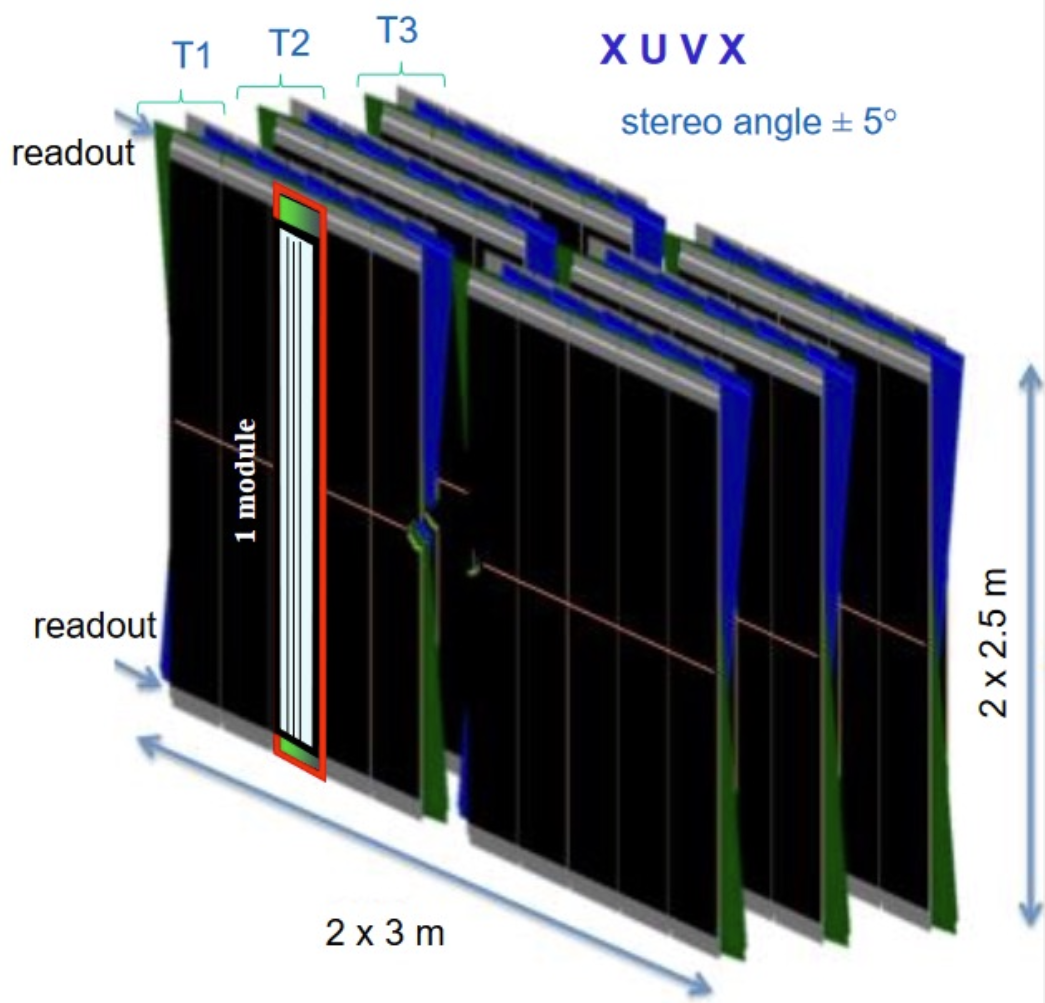}
    \includegraphics[width=0.48\linewidth]{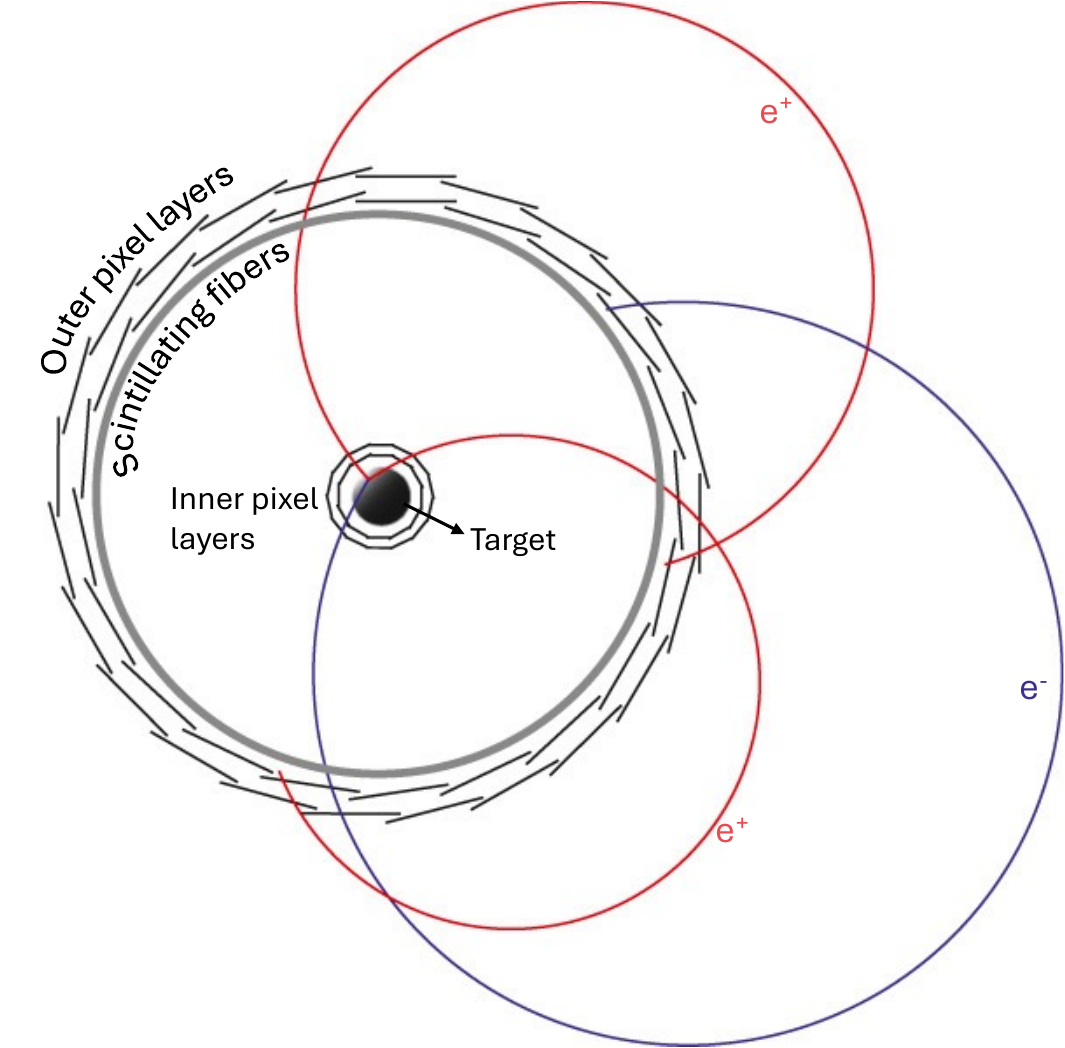}
    \caption{Left: LHCb SciFi tracker dimensions. Right: Mu3E central tracking system layout~\cite{bib:Mu3eSciFi3}.}
    \label{fig:SciFitracker}
\end{figure}

The LHCb SciFi tracker~\cite{bib:LHCbSciFi} consists of three stations, as illustrated on the left of Fig.~\ref{fig:SciFitracker}, each comprising four planes oriented in the X, U, V, and X directions. Each plane, supported by a $20$~mm-thick Nomex Core honeycomb base, includes six or eight layers of double-clad polystyrene scintillating fibers (SCSF-81)~\cite{bib:KuraraySciFi} with a diameter of $250$~\um, an excited electron decay time of $2.4$~ns, and an attenuation length of $3.5$~m. With precise alignment, the LHCb SciFi achieves a spatial resolution better than $70$~\um. The scintillation light is guided to the Hamamatsu silicon photomultiplyers (SiPM), MPPC S13552--H2017, which have a pixel size of approximately $60$~\um and located at both ends of the fibers. The signals are read out by the PACIFIC ASIC at a rate of $40$~MHz. The material budget per module is approximately $1$\%, dominated by six fiber layers, with the honeycomb base being the second largest contributor, as detailed in Table.~4.3 of Ref\cite{bib:LHCbSciFi}.

The Mu3E experiment employs the fast timing capabilities of the scintillating fiber to suppress background. The Mu3E SciFi tracker~\cite{bib:Mu3eSciFi1} is a cylindrical detector with a radius of $61$~mm, placed between an inner and outer double-layer silicon pixel trackers, as illustrated on the right of Fig.~\ref{fig:SciFitracker}. It consists of 12~sections of fiber ribbons, each comprising three~layers of SCSF-78 fiber~\cite{bib:KuraraySciFi} with a diameter of $250$~\um, an excited electron decay times of $2.7$~ns, and an attenuation length of $4$~m. In contrast to the LHCb SciFi design, the Mu3E SciFi has no protective structure, enabling an low material budget of $0.2$\%. It provides a spatial resolution of about $100$~um, and benefited by the fast excited electron decay time and double-ended readouts, achieves a timing resolution below $400$~ps~\cite{bib:Mu3eSciFi1,bib:Mu3eSciFi2}.

The LHCb and Mu3E experiments demonstrate the strong potential of SciFi technology for precision tracking. A SciFi could be deployed outside an inner silicon tracker to provide additional hits, while also suppressing background, such as pile up and synchrotron radiation, through its fast integral window. Another idea is to incorporate a SciFi envelope tracker outside a gaseous detector, where it could serve as a calibration layer for the gaseous detector.

\begin{figure}[h]
    \centering
    \includegraphics[width=0.5\linewidth]{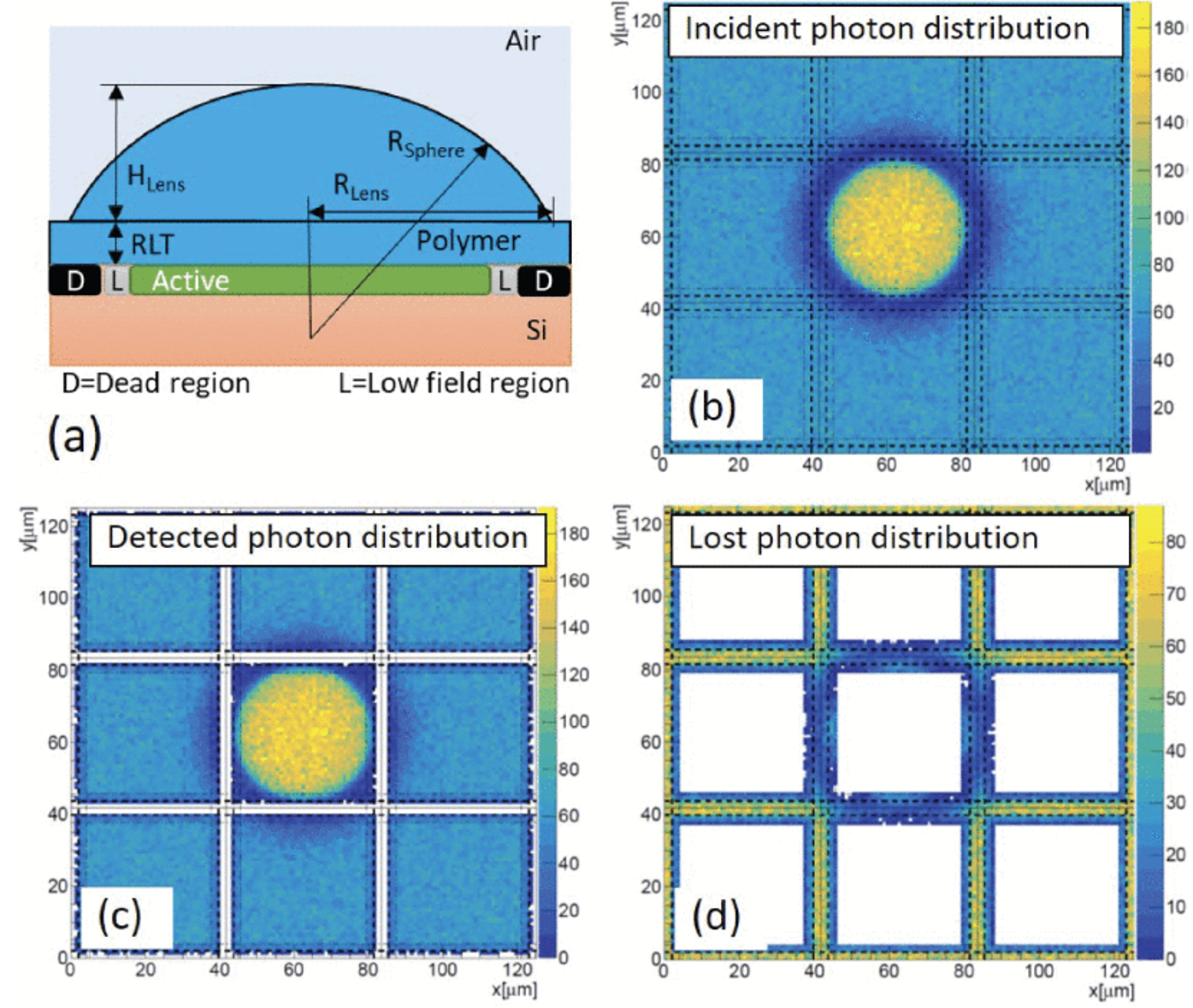}
    \caption{(a) Layout of the microlens-enhanced SiPM. (b) Incident photon distribution in simulation. The microlens is implemented only on the center pixel. (c) Detected photon distribution. (d) Lost photon distribution. The amount of lost photon around the center pixel with a microlens is noticeably lower compared to adjacent pixels without a microlens~\cite{bib:microlens}.}
    \label{fig:microlensSiPM}
\end{figure}

There is ongoing R\&D effort exploring the integration of microlenses on SiPM to improve photo-detection efficiency~\cite{bib:microlens}, thereby enhancing signal and hit efficiency. As shown in Fig.~\ref{fig:microlensSiPM}(a), dome-shaped microlenses are added on the surface of the SiPM. These microlenses are arranged in a checkerboard fashion that deliberately covers the dead areas of the SiPM, guiding photons that would otherwise be lost toward the active regions, as demonstrated in Fig.~\ref{fig:microlensSiPM}(b)--(d). Reference~\cite{bib:microlens} reports that microlenses can enhance photo-detection efficiency by $15$\%--$20$\%. Additional R\&D efforts are also aimed at signal enhancement by strengthening the radiation hardness of the fibers, and optimizing the cladding materials~\cite{bib:CERN-DRD4}. 

\subsection{Particle Identification}\label{sec:PID}
\subsubsection{The \textit{xpDIRC}: A Next-Generation Concept for Extreme-Performance DIRC Detector}

The Detection of Internally Reflected Cherenkov light (DIRC) technology has proven to be a highly efficient and compact particle identification (PID) method for modern collider experiments. Its successful operation in detectors such as the BaBar DIRC, the GlueX DIRC, and the Belle~II TOP has demonstrated reliable $\pi/K$ separation up to 3.5~\gevc. The high-performance DIRC (hpDIRC), developed recently for the ePIC detector at EIC, represents the major evolution of this technology. With its advanced focusing optics, small-pixel photon sensors, and fast timing electronics, the hpDIRC aims to achieve $\pi/K$ separation up to 6~\gevc. 

Building on this foundation, the \textit{extreme-performance DIRC} (xpDIRC) concept explores innovative optical geometries and readout configurations designed to extend performance even further—potentially enabling $\pi/K$ separation approaching 10~\gevc. The ongoing R\&D also targets cost and material budget reduction, as well as design flexibility to improve integration capability for Detector-2 at the EIC.  

\begin{figure}[h!]
  \centering
  \includegraphics[width=0.98\textwidth]{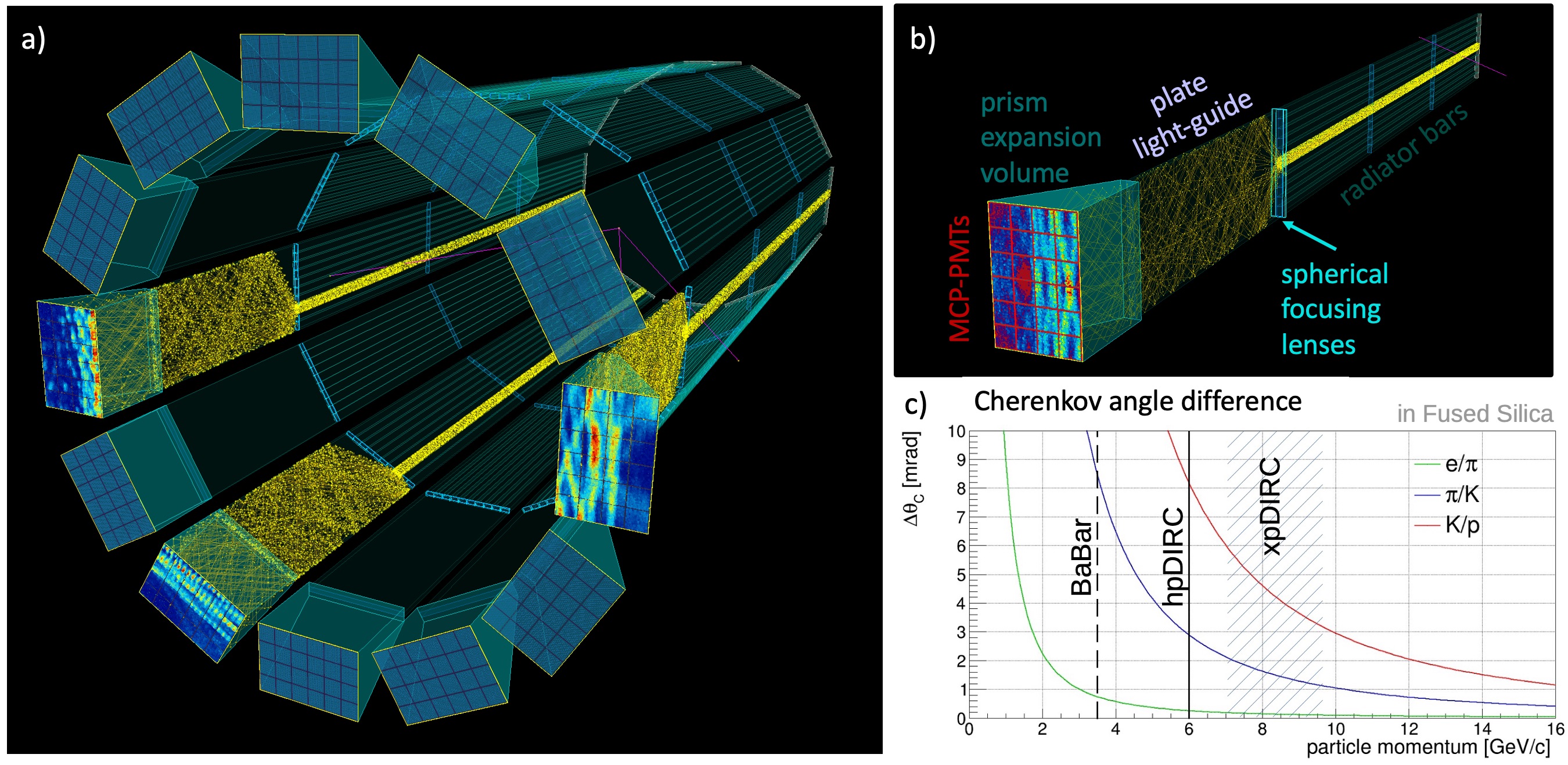}
  \caption{(a) Geant4 event display of the \textit{xpDIRC} with a wide plate acting as a light-guide. Yellow lines trace photon paths from three sample tracks (magenta). 
  (b) One xpDIRC module with key components labeled. 
  (c) Theoretical $\pi/K$ Cherenkov angle separation limits as a function of particle momentum, compared with BaBar DIRC and ePIC hpDIRC performance.}
  \label{fig:xpDIRC_event}
\end{figure}

The xpDIRC program began by systematically studying the theoretical limits of DIRC performance using the ePIC hpDIRC configuration as a reference. These investigations isolated the contributions of key factors such as chromatic dispersion, timing precision, and focusing aberrations. Motivated by both performance enhancement and cost optimization, the xpDIRC geometry replaces the narrow light-guide bars in the hpDIRC’s short section near the lens with wide plates, which act as extended expansion volumes coupled to the existing prism readout. Figures~\ref{fig:xpDIRC_event}~(a) and~(b) illustrate this new layout.

Initial simulations focused on the optimal geometry of the plate and the positioning of the focusing optics. Several designs were explored, including a long cylindrical lens placed either between the plate and prism or between the narrow bars and the plate, as well as a configuration with individual spherical lenses placed between each narrow bar and the wide plate. The latter was chosen as the current preliminary baseline design. Future studies will explore if this configuration can also significantly reduce the prism size and thus the overall sensor area—a major cost driver in DIRC construction.

While the xpDIRC baseline design currently employs MCP-PMT sensors, simulation results indicate that next-generation SiPMs may soon become a viable alternative. Historically, SiPMs were excluded due to their high dark-count rates (DCR). However, our studies incorporating the most recent SiPM DCR values suggest that acceptable performance may be maintained, particularly when combined with smaller prisms and optimized cooling. This development could mark the first practical implementation of SiPMs in a DIRC system, significantly reducing cost and adding valuable alternative to challenging market of acceptable single photon sensor candidates.

Furthermore, thinner radiator bars envisioned in the xpDIRC design could improve performance of following DIRC detector systems. When combined with high-resolution barrel calorimetry, such a configuration could improve photon energy resolution—crucial for DVCS measurements on nuclei—and increase electron identification purity in parity-violating DIS studies.

The initial phase of xpDIRC simulations has been conducted using the Geant4 DIRC framework—originally developed for the PANDA Barrel DIRC, validated in CERN test-beam campaigns, and now used for multiple DIRC projects worldwide. Various optical layouts were implemented and evaluated in terms of $\pi/K$ separation power relative to the ePIC hpDIRC baseline.

\begin{figure}[h!]
  \centering
  \includegraphics[width=\textwidth]{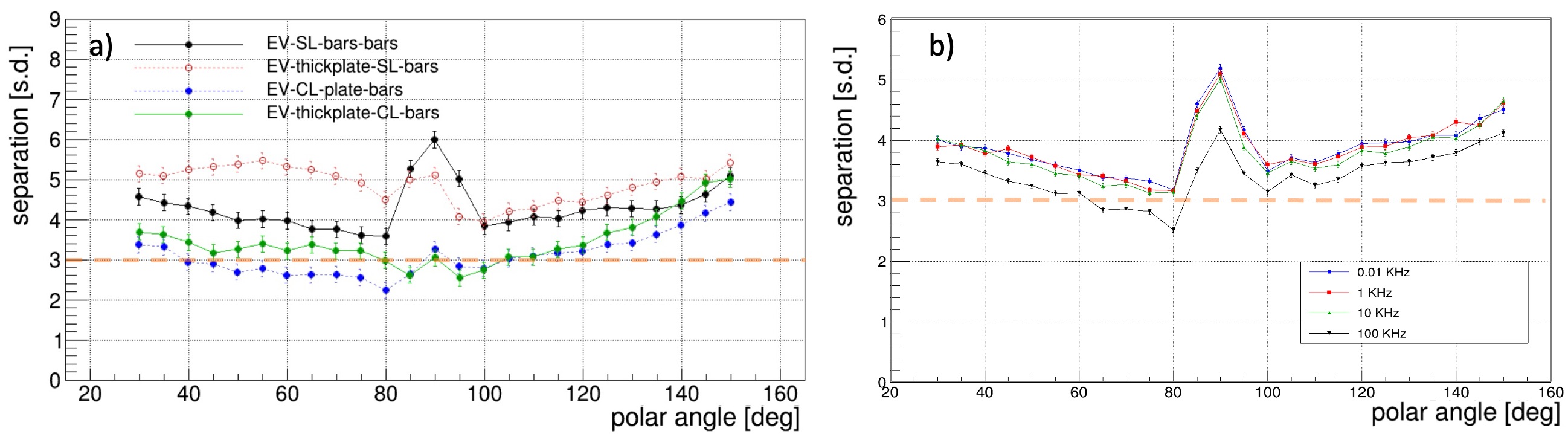}
  \caption{(a) $\pi/K$ separation power as a function of polar angle at 6~GeV/$c$ for different DIRC light-guide and focusing configurations. 
  (b) Impact of dark count rate (DCR) on $\pi/K$ separation for the ePIC hpDIRC geometry, illustrating robustness against increased DCR levels.}
  \label{fig:xpDIRC_performance}
\end{figure}

The reference hpDIRC configuration (black points in Fig.~\ref{fig:xpDIRC_performance}a) employs 35~mm-wide narrow bars throughout. The hybrid geometry (blue points) replaces these bars with a wide plate acting as a light-guide, reducing both fabrication cost and optical complexity. A more advanced xpDIRC geometry (green points) introduces a 3-layer cylindrical lens between narrow bars and a thicker plate, while the most innovative configuration (red open points) employs individual spherical lenses between each bar and the plate. This configuration achieves $\pi/K$ separation exceeding 3~s.d. across the full polar-angle range, surpassing hpDIRC performance and underscoring the xpDIRC’s potential for next-generation PID capability.

To assess the feasibility of SiPM-based readout, the impact of dark-count noise on PID performance was simulated (Fig.~\ref{fig:xpDIRC_performance}b). The baseline MCP-PMT configuration (blue points, 0.01~kHz/mm\textsuperscript{2}) was compared to realistic SiPM conditions ranging up to 100~kHz/mm\textsuperscript{2}. Even under these extreme noise conditions, the system retains sufficient separation power across most of the acceptance. These encouraging results suggest that SiPM implementation—once considered impractical—could soon be achievable, especially with modern cooling and annealing strategies.

All current xpDIRC studies are conducted within the validated Geant4 simulation framework. Upcoming work will further refine timing precision, chromatic dispersion mitigation, and focusing optimization. Leveraging synergies with the ongoing hpDIRC development for ePIC, the xpDIRC prototype is planned to be integrated into the Cosmic Ray Telescope (CRT) facility at Stony Brook University. A three-year prototype program, including procurement of new focusing lenses and wide plates, is under consideration for support through the DOE Generic R\&D program.

The \textit{xpDIRC} thus represents an exciting and realistic step toward a new generation of Cherenkov-based PID detectors—capable of higher performance, reduced cost, and potentially the first practical use of SiPMs in a DIRC system. Its early results demonstrate both strong technological potential and a clear path toward experimental validation.
        
\subsubsection{hpRICH}
\label{hpRICH}

Our GEANT simulation studies show that by making conservative assumptions about possible aerogel
and HRPPD improvements (as communicated by the Aerogel Factory and by Incom Inc., respectively), as well as optimizing an overall detector design,
one can build a high performance proximity focusing RICH (hpRICH) with a momentum range where a 3$\sigma$ $\pi$/K separation
can be achieved up to $\sim$15~GeV/c, which is more than a factor of two better than for ePIC
pfRICH. To be specific, we assumed the following configuration:

\begin{itemize}
    \item Two layers of 15~mm thick aerogel tiles with an average refractive index 1.014 (pion-kaon gap of 3~mrad at 15~GeV/c) in a so-called focusing configuration and a transmission length of 35~mm
\item 50 cm length of the expansion volume
\item HRPPDs with a 75\% geometric efficiency, double density pads (pitch 1.625 mm) and a
QE($\lambda$) curve with a peak value of 40\% at 450~nm 
\item Acrylic filter wavelength cutoff at 375~nm
\end{itemize}

Such a configuration has a well-balanced combination of emission point uncertainty (driven
by aerogel thickness), detection point uncertainty (determined by HRPPD pad size), and remaining
chromatic effects. Preliminary GEANT studies show that one can achieve about
2.5~mrad single photon resolution, with an average number of 8-10 detected photoelectrons.
This translates into a track-level Cherenkov photon resolution below 1~mrad, meaning $\sim 3\sigma$ $\pi$/K separation at 15~GeV/c.

Back-of-the-envelope calculations show that if such an hpRICH detector is installed 2.5~m away
from the interaction point (IP), and equipped with new generation HRPPDs with 6 micron diameter pore MCPs, in order to comfortably reach
single photon timing resolution of 20~ps or better, one can achieve $\pi$/K separation on a higher than 3$\sigma$ level up to
$\sim$6~GeV/c, provided a start timing reference has a similar resolution of 20ps/$\sqrt{10}$ $\sim$7-8~ps,
even without making use of a multi-photon Cherenkov light flash produced in the HRPPD
window. Therefore, time of flight functionality enables low momenta PID. Such a timing
reference can be achieved by installing either an HRPPD-based hpRICH in the electron-going
endcap or by using an HRPPD-based time of flight wall in this endcap and making use of a
scattered DIS electron (or other identified particles) timing in this acceptance.

Should timing resolution requirements be less demanding for this subsystem, an interesting alternative photosensor concept would be an MCP-PMT with an integrated TimePix4 readout~\cite{bib:timePix4}. hpRICH would not be able to take a full advantage of an ultra-high TimePix4 spatial resolution (55 micron pixels), however its ability to operate at a very low gain (down to few times $10^4$) would allow one to increase MCP-PMT lifetime by up to two orders of magnitude. Such a hybrid photosensor may also be of interest for xpDIRC where a moderate timing resolution on the order of several dozens of picoseconds is acceptable, but improved spatial resolution and resilience to a high instantaneous hit occupancy would certainly improve the performance.

As mentioned earlier in section \ref{ePIC-dRICH}, such a high performance proximity focusing RICH can be complemented by a single radiator gaseous RICH in the hadron-going endcap in order to provide a continuous hadron PID coverage up to momenta of 50~GeV/c and beyond.

\subsubsection{AC-LGAD/ToF}
\label{sec:aclgad}
Time-of-Flight (TOF) measurements are employed in high-energy nuclear physics experiments to provide particle identification generally up to $p \sim3$~\gevc. Detectors which can provide needed TOF capabilities normally have to be ``fast", meaning they must have timing resolution $<100$~ps, or better, depending on the experiment (e.g. what information is used in addition to the time information supplied by the dedicated TOF subsystem). Given the applications of AC-LGAD silicon technology outlined in Sec.~\ref{sec:B0}, this technology also has clear applications to a TOF system. This approach will already be employed in the barrel and forward regions of the ePIC detector to supply supplementary PID at lower momentum, where the RICH detectors tend to focus on higher momenta. 

Given the 4D nature of the AC-LGAD technology, one could also look to use AC-LGADs for more tracking layers than just a TOF layer in the main tracking volume of a second detector. This would allow for two independent layers to provide a ``start" and ``stop" time, and enable a self-contained TOF measurement with matched hits between two AC-LGAD layers. In ePIC, the requirement for timing has been set at around $\sim25$~ps of timing resolution from the AC-LGADs. This high requirement means that the power consumption and material budgets are less-optimal for inner-tracking layers in a primary tracking volume. However, as briefly discussed in Sec.~\ref{sec:B0}, advancements in the readout technology or study of monolithic sensors could allow for a lower material budget for the combined sensors and readout, and enable use of the 4D technology in more layers of the primary tracker. Additionally, this timing capability aids in the rejection of various background hits, which could prove useful as the EIC luminosity increases.

\subsubsection{TPC dE/dx}
\begin{figure}[h]
    \centering
    \includegraphics[width=\linewidth]{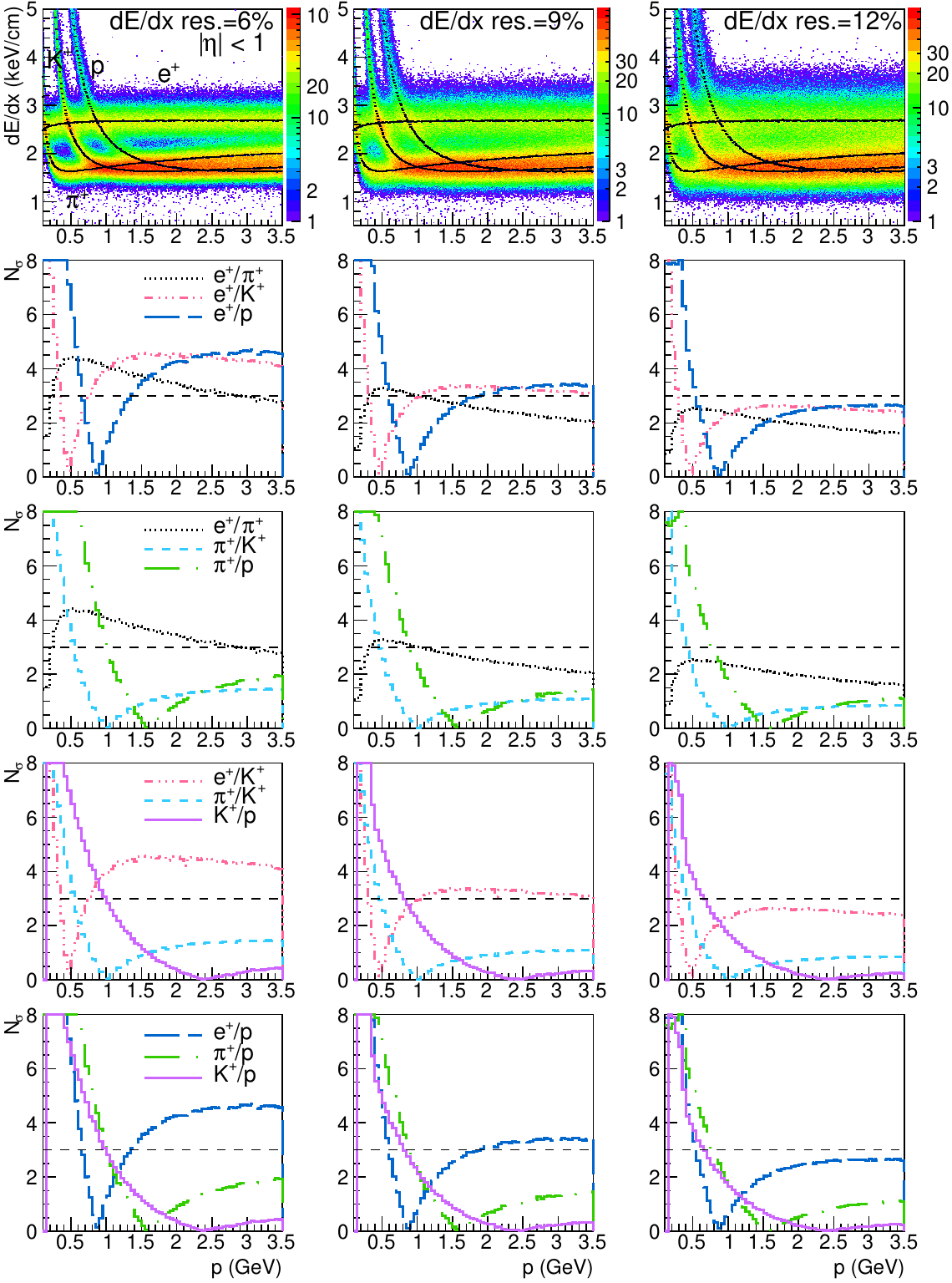}
    \caption{Top: $dE/dx$ v.s. momentum with different $dE/dx$ resolutions. Second to fifth rows: separation power of $e^\pm$, $\pi^\pm$, $K^\pm$, and $p,\bar{p}$ identifications, respectively. The horizontal dashed line indicates the $3\sigma$ separation.}
    \label{fig:det2_tpc_dedx}
\end{figure}

Particle identification(PID) using the TPC $dE/dx$ measurement is studied with the simulation setup described in Section~\ref{sec:det2_tpc_sim}, and the results are summarized in Table~\ref{tab:det2_tpc_dedx_nsig}. This study uses single-track events within the pseudorapidity range $-1<\eta<1$. The $dE/dx$ distributions for three different resolutions, $6$\%, $9$\% and $12$\%, are shown in Fig.~\ref{fig:det2_tpc_dedx}, along with the corresponding particle separation power ($N_\sigma$) as a function of momentum. The results demonstrates that electron identification ($e$ID) capability is highly sensitive to the $dE/dx$ resolution. Hadron identification is effective via $dE/dx$ up to $0.7$~\gevc. A TPC with a $dE/dx$ resolution of $6$\% can achieve $e$ID in the momentum range of $1.3$ and $2.9$~\gevc, whereas a worse resolution of $9$\% limits the $e$ID capability to $0.3$--$0.6$~\gevc. At $12$\% resolution, $dE/dx$ become ineffective for $e$ID and $\pi^\pm$ identification.

\begin{table}[h]
    \centering
    \caption{Momentum ranges where particle identification separation power exceeds 3$\sigma$.}
    \begin{tabular}{cccc}
    \hline
             & dE/dx resolution = 6\% & dE/dx resolution = 9\% & dE/dx resolution = 12\%\\ 
    \hline
    $e^{\pm}$ ID $\geq3\sigma$   & $1.35<p<2.9$ GeV/$c$  & /                    & /\\
    $\pi^{\pm}$ ID $\geq3\sigma$ & $0.2<p<0.5$ GeV/$c$ & $0.3<p<0.45$ GeV/$c$  & /\\
    $K^{\pm}$ ID $\geq3\sigma$   & $0.15<p<0.3$ GeV/$c$  & $0.15<p<0.25$ GeV/$c$ & $0.15<p<0.25$ GeV/$c$ \\
    $p,\bar{p}$ ID $\geq3\sigma$ & $0.15<p<0.65$ GeV/$c$  & $0.15<p<0.6$ GeV/$c$ & $0.15<p<0.55$ GeV/$c$\\
    \hline
    \end{tabular}
    \label{tab:det2_tpc_dedx_nsig}
\end{table}
\FloatBarrier

\subsubsection{Cluster Counting \texorpdfstring{$dN_{cl}/dx$}{dNcl/dx}}
In addition to $dE/dx$ measurements, cluster counting technique, denoted as $dN_{cl}/dx$, may offer an alternative due to its smaller intrinsic resolution than $dE/dx$. The relative uncertainty of $dE/dx$ depends on three factors, number of hits $N_{hit}$, the track length $L_{trk}$ in meter, and the ionizing gas pressure $P$ in standard atmosphere, and is given by
\begin{align}
\frac{\sigma_{dE/dx}}{dE/dx}&=\frac{0.41}{N^{0.43}_{hit}(L_{trk}P)^{0.32}}\label{eq:err_dEdx}\text{ .}
\end{align}
For cluster counting, the relative uncertainty is determined by the total number of clusters, which is the product of the clusters density per unit length $\delta_{cl}$ and $L_{trk}$, as described in
\begin{align}
\frac{\sigma_{dN_{cl}/dx}}{dN_{cl}/dx}&=\frac{1}{\sqrt{N_{cl}}}=\frac{1}{\sqrt{\delta_{cl}L_{trk}}}\label{eq:err_dNdx}\text{ .}
\end{align}

Using the TPC configuration described in Sec.\ref{sec:det2_tpc_sim} with an argon-based gas mixture at $1$~atm yields approximately $20$~clusters per centimeter. For a track at $\eta=0$, the effective path length through the TPC is $1.15-0.14=1.01$~m. Assuming a pad row pitch of $1$~cm results in about 100 hits per track. Substituting these values into eq.\eqref{eq:err_dEdx} and~\eqref{eq:err_dNdx} gives estimates of the relative uncertainties for both methods,
\begin{align}
\frac{\sigma_{dE/dx}}{dE/dx}&=5.5\%\text{ ,}
\end{align}
and
\begin{align}
\frac{\sigma_{dN_{cl}/dx}}{dN_{cl}/dx}&=2.2\%\text{ .}
\end{align}
These estimates highlight the improved precision achievable with cluster counting. However, determining the actual particle separation power requires detailed simulations to evaluate the $dN_{cl}/dx$ differences between particle species.

Achieving minimal uncertainty in cluster counting demands a sensor with a fine timing resolution to resolve individual clusters. For an argon-based gas mixture, a sampling rate of approximately $10$~ns is required~\cite{bib:clusterCountingArgon}, which is well within the capabilities of the TimePix3 chip, with a timing resolution of $1.56$~ns~\cite{bib:timePix3}.
\FloatBarrier
\subsection{Hadronic Calorimetry}

High–precision measurements of forward hadrons and jets are a key driver for the second EIC detector. In the hadron–going direction the HCAL must provide good energy resolution over a broad dynamic range while operating in a region of high radiation and large particle flux. The Yellow Report requirements call for a forward hadronic energy resolution of better than
\begin{equation}
    \frac{\sigma_E}{E} \lesssim \frac{50\%}{\sqrt{E/\mathrm{GeV}}}
\end{equation}
with a strong preference for $\sim 35\%/\sqrt{E}$ or better in order to fully exploit exclusive and semi–inclusive measurements. Conventional steel–scintillator sampling calorimetry is unlikely to reach this performance without large depth and cost; typical sampling HCALs achieve $\sim 60\%/\sqrt{E}$ for single hadrons. This motivates an ``upgraded'' forward HCAL concept that exploits dual readout of Cherenkov and scintillation light to approach $\sim 30\%/\sqrt{E}$ performance.

\subsubsection{Limitation of conventional sampling HCALs}

In non–compensating calorimeters the dominant contribution to the hadronic energy resolution arises from event–by–event fluctuations of the electromagnetic shower fraction $f_{\rm em}$, driven primarily by $\pi^0 \to \gamma\gamma$ production. The visible signal is
\begin{equation}
S \propto E[f_\mathrm{em}+h/e(1-f_\mathrm{em})]
\end{equation}
where $(h/e) < 1$ encodes the reduced response to the non–electromagnetic component. Fluctuations of $f_{\rm em}$ both broaden the response and induce non–linearity. Even with optimized sampling fraction and absorber, improving beyond $\sim 45$–$50\%/\sqrt{E}$ is extremely challenging without achieving true compensation ($h/e \approx 1$).
In the forward region, where space is constrained and radiation levels are high, increasing depth or segmentation alone is not sufficient. A qualitatively different handle on $f_{\rm em}$ is needed.

\subsubsection{Dual–readout principle: Separating scintillation and Cherenkov light}
Dual–readout calorimetry addresses this problem by measuring, in the same shower, two optical signals with different sensitivity to the electromagnetic and non–electromagnetic components. In its simplest form one measures \cite{Akchurin:2005yv}:
\begin{align}
S &\propto E \left[f_{\rm em} + (h/e)S(1-f_{\rm em})\right] , \\
C &\propto E \left[f_{\rm em} + (h/e)C(1-f_{\rm em})\right] 
\end{align}
where $S$ is dominated by scintillation light from all charged shower particles, while $C$ is Cherenkov light, produced predominantly by the relativistic component (essentially the electromagnetic part for typical thresholds). For suitable choices of active media one has $(h/e)C \ll (h/e)S$, so that the ratio $C/S$ is strongly correlated with $f{\rm em}$. Event by event, one can solve these two equations for $f{\rm em}$ and $E$, thereby removing the dominant source of fluctuations in the response.

\begin{figure}[h]
    \centering
    \includegraphics[width=0.48\linewidth]{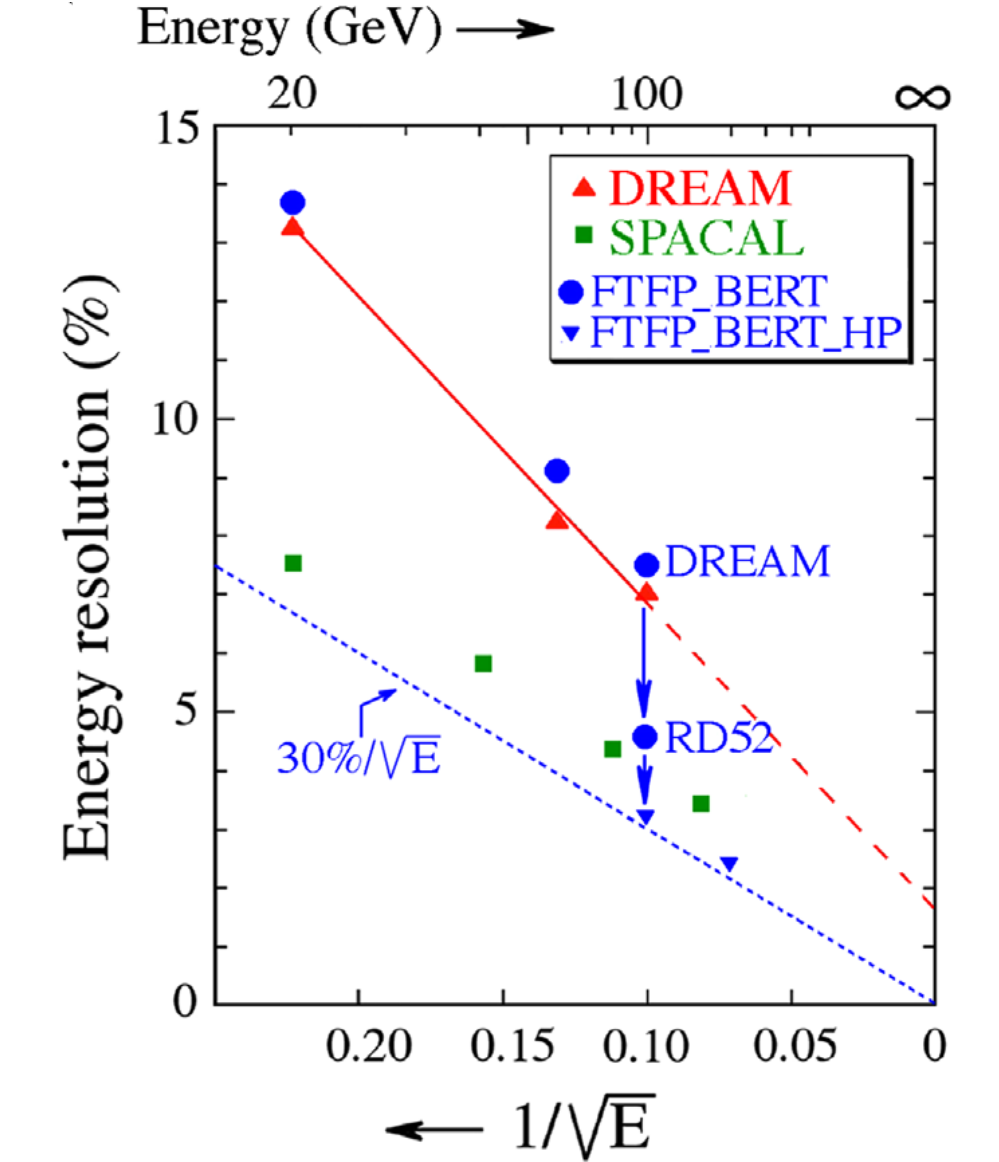}
     \caption{Experimental data on hadronic per-
formance compared to GEANT4 simulations.
The experimental data obtained with the original DREAM fiber
calorimeter are compared with simulations using the standard FTFP-BERT
hadronic simulation package for the geometry of that detector. The
improvement expected for a larger detector with the RD52 geometry is also depicted, both for the standard FTFP-BERT package and for the high-precision version of this package. For comparison, the record setting experimental data
reported by SPACAL are also shown, as well as a curve representing an energy resolution of 30\%/$\sqrt{E}$. From \cite{Wigmans:2016zfc}.}
    \label{fig:hcal-resol}
\end{figure}

The DREAM and RD52 programs at CERN and elsewhere have demonstrated this technique in copper–fiber and crystal/glass prototypes. A copper–fiber dual–readout module achieved hadronic resolutions of order
\begin{equation}
    \frac{\sigma_E}{E} \lesssim \frac{30-35\%}{\sqrt{E/\mathrm{GeV}}} \oplus (1-2)\%
\end{equation}
with an essentially linear response up to several hundred GeV, a dramatic improvement over the $\sim 60\%/\sqrt{E}$ resolution obtained when using the scintillation channel alone \cite{Lacava:2011zz}. Recent studies of dual–readout for homogeneous crystal calorimetry indicate that resolutions of $\sim 30\%/\sqrt{E} \oplus 2\%$ are achievable in realistic detector configurations.
Figure \ref{fig:hcal-resol} from \cite{Wigmans:2016zfc} summarizes the situation concerning the hadronic energy resolution, for single pions. 
A simple analytic expression for the dual–readout resolution in terms of sampling, photostatistics, and residual $f_{\rm em}$ fluctuations has been derived and validated against detailed simulations; it demonstrates that the sampling term can realistically be pushed to $\sim 25$–$30\%/\sqrt{E}$ for hadrons while maintaining a small constant term \cite{eno2025resolutiondualreadoutcalorimeters}.  This defines a natural performance goal for a forward EIC HCAL.

\subsubsection{Cherenkov vs. scintillation in high–density glass}

For an EIC detector, a promising realization of the dual–readout concept is based on high–density glasses that simultaneously provide scintillation and Cherenkov light, with the two components separated either spectrally or temporally. The ongoing SBIR/STTR program led by Tanja Horn from CUA and collaborators has developed so–called C/S glass, building on the earlier SciGlass work for electromagnetic calorimetry (see \ref{sec:tanja}).

Parallel R\&D in the CEPC and CALICE communities on scintillating glass–tile HCALs (primarily for particle–flow) has already shown that high–density glass tiles can deliver excellent light yield and uniformity in a sampling HCAL configuration \cite{Du:2022cxr}.

Separation of the two light components can be achieved in multiple ways:

\begin{itemize}
\item \textbf{Spectral separation}: Cherenkov light is concentrated in the near–UV/blue, while the glass scintillation spectrum can be shifted to longer wavelengths via dopants. Using dichroic filters or wavelength–selective coatings, one can route the two components to separate SiPMs.
\item \textbf{Timing separation}: Cherenkov light is prompt (tens of ps), while the scintillation typically has decay times of a few to tens of ns. With modern NUV–HD SiPMs and fast electronics, sub–100~ps timing on individual channels is achievable, enabling a time–based decomposition of the early Cherenkov peak and the slower scintillation tail.
\end{itemize}

An attractive option for the forward HCAL is thus a sampling structure of steel (or Cu/W) absorber plates interleaved with tiles of high–density C/S glass, read out on two channels per tile (C and S). This can be complemented by longitudinal segmentation and modest transverse granularity to preserve particle–flow performance in the overlap region with the barrel.

\subsubsection{Expected performance and impact on the EIC physics program}

Using the recent dual–readout resolution formula as a guide, one can estimate the performance of such a C/S–glass based forward HCAL. For reasonable sampling fractions and tile thicknesses, and assuming Cherenkov and scintillation light yields at the levels already demonstrated in existing prototypes, the sampling term is expected to be in the range of 25-30\% with a constant term $b \lesssim 2$–$3\%$, dominated by non–uniformities and calibration \cite{eno2025resolutiondualreadoutcalorimeters}.

 This would represent a factor of two improvement in resolution over a conventional steel–scintillator HCAL of similar depth, and would meet or exceed the Yellow Report performance targets in the forward region.
For the EIC physics program, such an improvement translates directly into:
\begin{itemize}
\item reduced missing–energy tails and better exclusivity cuts in deeply–virtual and photoproduction channels,
\item improved jet energy and mass resolution for forward jets, with direct impact on studies of transverse–momentum–dependent and 3D structure functions.
\end{itemize}
The dual–readout concept also naturally synergizes with high–granularity and precision timing developments, opening the possibility of a forward HCAL that combines excellent energy resolution with imaging capability and $\mathcal{O}(10~\mathrm{ps})$ timing for pileup and background suppression \cite{akchurin2024highgranularitydualreadoutcalorimeterevolution}.

\subsection{Electromagnetic Calorimetry}

\subsubsection{SciGlass}
\label{sec:tanja}
High-performance scintillator materials are needed for particle identification and measurements of energy and momentum of electromagnetic particles in modern nuclear physics experiments. As an example, the US Electron-Ion Collider, a unique collider with diverse physics topics, requires electromagnetic calorimetry enabling high-quality electron identification and detection in the momentum range of 0.3 to tens of GeV~\cite{ABDULKHALEK2022122447}. The highest resolution in electromagnetic calorimeters can be provided by homogeneous materials, e.g., lead tungstate crystals. Inorganic glass scintillators have been investigated as an attractive and cost-effective alternative to crystals, that is also easier and faster to manufacture in mass production. Here, we summarize progress and status in the fabrication and characterization of recent scintillating glass samples on both test bench and beam tests. 

Glass scintillators have been investigated over the last five decades for various applications beyond fundamental science, e.g., in industry, for medical diagnostics~\cite{LECOQ2016130} and in oil well logging. However, the requirements for glass in industrial applications differ from the needs for scientific application at an EIC, in particular in terms of light yield and uniformity of energy resolution, size, timing, and radiation resistance. R\&D is ongoing in industry and for scientific instrumentation~\cite{Auffray_2015,Auffray:2158947,deFaoite_2015,7581952,DORMENEV2021165762, DORMENEV20101082,Novotny_2019, Dorenbos:1996snm}. Some of the most promising materials investigated are Cerium-doped hafnate glasses, doped and undoped silicate glasses, and nanocomposite scintillators. All of these have various shortcomings that include lack of uniformity and macro defects, as well as limitations in radiation length, density, radiation resistance, and timing. One of the most recent efforts is DSB:Ce, which is a Cerium-doped glass nanocomposite, as well as its optimized form, a Ce doped Ba-Gd silica glass. Small samples of this material have been shown to be in many aspects competitive with PWO. However, the sample sizes produced to date are limited and the issue of macro defects, which can become increasingly acute on scale-up, and radiation length remain. Scintilex, LLC has developed a new family of glass scintillators (SciGlass) that have comparable performance to current nanocrystalline glass ceramic scintillators but have considerable advantages in terms of simplified manufacturing processes and ease of scale-up. These glasses are based on barium silicates, which provide a natural high-density durable base, but with a tailored mix of property-modifying additives that improve performance and facilitate large-scale production. Scintilex has applied this approach to develop scintillating glasses doped with a range of elements that allow tuning of the scintillation light. Scintilex routinely produces rectangular samples with lengths up to about 40cm and has been exploring further scale-up. In comparison, other recent efforts on scintillating glass, e.g., those recently produced in China by the Large Area Glass Scintillator Collaboration that was established in October 2021 have been limited in length to 10 cm. Other scintillating glass, e.g., that produced by Schott/Germany in the context of the CERN DRD6 have also faced scale-up challenges and no large-size blocks have been produced (longest sample was 15 cm long).

The most recent samples produced by Scintilex have a transmittance of ~70~\% at ~500nm approaching PWO in the region of interest for physics detectors and improved light yield uniformity compared to earlier samples. The transmittance of the new samples is generally suitable for physics detectors, as is the light yield uniformity in combination with advanced computational methods. Further improvements of the transmittance, and in particular in the longer wavelength region, may be beneficial in future work on SciGlass.
The latest SciGlass blocks have an induced radiation absorption coefficient, dk~1 m$^{-1}$ for integrated doses of 30 Gy accumulated at about 0.5 Gy/min and thus fulfill the radiation hardness requirements of the EIC experiments. SciGlass shows signs of radiation damage at 100 Gy and significant radiation damage at 1000 Gy. The radiation resistance of SciGlass is comparable to that of other Ce-based scintillation glasses. In the context of the EIC requirements SciGlass is radiation hard up to 30 Gy. For radiation hardness comparable to PWO, where saturation of the induced radiation coefficient at 1-2 m$^{-1}$ is reached for 100-150 Gy, further optimization of SciGlass and other Ce-based scintillating glasses would be needed.

The decay time distribution of SciGlass is characterized by a fast component of (83.8 +- 1.9) ns with a weight of 81~\% and a slow time component of (846.5 +- 0.2) ns with a weight of ~19~\%. This confirms that for dedicated light yield measurements a long time window should be used. Detailed studies like energy resolution measurements with detector prototypes – as those we have been carrying out for EIC - can help determine if the observed timing is an issue for the experiment at hand. Another avenue of investigation may take advantage of the Cherenkov component of SciGlass.

Initial beam tests have been carried out to evaluate the performance of SciGlass at GeV scale. These tests featured a comparison of PMT and MPPC (SiPM) readout to determine if MPPCs could be a suitable photosensor for SciGlass. The measured total energy deposition measured with a 3x3 array and electron energy of 4.7 GeV confirmed that MPPCs can be used as a readout method for SciGlass. The next step towards a SciGlass detector prototype beam test with was the evaluation of the on-beam performance of the full readout chain which included SciGlass blocks, photo-sensors (MPPCs), front-end (preamplifiers and bias boards), and back-end (flash-ADCs, triggered and streaming DAQ). A successful test would confirm a suitable readout chain so that the energy resolution measured would allow for true studies of the SciGlass properties (not predominated by the electronics). The beam test performed in 2023 underlined the importance of optimizing the full readout chain for scintillating glass and produced new front-end electronics with good stability.

To observe the response of SciGlass to hadrons we carried out a beam test with low-energy protons ranging from 80 MeV to 220 MeV. These first data with a hadron beam will allow for further characterizing of the glass as a calorimeter material and may also be used to give an indication of the Cherenkov contribution to the light yield. The latter may be of interest for improving hadron calorimeter performance. We also carried out a beam test with low-energy electrons ranging from 150 MeV to 630 MeV and analyzed the data. The physics results from the SciGlass characterization studies and results from the completed beam tests were published in Ref~\cite{Horn:2025hnz}. 

\subsubsection{LYSO}

 For the last decades the LYSO (lutetium-yttriumoxyorthosilicate) heavy crystal scintillators have been developed which are now widely considered for particle physics experiments, due to high light yield (x200 times of PWO crystals), fast decay time (40 ns), and superb radiation hardness (x5 times larger than PWO). These make them an obvious technology candidate for high precision electromagnetic calorimetery, particularly at low energy, and time of flight measurements. Its typical characteristics are: density 7.4 $g/cm^3$, radiation length $X_0$ = 1.14 cm and Moliere radius $R_M$ = 2.07 cm.

 The high light output of LYSO crystals enables the low energy photon and electron measurements down to 1 MeV and even below. The energy resolution of $\sigma_E/E<1\%/\sqrt{E}\oplus 1\%$ has been routinely achieved. The last generation crystals with high uniformity demonstrated further improved energy resolution of $\sigma_E/E=1.5\%$ for 70 MeV positrons, and time resolution of $\sigma_t$=110 ps for 30 MeV positrons (arXiv:2409.14691), dropping to $<$100 ps at E$>$50 MeV, and even down to 30 ps for 10 GeV photons for a particular calorimeter design (Conf. Series 928 (2017) 012023). 

 An array of a few mm thick LYSO crystals can be used for time-of-flight measurements for charged particles. Such a thickness leads to a few MeV energy deposit by a MIP particle that provides light output up to a few thousand photoelectrons, depending on photosensor type and configuration. CMS collaboration achieved 30 ps time resolution for MIP particles for a proto-type module for the MIP Timing Detector to be built of 3.75 mm thick LYSO crystals with SiPM readout (arXiv:2504.11209). 

LYSO crystals can tolerate harsh radiation conditions (up to $\sim$10 Mrad) which makes them a strong candidate for measurements close to the beam line. Another advantage of LYSO scintillators is their small temperature dependence of light output of $-0.2 \%/^{\circ}C$ (x10 times smaller than for PWO crystals), which allows to relax the requirements for temperature stability of the detector.

\subsection{Muon Detection}
While the ePIC detector can identify muons using the electromagnetic and hadronic calorimeters, the potential inclusion of a dedicated muon detector in a second detector has been investigated. Although not currently part of the ePIC baseline design, such a system could provide complementary capabilities. We have explored dedicated muon detector options for both the barrel and, potentially, the forward region—either as additions to or replacements for the hadronic calorimeters. In the barrel, where the hadronic calorimeter is primarily used for neutral particle detection rather than precise energy measurements, a muon detector based on the Belle-II KLM design may be a viable replacement. In contrast, integration in the forward region is more challenging due to two key constraints: the hadronic calorimeter is essential, and space is limited by the current interaction region (IR) design. This section discusses the advantages of the di-muon channel over the di-electron channel for quarkonium reconstruction (Section~\ref{benefit_di-muon_channel}), evaluates muon identification capabilities of the current ePIC design (Section~\ref{ePIC_muon_capabilities}), and presents simulation studies based on the Belle-II KLM detector (Section~\ref{KLM}). 

\subsubsection{Benefit} 
\label{benefit_di-muon_channel}
\begin{figure}[h]
    \centering
    \includegraphics[width=0.48\linewidth]{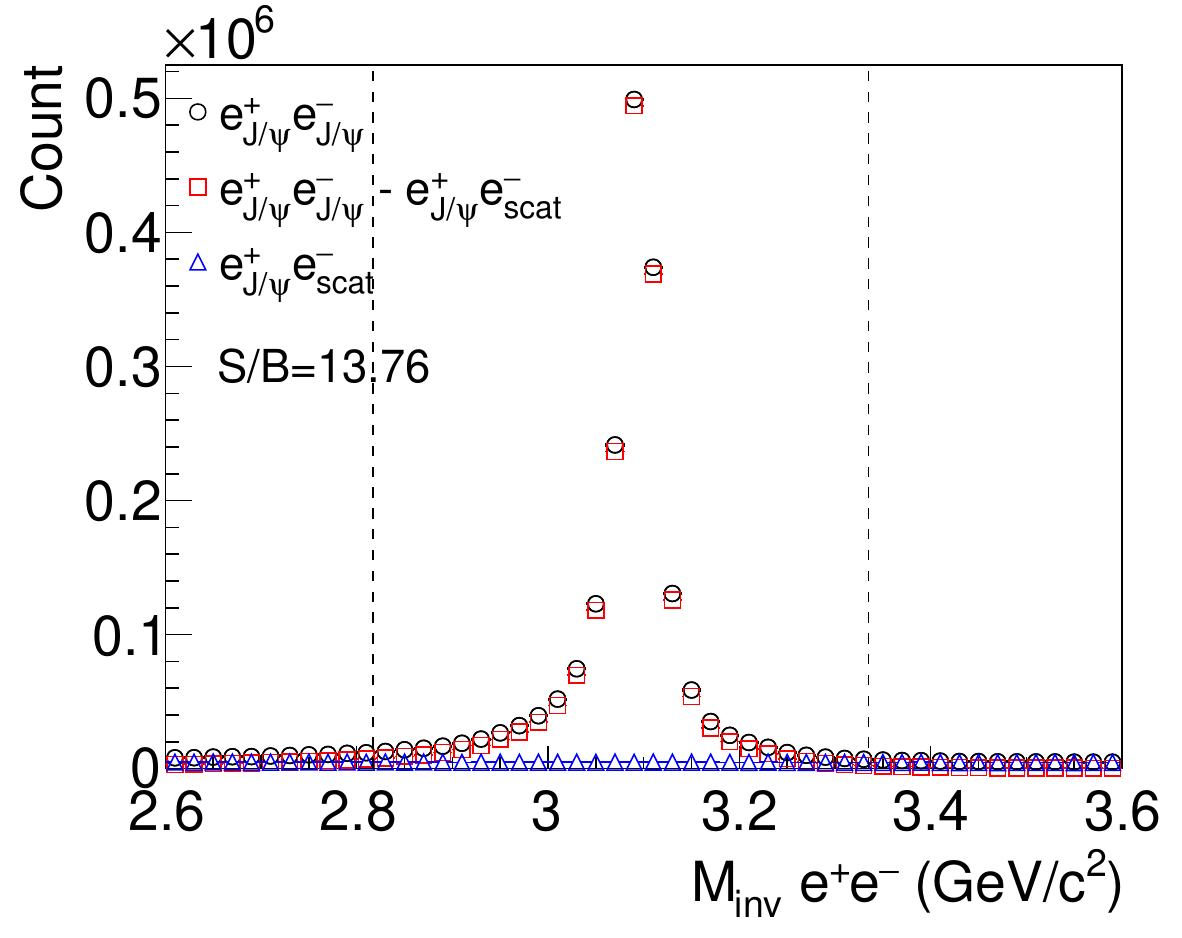}
    \includegraphics[width=0.48\textwidth]{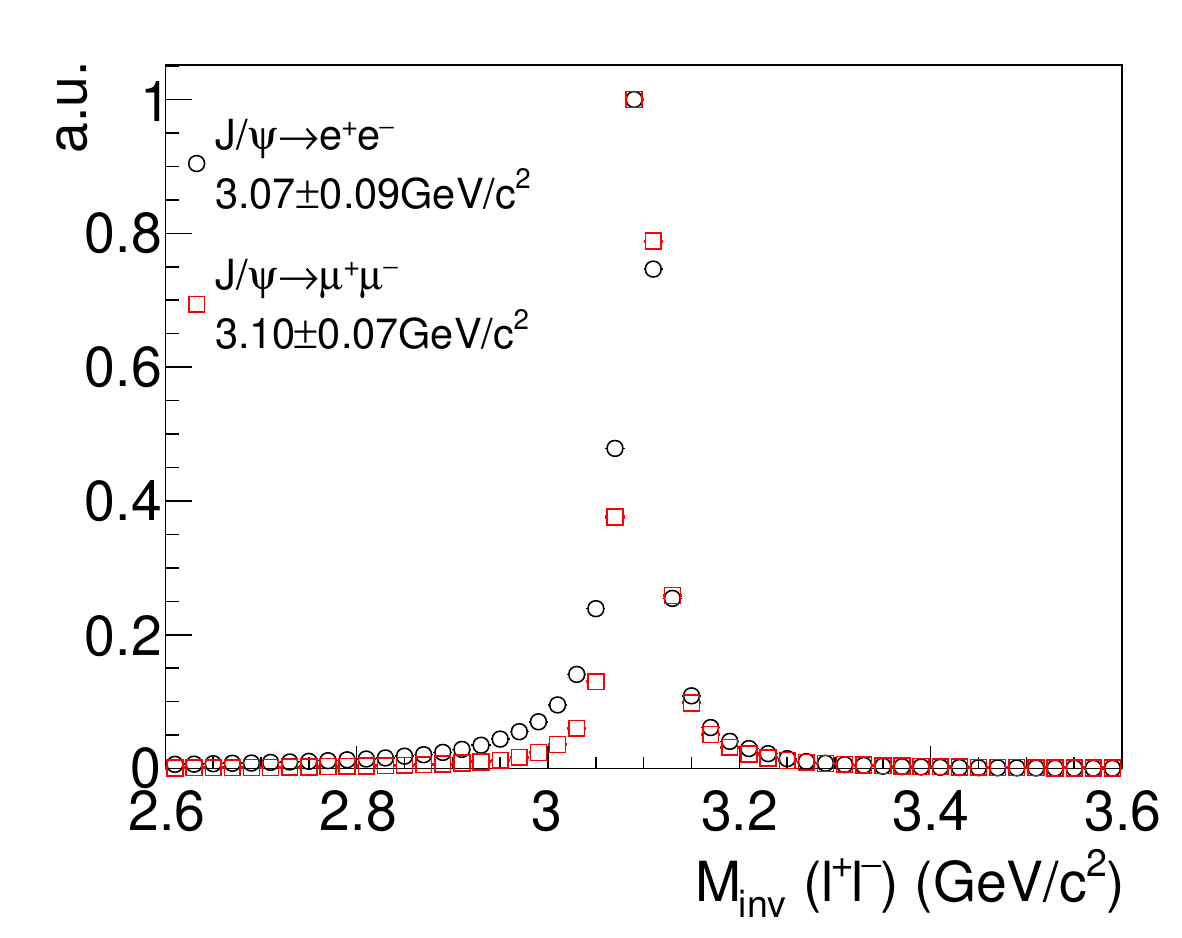}
    \caption{Left: Invariant mass spectra of dielectrons. Right: Invariant mass spectra of dileptons. Combinatorial background from scattered electrons is excluded. The spectra on the right are scaled to the same height for shape comparison.}
    \label{fig:invMass_ee_vs_mumu}
\end{figure}

Simulation using the ePIC tracking design were conducted to evaluate the benefits of muon identification in $J/\psi$ reconstruction. The study is based on exclusive events without other physics processes. True PID was applied, meaning that the only background present in the invariant mass spectra arises combinatorial pairs of $J/\psi$-decayed positrons and scattered electrons. The dielectron invariant mass spectra  shown on the left of Fig.~\ref{fig:invMass_ee_vs_mumu} demonstrate that this combinatorial background is negligible, yielding a signal-to-background ratio of $13.76$. However, in more realistic inclusive events, the combinatorial background is expected to be higher.

The right panel of Fig,~\ref{fig:invMass_ee_vs_mumu} compares the invariant mass spectra from dimuons and dielectrons channels. The dielectron spectrum excludes scattered electron backgrounds for clarity. Both channels give a clear $J/\psi$ mass peaks at about $3.09$~GeV/$c^2$. However, the dielectron channel shows stronger lower mass tail, attributed to the larger bremsstrahlung energy losses experienced by electrons and positrons compared to muons and anti muons.

\begin{figure}[hb]
    \centering
    \includegraphics[width=\textwidth]{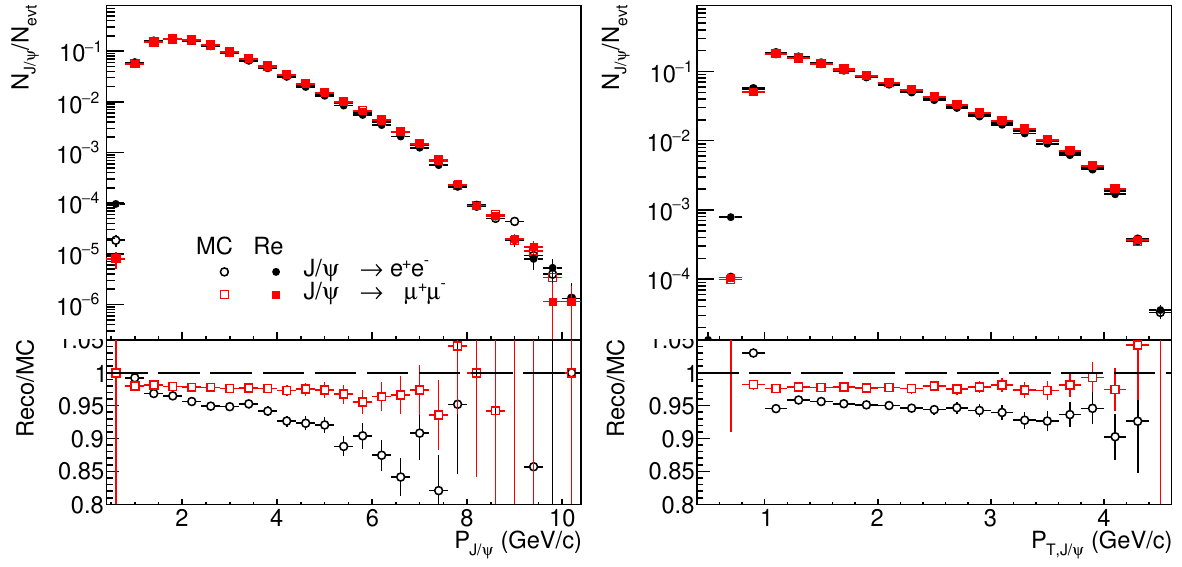}
    \caption{Reconstructed $J/\psi$ momenta (left) and transverse momenta (right).}
    \label{fig:jpsiMom_ee_mumu}
\end{figure}

Figure~\ref{fig:jpsiMom_ee_mumu} shows the reconstructed $J/\psi$ momentum and transverse momentum spectra, as well as the ratio of reconstructed to true values, for both dilepton channels. While the dimuon channel exhibits a flat reconstruction ration around $95$\%, the dielectron channel shows a decreaseing trend with increasing $J/\psi$ momentum and transverse momentum. This trend likely reflects energy loss due to bremsstrahlung, where high momentum $J/\psi$s are reconstructed with lower momenta. This miscontruction requires correction through unfolding techniques.

These simulation results show that the muon identification has a positive impact on $J/\psi$ reconstruction by eliminating ambiguity with scattered electron and mitigating the effects of bremsstrahlung. 



\FloatBarrier

\subsubsection{ePIC-like calorimeters} \label{ePIC_muon_capabilities}
As the second EIC detector is being considered with a dedicated barrel muon system, we extend our study to explore the feasibility of adding dedicated muon detectors in the forward region. Such detectors are not part of the current ePIC baseline design, and would thus provide a complementary capability. However, integrating a forward muon detector presents significant challenges due to (i) the requirement for hadron calorimetry in this region and (ii) limited space imposed by the current interaction region (IR) design.

To assess the forward muon identification performance using existing ePIC subsystems, we conduct simulation studies based on ePIC-like calorimeters and PID detectors. In particular, we evaluate the ability to identify minimum ionizing particle (MIP)-like signals in the EMCAL and HCAL, which can serve as signatures of muons. For the EMCAL, we utilize the total energy deposition and the total number of hits; for the HCAL, we exploit both the energy deposition and hit distribution across individual layers to improve discrimination between muons and pions.

We generated 10,000 single-particle events for both muons and pions over a range of momenta and angles. The dataset was split evenly, with one half used to train a Multilayer Perceptron (MLP) neural network implemented in ROOT TMVA, and the other half used for validation. After training the classifier to distinguish muons (signal) from pions (background), we applied it to a mixed sample of 2 million muon and pion events to evaluate the muon identification efficiency. This study also serves as a case example of applying machine learning techniques to enhance particle identification in forward detector regions.

\begin{figure*}[h]
    \centering
    \includegraphics[width=\linewidth]{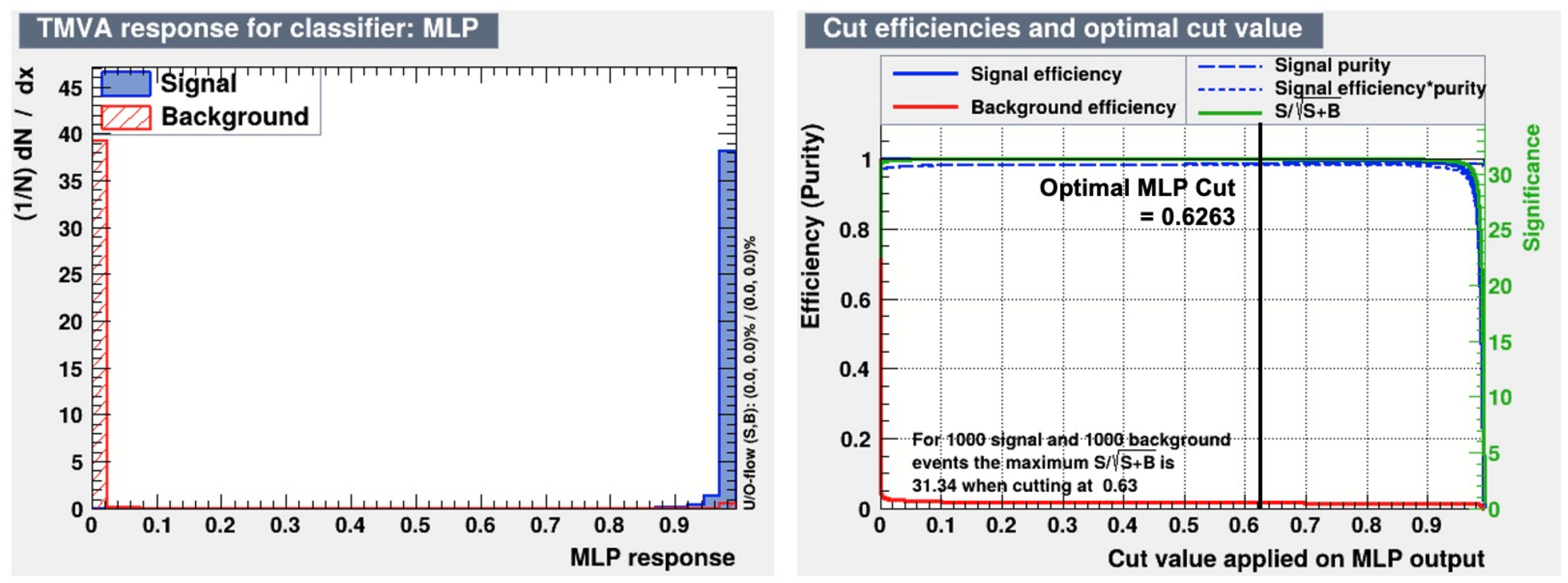}
    \caption{Left: Distribution of the MLP classifier output for signal (muons, blue histogram) and background (pions, red histogram). The small overlap of the red histogram beneath the blue indicates MIP-like pion background, where pions mimic muon behavior. Right: Signal and background efficiencies as a function of the MLP output cut value. The goal is to maximize signal efficiency while minimizing background efficiency. The optimal cut value was determined to be 0.6263; events with an output above this threshold are classified as signal, while those below are treated as background.}
    \label{fig:tmva}
\end{figure*}

Based on input variables derived from detector responses in the EMCAL and HCAL, machine learning techniques were employed to determine the discrimination distribution and optimal selection cut, as shown in Fig.~\ref{fig:tmva}. In the left panel, the small overlap of the background (pion) histogram beneath the signal (muon) histogram reflects MIP-like pion events, where pions mimic the behavior of muons. These events are intrinsically difficult to distinguish. The right panel shows the signal and background efficiencies as a function of the MLP output cut value. The aim is to maximize signal efficiency while minimizing background contamination. An optimal cut value of 0.6263 was selected; events with MLP output above this threshold are classified as signal, while those below are treated as background.

\begin{figure*}[h]
    \centering
    \includegraphics[width=0.6\linewidth]{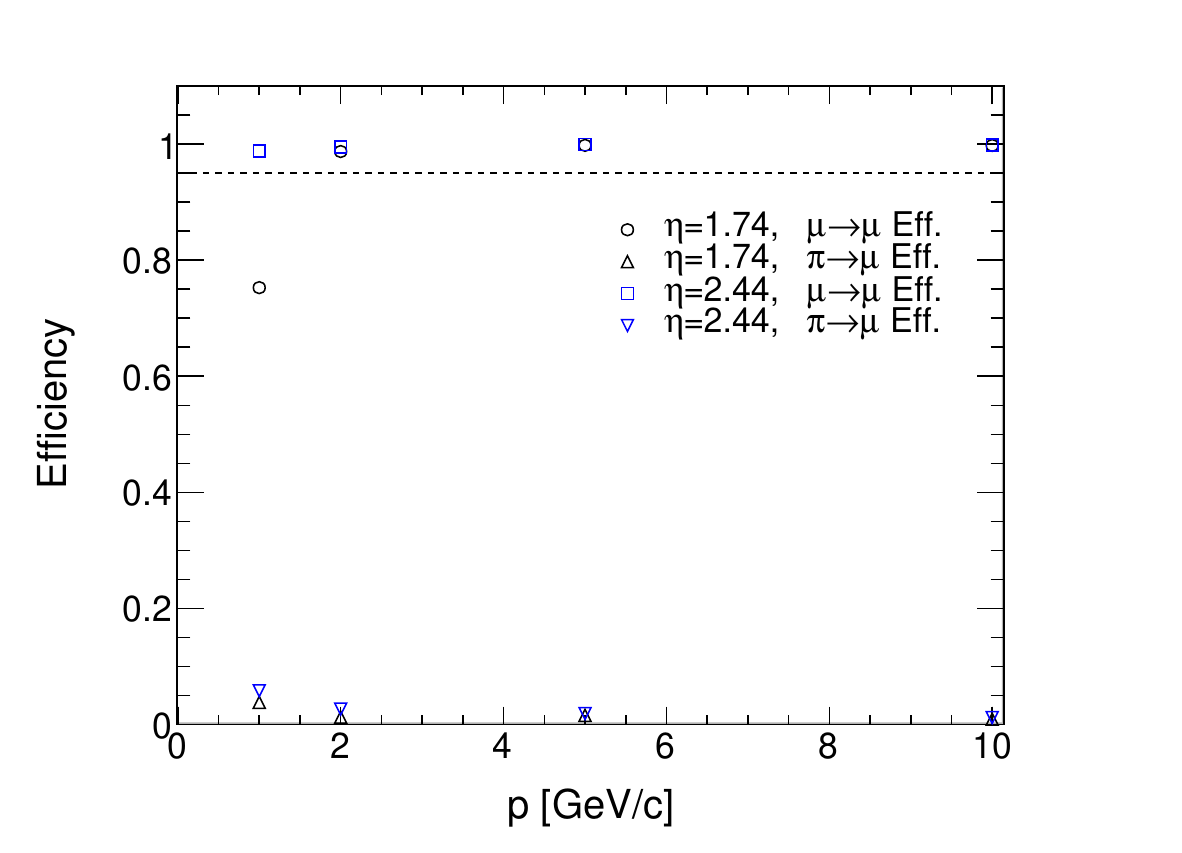}
    \caption{Muon identification (muID) efficiency and misidentification (mis-ID) rate as functions of momentum at two different pseudorapidities: $\eta = 1.74$ and $\eta = 2.44$. The muID efficiency represents the probability of correctly identifying a muon, while the mis-ID rate corresponds to the probability of a pion being misidentified as a muon. Results demonstrate high muID efficiency and low pion misidentification, particularly at forward pseudorapidity.}
    \label{fig:muID_eff}
\end{figure*}

\begin{figure*}[h]
    \centering
    \includegraphics[width=\linewidth]{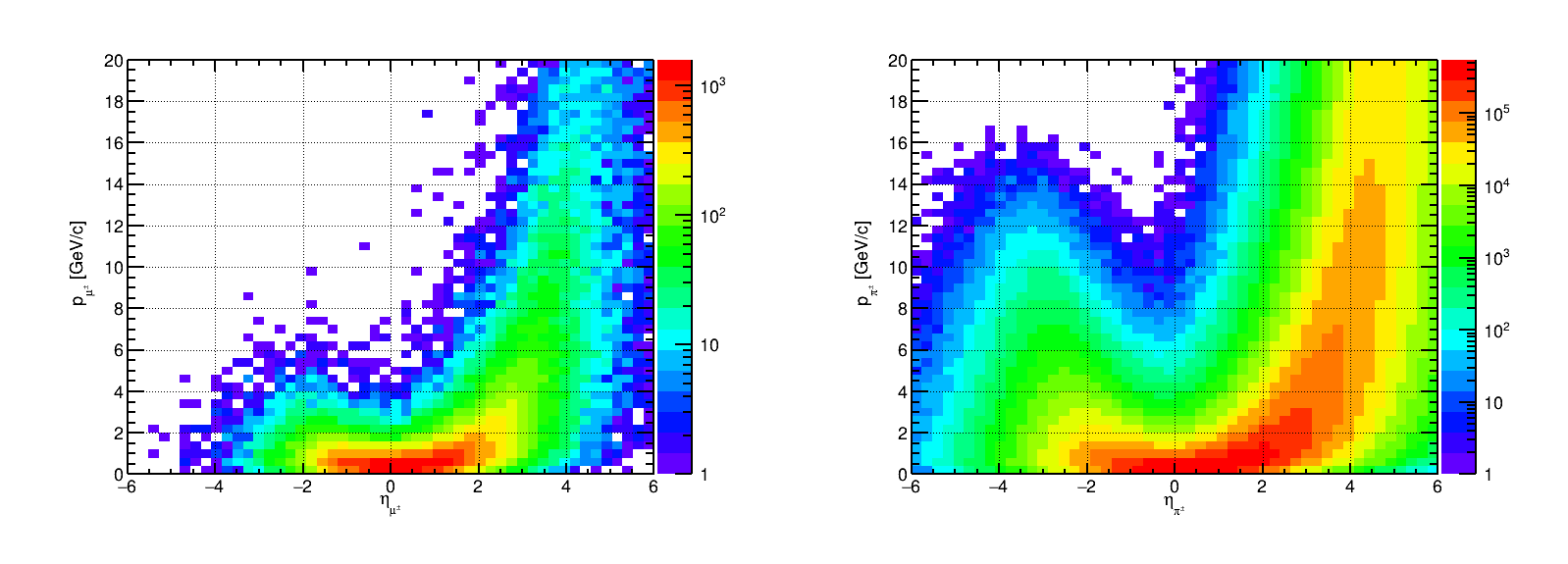}
    \caption{Two-dimensional histograms of momentum versus pseudorapidity from 10 million Pythia-generated $ep$ collisions at 18$\times$275~GeV$^{2}$. (Left): Muons; (Right): Pions. The number of muons and pions in each momentum–pseudorapidity bin was extracted and used to derive a scaled cross-section, which was applied to the simulated muon and pion samples for a more realistic evaluation of muon identification performance.}
    \label{fig:cross_section_mu_pi}
\end{figure*}

Figure~\ref{fig:muID_eff} presents the muon identification efficiency (true muons classified as muons) and the misidentification rate (pions classified as muons) as functions of momentum at two different pseudorapidities. In the forward (hadron-going) region, the pion misidentification rate is below 1\% for the given sample size. However, when considering the relative cross sections of muons and pions, as shown in Fig.~\ref{fig:cross_section_mu_pi}, identifying single muons using calorimeter information alone remains challenging. In contrast, muon pair reconstruction, such as from $J/\psi \rightarrow \mu^{+} \mu^{-}$ decays, may offer better background suppression through invariant mass reconstruction.

In addition to calorimeter-based identification, we investigated the potential for particle identification (PID) using dedicated ePIC PID subsystems, specifically the dRICH and forward time-of-flight (TOF) detectors, with a focus on low-momentum muons and pions. This is motivated by the fact that low-momentum particles in the forward region may not reach the EMCAL or HCAL. Given the similar masses of muons and pions, a separation of at least 3$\sigma$ is still required for reliable PID.

\begin{figure*}[h]
    \centering
    \includegraphics[width=\linewidth]{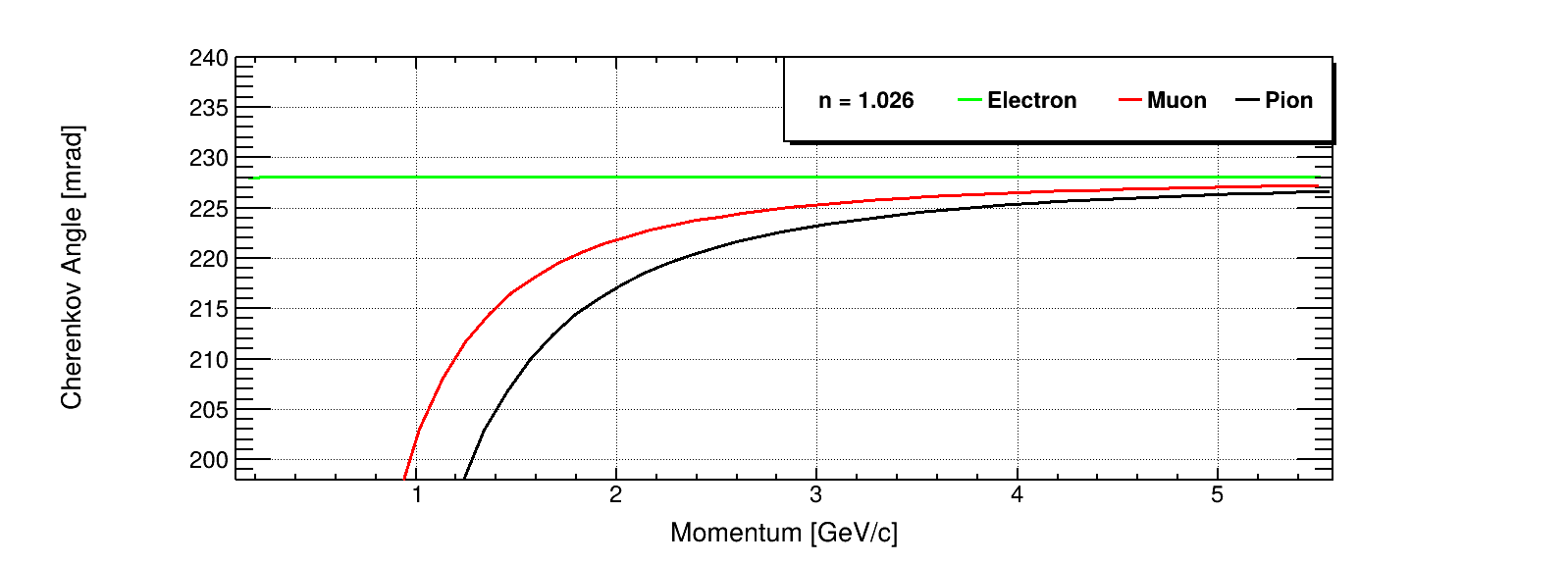}
    \caption{Cherenkov angle as a function of momentum for electrons (green), muons (red), and pions (black), calculated using a refractive index of $n = 1.026$ corresponding to the aerogel radiator in the dRICH. This estimation illustrates the separation power of the dRICH at low momenta, particularly for distinguishing muons from pions.}
    \label{fig:drich_separation_power}
\end{figure*}

To evaluate the PID performance, we performed theoretical estimations based on current ePIC detector parameters. Figure~\ref{fig:drich_separation_power} shows the Cherenkov angle calculated using the approximation $\theta^{2} \approx 2(n - 1) - \frac{m^{2}}{p^{2}}$, assuming a refractive index of $n = 1.026$. The resulting difference in Cherenkov angles between muons and pions suggests a separation limit up to approximately 3.5~GeV/$c$, indicating that the dRICH could provide useful discrimination below this momentum. This estimate assumes a track-level ring resolution of 0.5~mrad with 15 detected photons.

\begin{figure*}[h]
    \centering
    \includegraphics[width=\linewidth]{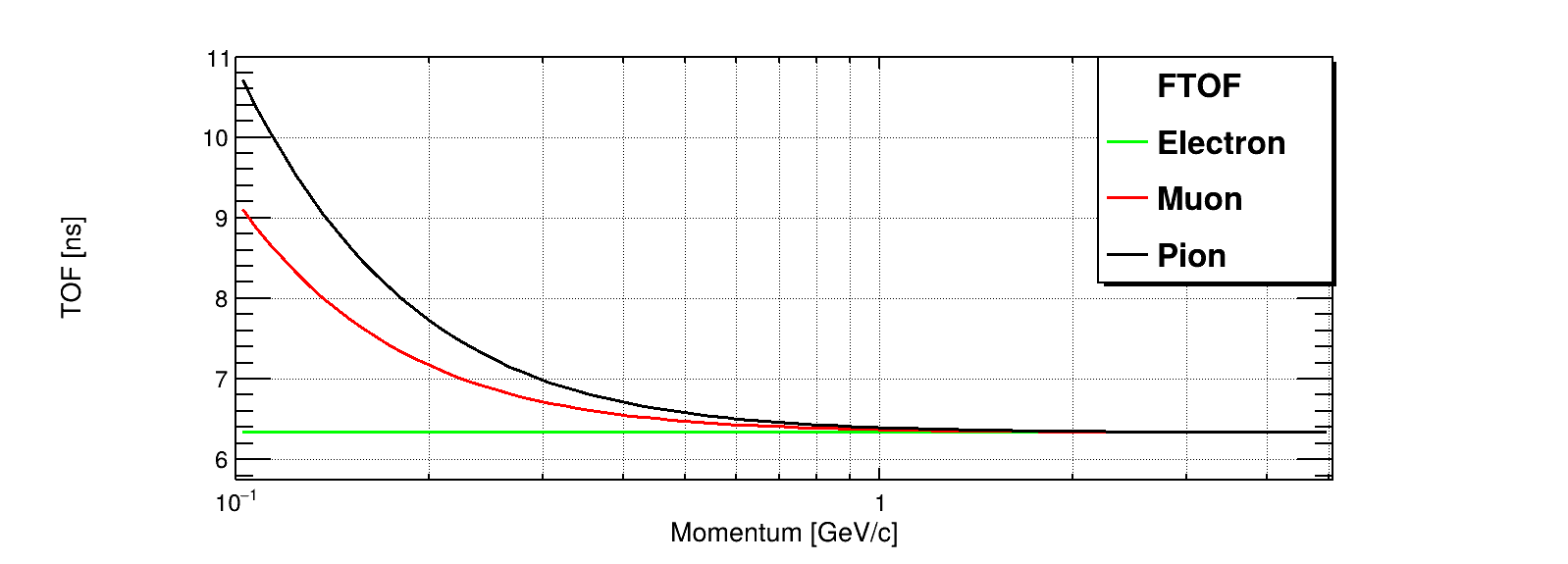}
    \caption{Time-of-flight as a function of momentum for electrons (green), muons (red), and pions (black), calculated assuming a straight-line path from the interaction point to the forward TOF plane ($z_{\textrm{min}} = 185$ cm, $r_{\textrm{max}} = 60$ cm). This estimation illustrates the separation power of the forward TOF (FTOF) detector at low momenta, particularly for distinguishing muons from pions.}
    \label{fig:fTOF_separation_power}
\end{figure*}

Figure~\ref{fig:fTOF_separation_power} shows the estimated time-of-flight difference between muons and pions based on a straight-line path from the interaction point to the forward TOF plane, with $z_{\mathrm{min}} = 185$~cm and $r_{\mathrm{max}} = 60$~cm. Assuming a TOF resolution of 30~ps, the separation limit is found to be approximately 0.5~GeV/$c$.

These estimates suggest that limited muon–pion separation is achievable at low momentum using current ePIC PID detectors in the forward region. However, a second EIC detector with upgraded PID systems could offer improved separation capabilities for muons and pions at low momentum.

\subsubsection{KLM for hadron/muon detection} \label{KLM}
As described above, an area in which a potential 2nd detector at the EIC can be complementary to the project detector ePIC is the muon system. Here we describe a proposed dedicated muon system for the 2nd detector that consists of active scintillator layers interleaved with passive iron layers that also serve as the magnet flux return of the central solenoid. Details can be found in Ref~\cite{Kelleher:2025yem}. This follows a similar design at Belle and Belle II detectors for their respective $K_L$ and $\mu$ id (KLM) systems. In addition to excellent muon identification, horizontal and vertically segmented readout will provide enough granularity to provide, in conjunction with ML/AI based reconstruction, competitive hadronic calorimetry. 
In contrast to the existing system at Belle II, direct readout with SiPMs will provide excellent timing information. This will enable a more compact design as it can be used to determine hit position along a strip and thus removes the need for an orthogonal strip layer. It can also be used to determine hadron momenta using Time-of-Flight.
Such a system would support a physics program using muons, as described earlier, as well as with jets. At the EIC about one third of jets contain neutral hadrons which makes their detection and energy determination important.

For the purpose of studying such a system, a full DD4HEP implementation was simulated. The baseline design has an inner radius of 1770~mm and an outer radius of 2835.4~mm with a KLM thickness of 1065.4~mm. R\&D on the KLM is the focus of another project investigated by Duke University, the University of South Carolina, and Indiana University and funded under U.S. DOE Field Work Proposal JLAB-NP-13, ``Generic EIC-related Detector Research and Development Program Description Under the Auspices of the Electron-Ion Collider Program Managed by Thomas Jefferson National Accelerator Facility", Proposal EICGENR\&D2023\_18. In that project, an optimization of the KLM for different objectives is discussed. This and more details will discussed in the corresponding forthcoming publication. Below we discuss results with a baseline design using fourteen layers of steel, interleaved with EJ-204 scintillator with a rise time of 1.8~s and S14160-4050H SiPMs, covering 64\% of the scintillator. A photodetection efficiency of 50\% is used. The carrier and amplifier board is based on the the boards developed for the HELIX experiment~\cite{Park:2021oic}.The effect of noise is not simulated yet. 

\begin{figure}[h]
\includegraphics[width=0.49\textwidth]{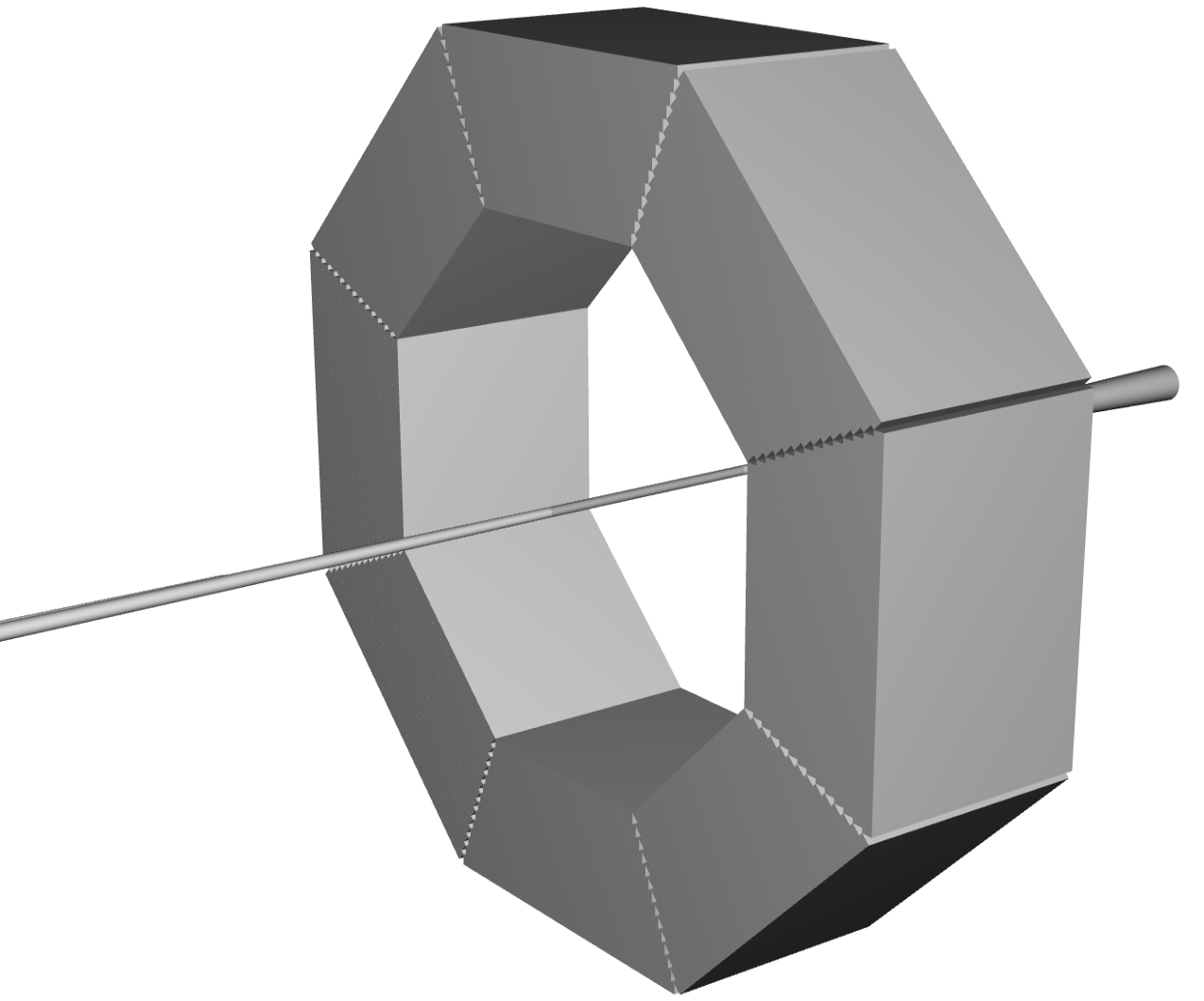}
\includegraphics[width=0.49\textwidth]{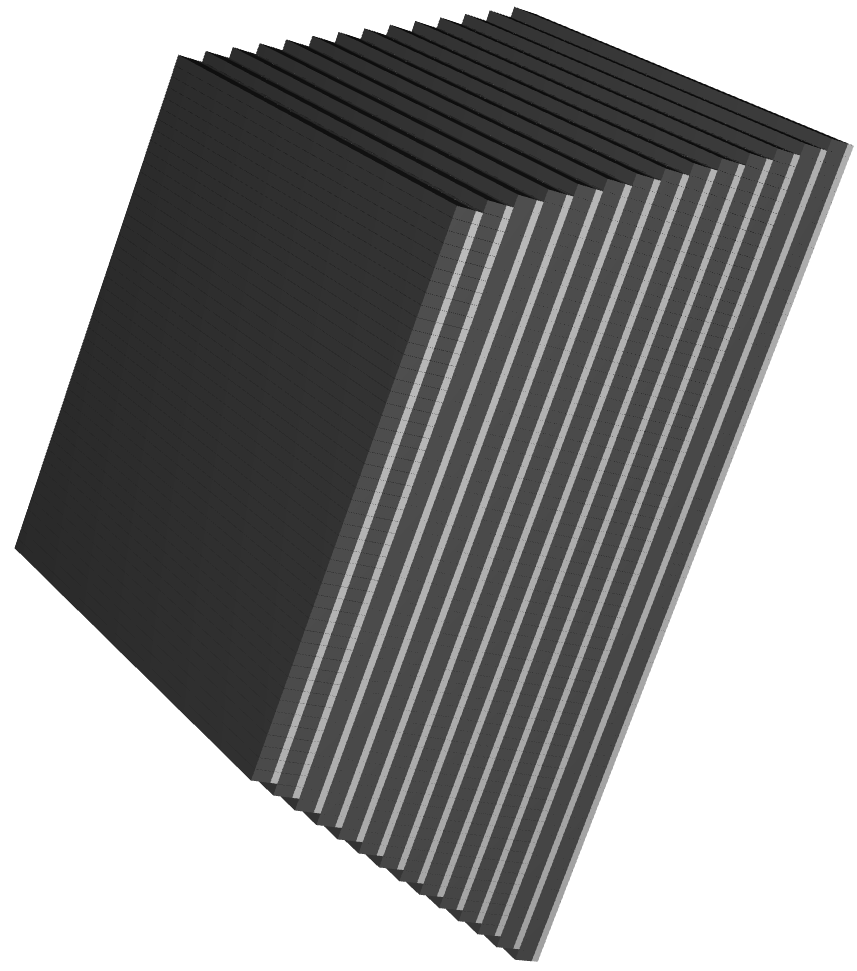}
\caption{Octagonal arrangement of sectors around the beampipe (left) and iron/scintillator sandwich structure of one sector (right)\label{fig:klmDesign}.}
\end{figure}

\begin{figure}[h]
\centering
\includegraphics[width=0.95\textwidth]{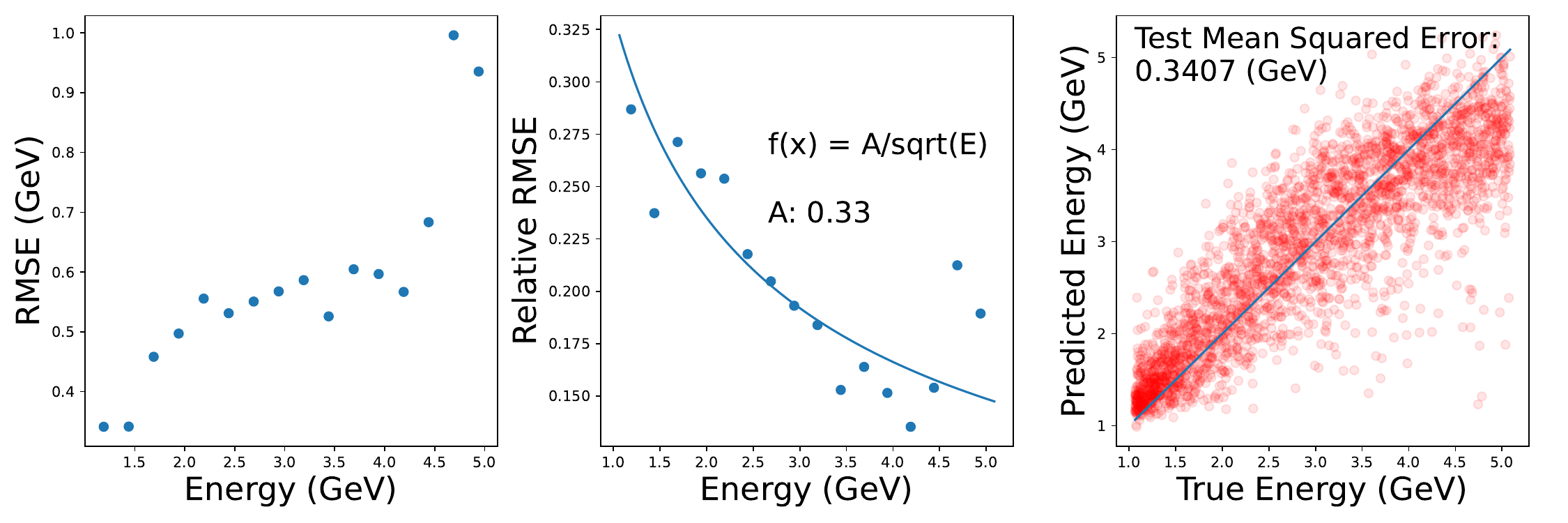}
\caption{\label{fig:klmERes} Energy resolution for neutrons for the baseline design.}
\end{figure}

\begin{figure}[h]
\centering
\includegraphics[width=0.49\textwidth]{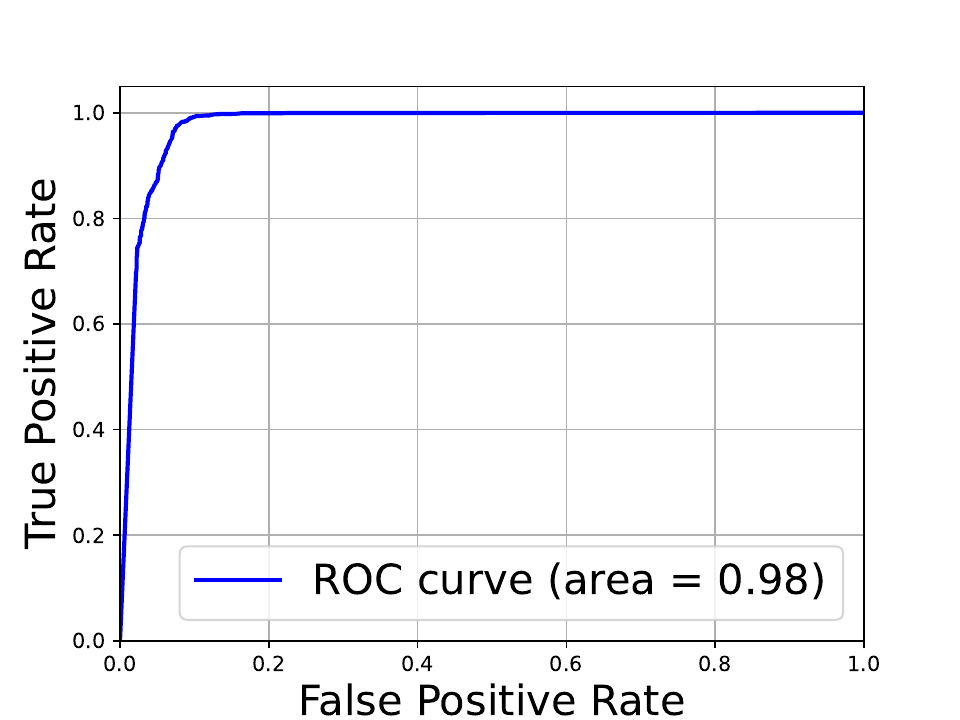}
\includegraphics[width=0.49\textwidth]{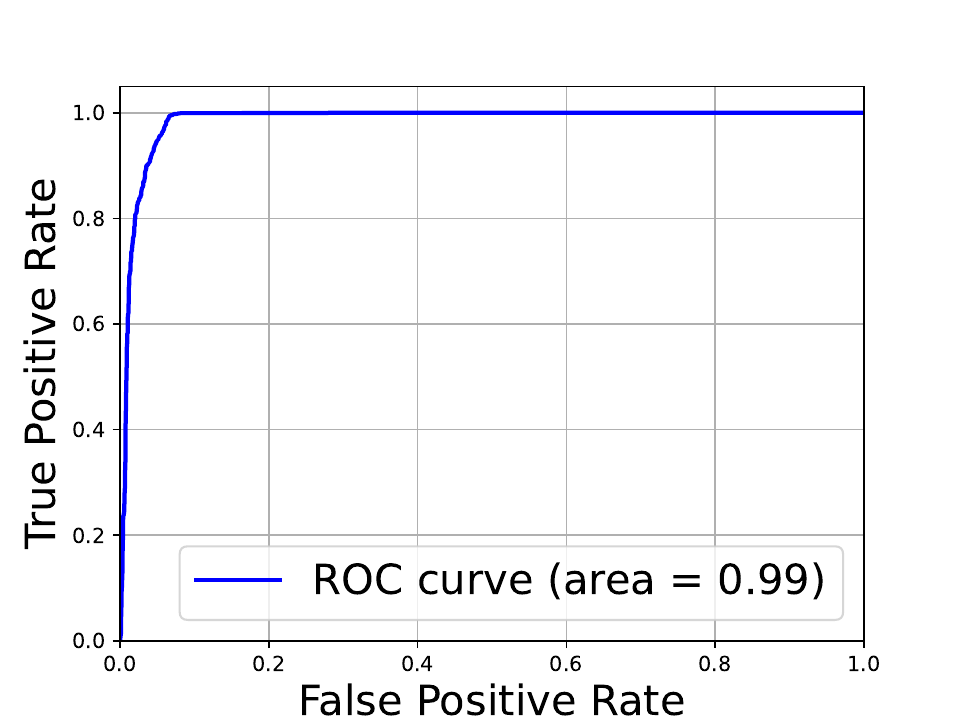}
\caption{ROC curves for MuID performance with baseline design using a GNN method for 1 GeV muons (left) and 5 GeV muons (right).\label{fig:ldrdMuonPID}}
\end{figure}

The circuit is simulated with a rise time of 1.1~ns, fall time of 15~ns. We use  A steel scintillator split of 73\% to 27\%, which is sufficient for the magnetic flux return for a 1.5~T central magnet and enable muon ID over a wide momentum range. Scintillator strips of 1.5~m length, 3~cm width and 2cm thickness are assumed in the baseline design. 
The layout of the detector around the beamline is shown in Fig.~\ref{fig:klmDesign}.
The simulation with optical photons generated in the scintillator and propagated through the readout, including digitization, shows that a timing resolution of about 100~ps can be reached. This corresponds to a spacial resolution along the strip of about 2~cm, comparable with the information from an orthogonal layer. Using ToF, this resolution would result in a momentum resolution of about 15\% for $K_L$ mesons with a momentum of about 1.2 GeV, whereas the same resolution can be reached for neutrons at a momentum of 1.8~GeV.
Figure~\ref{fig:ldrdMuonPID} shows the performance of the system for muon ID which reaches 0.99 in the area under the ROC curve. Figure~\ref{fig:klmERes} shows the energy resolution reached in simulation which reaches around $\nicefrac{0.33}{\sqrt{E}}$.


\FloatBarrier
\section{Needed R\&D}

The success of the EIC has been built upon a strong foundation of a generic detector research and development (R\&D) program that ran from 2011 until 2021, when it was superseded by a project-focused program aimed at reducing the risks associated with the baseline technologies chosen for ePIC. Over the course of a decade, the generic R\&D program proved to be an essential driver of innovation, enabling the development of advanced technologies that now form the backbone of the ePIC detector. By fostering a broad, exploratory environment, the program not only provided solutions to immediate technical challenges but also seeded novel concepts that ultimately enhanced the scientific capabilities of the EIC.

Beyond hardware development, the program supported a variety of vital projects, such as machine background studies and simulation software, which enabled a more accurate definition of the physics requirements. Perhaps most importantly, the program helped build a strong and cohesive community around the EIC. This is clearly reflected in the large number of scientists who participated in the R\&D program and who now form the core of the detector subsystem groups in ePIC.

As we look toward the prospect of a second detector, the role of a vibrant and sustained R\&D effort becomes even more critical. With sufficient lead time before construction, such a program offers the unique opportunity to move beyond incremental improvements and to pursue transformative technologies that can redefine the scope of what is experimentally achievable. Areas such as next-generation tracking, calorimetry, particle identification, and integration with modern computing and AI-driven data acquisition can all benefit from dedicated investment in exploratory R\&D.

Importantly, the need for such a program is echoed by parallel efforts in the global high-energy physics community. Current R\&D activities for the FCC-ee, the proposed high-luminosity electron–positron collider in Europe, demonstrate striking similarities to the requirements of an EIC detector. Both machines demand ultra-precise tracking, highly granular calorimetry, and advanced particle identification in a high-rate environment. This overlap underscores the timeliness of investing in detector R\&D for the EIC, as cross-fertilization of ideas between these communities can accelerate innovation and broaden the impact of new technologies.

Recent U.S. efforts provide another important lesson. At Jefferson Lab, a new two-year generic detector R\&D program was launched with the goal of supporting both the EIC and broader nuclear physics initiatives. Unfortunately, this program was terminated by the Department of Energy after only two years, before its full potential could be realized. The early termination of that effort highlights the need for long-term continuity in R\&D support if we are to nurture transformative technologies and ensure the readiness of the community to deliver cutting-edge detectors.

Examples of R\&D projects from this short-lived second round, many of which were aimed at novel technologies particularly relevant for a future second detector, included:
\begin{itemize}
\item Generic glass scintillators for EIC calorimeters
\item Continuation of EIC KLM R\&D
\item Pressurized RICH
\item Z-Tagging Mini DIRC
\item GridPIX detector with low-mass and high-efficiency cooling
\item Development of high-precision and eco-friendly MRPC TOF detector for the EIC
\item Development of double-sided thin-gap GEM-$\mu$RWELL for tracking at the EIC
\item A fast-timing MAPS detector for the EIC
\item Large-area monolithic active pixel sensors
\item Scintillator fiber trackers for the ZDC and off-momentum detectors
\item Photonics-based readout and power delivery by light for superconducting nanowire detectors for the EIC
\end{itemize}
and many more.
This portfolio illustrates both the breadth and ambition of detector R\&D needed for the EIC. It also demonstrates that, given the proper support, the U.S. community is well-positioned to explore bold new ideas that can significantly advance the capabilities of a second detector. By contrast, in Europe, ongoing R\&D programs for the FCC-ee are sustained over longer timescales, reflecting a recognition that transformative detector technologies require stability and continuity. The parallel challenges of FCC-ee and the EIC—such as ultra-precise tracking, advanced calorimetry, and timing—highlight that only through consistent, long-term investment can the community deliver the breakthroughs essential for the next generation of collider experiments.

Equally important, a second detector R\&D program will serve as a powerful catalyst for community building. By engaging a diverse set of institutions, disciplines, and early-career researchers, it will expand the reach of the EIC project, ensuring a broader participation base and fostering the next generation of detector scientists. This inclusive environment of innovation and collaboration is essential not only for the technical success of a second detector but also for the long-term vitality of the EIC scientific community.
In this sense, the R\&D program should be seen not as a peripheral activity, but as a {\emph central} pillar in the realization of a second detector—both in terms of technical readiness and in strengthening the collaborative fabric of the community that will bring it to life.

In this chapter, we discuss a selection of potential R\&D projects that emerged from our LDRD efforts.
This list is by no means exhaustive, but it highlights several key technologies that are expected to play a major role in a second detector and that will require substantial dedicated R\&D:
\begin{itemize}
\item \textbf{Monolithic LGADs}: Combining the excellent timing resolution of LGADs with the integration and compactness of monolithic sensors would enable unprecedented precision in both space and time, particularly valuable for vertexing and pileup mitigation.
\item \textbf{Ultra-thin solenoids}: Reducing the thickness and material budget of superconducting solenoids would improve detector acceptance and performance, while also lowering backgrounds from secondary interactions.
\item \textbf{Novel technologies to reduce services (power, data)}: Power delivery and data transmission are increasingly limiting factors in large-scale detectors. Innovative approaches, such as wireless links, optical power transfer, or advanced low-power electronics, could dramatically reduce infrastructure needs and simplify integration.
\item \textbf{MAPS with extremely fast timing}: Monolithic Active Pixel Sensors (MAPS) have already transformed tracking, but extending their capabilities to include precise timing would allow for 4D tracking, enhancing pattern recognition and pileup suppression in the high-rate EIC environment.
\item \textbf{Breaking the 30\% barrier in hadronic calorimetry}: Achieving significantly better energy resolution for hadronic showers would be transformative for jet physics and flavor tagging, opening new windows for precision measurements at the EIC.
\end{itemize}

As physics requirements grow increasingly demanding, the technologies needed to meet them inevitably become more complex. Consequently, the costs and resources required for their development rise significantly. Some of these key items have already reached a level where their development is so cost- and labor-intensive that nuclear physics (NP) R\&D alone cannot sustain them. The costs are prohibitive at the scale of NP’s capabilities, making international collaboration and cross-field partnerships (e.g., between NP and HEP) essential for success.

Ultimately, advancing these technologies is not an end in itself but a direct enabler of the ambitious physics program of a second EIC detector—enhancing capabilities in heavy-flavor tagging, jet tomography, and precision measurements of hadronic structure that will fully exploit the unique opportunities of the EIC.

\subsection{GridPix}
\label{sec:gridpix}
Depending on the background rate, event activity, and magnetic field, low-momentum tracking could be affected by the distortions due to space charge present in the TPC volume. 
To simulate realistic track reconstruction performance one needs to measure detector performance for each considered gas mixture. For correct estimate of the IBF produced in the TPC based on GridPix technology, accurate measurements have to be performed together with detailed simulations. 

The GridPix readout uses the 256 by 256 pixelated Timepix chip. Electronics used to read it out, consuming 1.2--5.4 kW of power, occupancy dependent. Conventional water cooling is bulky and needs a lot of material. EIC events are asymmetric by design, thus forward direction physics observables are sensitive to material budget. In addition, water cooling is not uniform, which is important for single-electron counting as gas amplification is temperature dependent. The possibility of developing and testing ``modern" $CO_{2}$ cooling will guarantee the extremely low mass detector special for low-momentum particles reconstruction and PID including the best track finding for a barrel setup. Corresponding R\&D is currently undergoing at Yale University and Purdue University~\cite{GridPixSoW2022}.
\subsection{A Mini-DIRC for Z-Tagging Nuclear Fragments}

The proposed ion beamline optics for a second EIC detector in IR-8 include a high dispersion ($0.48\text{ m}/100\%$) focus $\sim45$~m downstream of the Interaction Point (IP)~\cite{bib:EICprojectWiki}
Given the expected beam size at the high dispersion focus, beam insertion trackers (``Roman Pots'') located near the focus can track any ion whose magnetic rigidity ($K=P/Z$) differs from the beam rigidity by more than $1\%$. The concept of the mini-DIRC is to identify the atomic number  $Z$ of any beam fragment, from protons to uranium by measuring the absolute Cherenkov light yield emitted in a thin silica (SIO$_2$) radiator.

An ion of charge $Ze$ with velocity $c\beta$ passing through a medium with index of refraction $n(\lambda)$ will emit light with a absolute spectral distribution of 
\begin{align}
\frac{d^2N_\gamma}{dx d\lambda} = 2\pi \frac{\alpha_{QED} Z^2}{\lambda^2} \left[ 1 - \frac{1}{n^2(\lambda)\beta^2}\right]
\end{align}
The key feature of this distribution is the $Z^2$ dependence on the incident ion charge. Note that all ion fragments reaching the downstream focus have essentially the same velocity as the original beam. With such a high yield of photo-statistics, the difference in light yield between a ions of charge $Z$ and $Z-1$ will be significantly greater than the fluctuations in light yield, thus enabling identification of the atomic species of fragments event-by-event.

\begin{figure}[h]
\begin{center}
\includegraphics[width=0.38\textwidth,trim={4cm 0 3cm 0},clip]{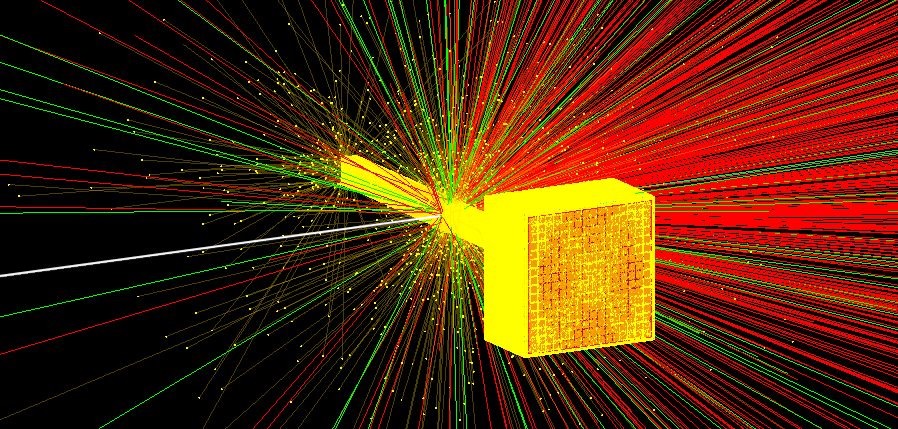} \hfill
\includegraphics[width=0.60\textwidth,trim={0 0 2mm 0},clip]{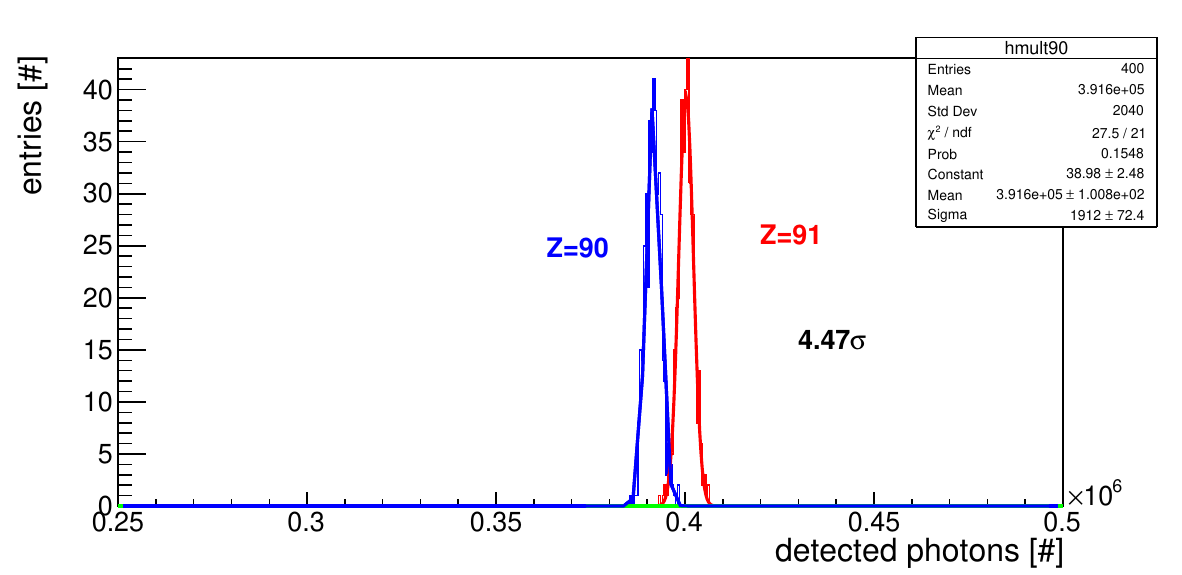}
\caption{\textbf{Left: } Single event display of $Z=90$ ion of momentum 100~GeV/u (white) passing through center of $6\times 10\times 150\text{ mm}^3$ SiO$_2$ radiator.
The photosensor array is a $16\times 16$ array of 3~mm SiPM pixels. 
Optical photons are in yellow, secondary electrons and gammas are  in red and green. 
\textbf{Right: } Photo-electron yield distributions for $Z=90$ and $Z=91$ ions.}
\label{fig:Ztag}
\end{center}
\end{figure}

\paragraph{GEANT4 Simulations}
The GSI DIRC simulation code was adapted to a specialized {\tt minidirc} code~\cite{bib:minidirc} 
for detailed simulations of miniDIRC geometry, light collection and photo-sensors. A MiniDIRC design is illustrated in Fig.~\ref{fig:Ztag}, which also details the geometry. The $10\sigma$ beam envelope at the second focus is 5~mm, so the beam would pass this distance from the far end of the radiator bar. The single-event display in Fig.~\ref{fig:Ztag} shows secondary electrons and photons (above default GEANT4 threshold), and tracks $10\%$ of all Cherenkov photons (including those generated by secondary electrons). The total projected number of detected photoelectrons in the simulation of an ion $Z$ of momentum 100~GeV/u is $\sim 48 Z^2$.  This yield is independent of the impact point of the ion, if the radiator is built to DIRC specifications. The projected photon detection statistics for $Z=90$ and $Z=91$ ion fragments at 100~GeV/u are  shown in Fig.~\ref{fig:Ztag}, clearly showing a $\sim 4.5\sigma$ separation of the two atomic species. Note that as the fragment ion charge decreases, the relative separation of the $Z$ and $Z+1$ light yield distributions grows.
\paragraph{Future Work}
For any given beam species ${}^AZ$, there are potentially ion fragments $Z'$ tracked at the high dispersion focus for $1\le Z'\le Z$. The expansion volume provides a proximity focus of the Cherenkov light cone. This focus effect is not relevant for the $Z$-tagging properties of this detector. Future work will consider strategies for mitigating pixel saturation effects in the photo sensor (including defocussing the light) while maintaining the full dynamic range of $Z'$ identification. Additional simulations will also evaluate the impact of relaxing the specifications of face parallelism and corner squareness.

\section{Overlap with Other Projects}
The detector technologies discussed in this report are not being developed in isolation. Many of them are also central to current and planned collider experiments, creating natural opportunities for shared
R\&D, common test-beam programs, and cross-validation of performance in different environments. In this section we highlight, in particular, the overlap with Belle~II at SuperKEKB and with the detector concepts being developed for FCC-ee. In both cases, the timelines and technical needs are, at least in parts,  aligned with those of a second EIC detector, so coordinated development can reduce
risk and cost while broadening the scientific impact of key technologies.

\subsection{BELLE-II} 

Belle~II is a high-luminosity $e^{+}e^{-}$ experiment operating at SuperKEKB, optimized for
precision studies of heavy flavor, $\tau$ physics, and searches for physics beyond the Standard Model~\cite{Belle2TDR,Belle2UpgradeCDR}. 
Although its physics program is distinct from that of the EIC, the detector shares many of the same
building blocks: a low-material inner vertex detector, a large-volume gaseous tracker, powerful
charged-hadron PID based on Cherenkov and timing information, and a dedicated $K_{L}$/muon
(KLM) system. As a result, several of the technologies considered for a second EIC detector
have direct counterparts at Belle~II, and in some cases Belle~II already provides an operational
or near-term upgrade testbed.

\begin{figure}[h]
\begin{center}
\includegraphics[width=0.75\textwidth]{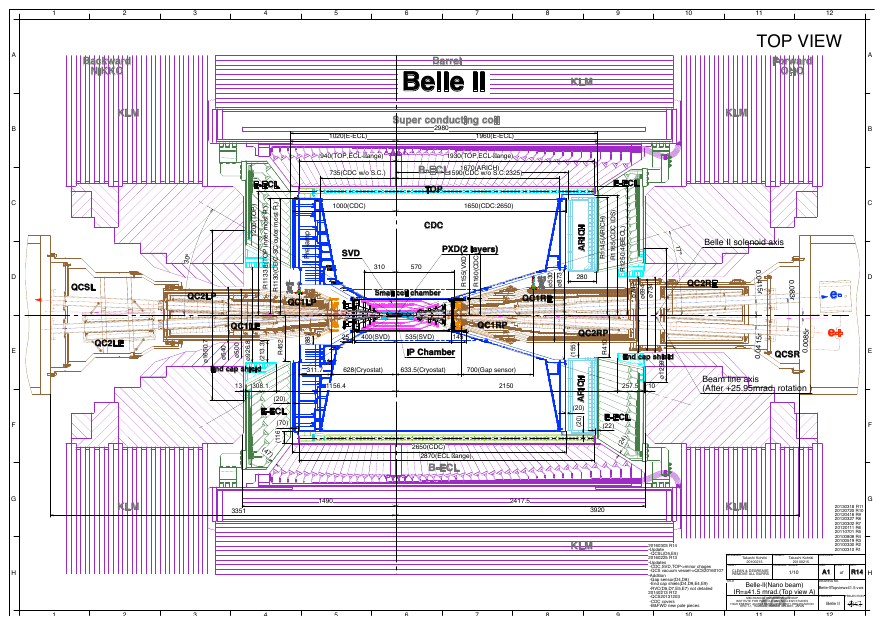}
\caption{Belle II detector top view.}
\label{fig:belle2topview}
\end{center}
\end{figure}

A prime example is the KLM system, which in Belle~II consists of scintillator strips embedded
in the iron flux-return yoke and read out by wavelength-shifting fibers and photosensors. The
KLM provides efficient muon identification and $K_{L}$ detection over a large solid angle while
doubling as the magnetic flux return. The KLM-inspired barrel muon concept discussed in
Sec.~\ref{KLM} follows a very similar philosophy: alternating layers of iron absorber and plastic
scintillator, with segmentation and timing good enough to support both muon identification
and coarse hadronic calorimetry. In a second EIC detector, such a system would complement
the calorimetric information for jets, enhance quarkonium measurements in the di-muon
channel, and provide redundancy for lepton identification. Joint development of scintillator
modules, SiPM-based readout, and reconstruction algorithms---including ML/AI-based pattern
recognition in a highly segmented iron-scintillator stack---would benefit both Belle~II KLM
upgrades and an EIC KLM, and can be organized around common prototypes and test-beam
campaigns.

There is also strong overlap in Cherenkov-based PID. Belle~II employs a time-of-propagation
(TOP) detector in the barrel and an aerogel RICH (ARICH) in the forward endcap, relying on
fast single-photon sensors and precise timing to achieve excellent $\pi/K$ separation up to a few
GeV/$c$. The xpDIRC and hpRICH concepts outlined in Sec.~\ref{sec:PID} build directly on the same
DIRC/TOP technology base, pushing toward higher momentum reach and improved robustness
through optimized optics and advanced photon detectors~\cite{xpDIRC,DRD4}. Operational
experience from the Belle~II TOP and ARICH systems---including photon-detector aging,
rate capability, calibration strategies, and timing/optics systematics---is directly relevant for
a second EIC detector. Conversely, EIC-driven R\&D on alternative photon sensors (e.g.\ modern
SiPMs or HRPPDs), compact readout geometries, and deep sub-100~ps timing could inform
future Belle~II upgrades or successor $e^{+}e^{-}$ flavor factories. The ECFA detector R\&D
roadmap already identifies Belle~II and the EIC as part of a common ecosystem for Cherenkov
and fast-photon-detector development.

Perhaps the most direct technology overlap is in gaseous tracking. Belle~II currently relies on a
large central drift chamber, but aging and occupancy considerations have motivated studies of a
TPC-based replacement for the central tracker~\cite{Belle2UpgradeCDR}. Recent work explores
a concept in which the inner volume is occupied by silicon pixel layers, while the remaining
radial space is filled with a single TPC drift volume read out on the backward end with a
GridPix-style pixelated Micromegas/Timepix3 readout and an Ar-based gas mixture, complemented
by fast timing layers (e.g.\ LGADs) at small radii for triggering and low-momentum PID.%
\footnote{See, for example, recent Belle~II TPC concept studies presented by P.~Lewis.}
This is strikingly similar to the mixed-technology tracking concepts evaluated for the second
EIC detector in Sec.~\ref{sec:PingTracking}, which combine an inner MAPS or fast silicon system with an outer
TPC instrumented with GridPix readout~\cite{GridPixQuad,GridPixIntro,PixelTPC}. The
challenges are also parallel: ion-backflow control and space-charge distortions, diffusion and
cluster separation in the chosen gas, heat removal and services for a fine-pitch pixel readout,
and synchronization with fast timing layers.

Because both Belle~II and the EIC aim to exploit TPC-based d$E$/d$x$ and cluster-counting
for PID (Secs.~4.4.4–4.4.5), there is a clear opportunity for shared simulation and hardware
R\&D on topics such as optimized gas mixtures, pixel pitch and shaping times, and reconstruction
algorithms capable of resolving individual clusters in high-rate environments. Test-beam campaigns
with common GridPix modules and Timepix3 readout, as already pursued in the EIC GridPix
program (Sec.~\ref{sec:gridpix}), could be designed to answer questions of interest to both experiments.

Finally, both Belle~II and the second EIC detector concept depend on fast timing layers based
on technologies such as LGADs. At Belle~II these layers are being considered primarily as a
solution to triggering and pattern-recognition challenges in a TPC-based central tracker, while
at the EIC they also serve as stand-alone time-of-flight PID systems and as a handle on beam-
related backgrounds (Sec.~\ref{sec:aclgad}). Nevertheless, the underlying device and electronics R\&D
is very similar: thin sensors with $O(20\text{–}50)$~ps resolution, low noise, and acceptable
radiation tolerance. Coordinated development of LGAD sensors, front-end ASICs, clocking,
and calibration procedures can therefore serve both communities.

In summary, Belle~II provides both an operational detector and an active upgrade program
whose technologies closely mirror many of the key subsystems envisioned for a second EIC
detector. The KLM, Cherenkov PID, gaseous tracking, and fast timing systems all offer natural
points of contact where joint R\&D can reduce duplication, accelerate progress, and ensure that
solutions developed for one experiment are robust and versatile enough to be adapted to the
other.

\subsection{Overlap Between FCC-ee and EIC Detector Requirements and Technologies} 

\begin{figure}[h]
\begin{center}
\includegraphics[width=\textwidth]{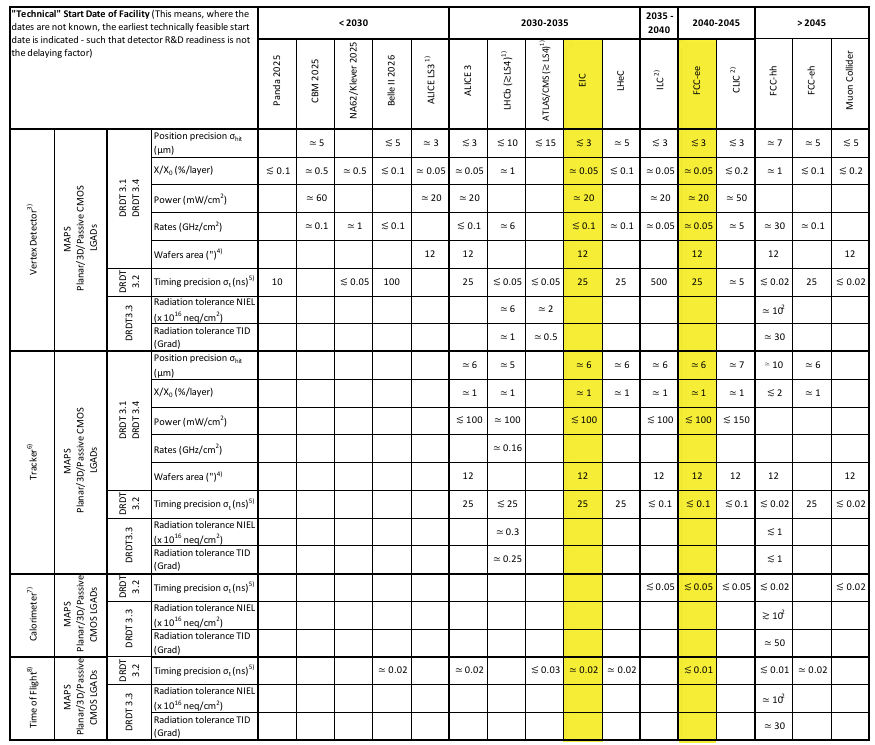} 
\caption{This matrix shows the requirements of detector systems listed at various facilities. Highlighted in yellow are EIC and FCC-ee that show remarkable similarities especially in tracking~\cite{Group:2021edg}.}
\label{fig:FCCReq}
\end{center}
\end{figure}

Although the physics programs of the Electron--Ion Collider (EIC) and the Future Circular Collider in its electron--positron mode (FCC-ee) are very different, their detector concepts reveal substantial areas of commonality. Both experiments must address demanding requirements in tracking, vertexing, calorimetry, timing, and particle identification, albeit for different physics motivations. Recognizing these overlaps provides an opportunity for coordinated research and development efforts, allowing the two communities to share advances and reduce duplication. Figure~\ref{fig:FCCReq} shows the detector requirement matrix that was featured in the 2021 ECFA Detector Research Roadmap~\cite{Group:2021edg}. The table reports on the horizontal axis the facilities while the vertical axis lists the quantities with the most demanding specifications, such as the spatial and temporal precision, power consumption, material contribution, and radiation tolerance, that must be achieved. Highlighted are EIC and FCC-ee showing very similar requirements.

\subsubsection{Common Detector Requirements}

A first area of overlap lies in the design of tracking and vertexing systems. Both the EIC and FCC-ee rely on ultra-low mass tracking with fine granularity to deliver precise measurements of charged particle trajectories. At the EIC, precision vertexing is needed to identify heavy-flavor hadrons and to study partonic dynamics, particularly in polarized beams and $e$+A collisions. At FCC-ee, the challenge is driven by the need for impact parameter resolutions significantly beyond those achieved at the LHC, so that rare decays and Higgs recoil analyses can be performed with unprecedented precision. In both cases, lightweight mechanical structures, advanced cooling techniques, and careful control of the material budget are decisive factors. 

\begin{figure}[h]
\begin{center}
\includegraphics[width=\textwidth]{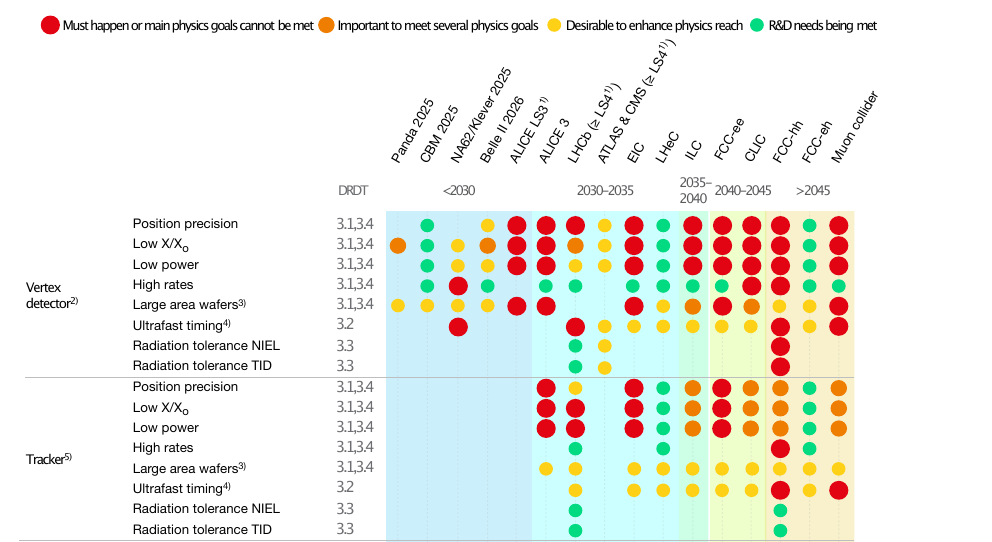} 
\caption{Detector matrix summarizing the requirements
for future solid state detectors. The table reports on the horizontal axis the facilities while the vertical axis lists the quantities with the most demanding specifications, such
as the spatial and temporal precision, power consumption, material contribution, and
radiation tolerance, that must be achieved. The colour coding is explained in the text. From \cite{Group:2021edg}.}
\label{fig:FCCTrack}
\end{center}
\end{figure}

Figure~\ref{fig:FCCTrack} compares the key tracking requirements for solid state tracking detectors for various facilities. The color coding is linked not to the intensity of the required effort but to the potential impact on the physics program of the experiment: Must happen or main physics goals cannot be met (red, largest dot); Important to meet several physics goals (orange, large dot); Desirable to enhance physics reach (yellow, medium dot); R\&D needs being met (green, small dot); No further R\&D required or not applicable (blank). As Figure~\ref{fig:FCCReq} this plot was featured in the 2021 ECFA Detector Research Roadmap~\cite{Group:2021edg}

\begin{figure}[h]
\begin{center}
\includegraphics[width=\textwidth]{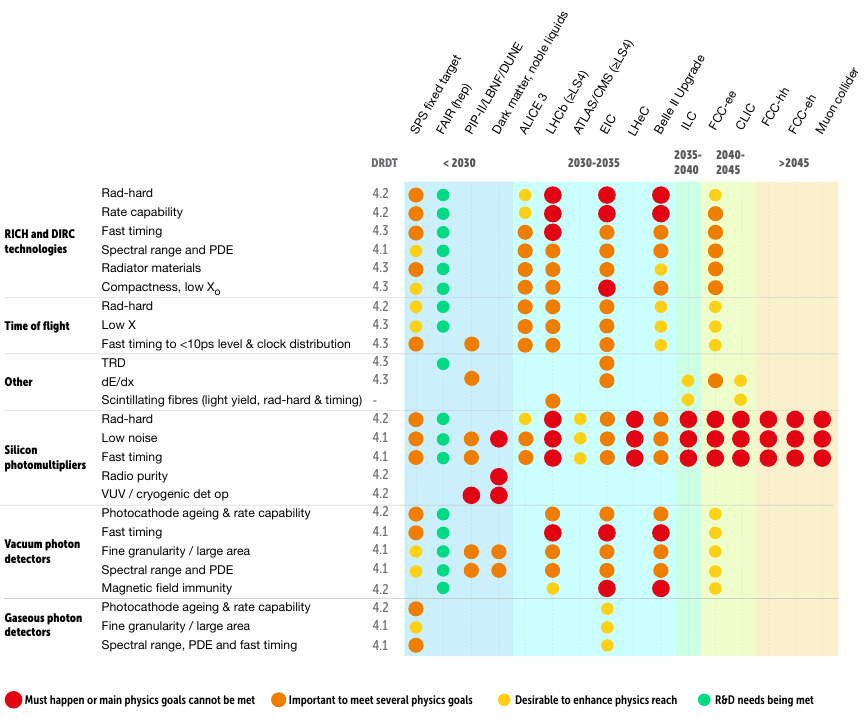} 
\caption{Schematic timeline of categories of experiments employing PID and photon detectors. From ECFA report \cite{Group:2021edg}.}
\label{fig:FCCPhoto}
\end{center}
\end{figure}

Particle identification also constitutes a shared priority. At the EIC, the ability to separate pions, kaons, and protons across a wide momentum and rapidity range has to be a key feature of any EIC detector. At FCC-ee, clean identification of leptons and hadrons underpins precision electroweak measurements and flavor physics analyses. Although the momentum regimes and physics drivers differ, both projects depend on complementary PID techniques such as Cherenkov detectors (DIRC and RICH), precise time-of-flight measurements with tens-of-picosecond resolution, and d$E$/d$x$ information from tracking detectors. In both contexts, the development of fast photon detectors and timing layers enhances the available toolset. Figure \ref{fig:FCCPhoto} depicts the schematic timeline of categories of experiments employing PID and photon
detectors. The color scheme is the same as in  Figure \ref{fig:FCCTrack}. The overlap between FCC-ee and the EIC are substantial in these areas with the exception of gaseous photon detectors (e.g. HRPPDs).

Calorimetry is another domain where requirements converge. FCC-ee demands calorimeters with extremely fine granularity, optimized for particle-flow reconstruction, in order to achieve the jet energy and photon resolutions commensurate with its enormous statistical datasets. The EIC also benefits from highly segmented calorimetry, particularly in its electromagnetic section, where separation of electrons and photons at low scattering angles is essential. Moreover, hermetic calorimetric coverage extending to forward and backward regions allows reconstruction of event kinematics across the full range of Bjorken-$x$. Both experiments thus converge on imaging calorimetry concepts, where silicon--tungsten electromagnetic calorimeters and high-granularity hadronic calorimeters play central roles.

Fast timing capabilities are increasingly important in both designs. For the EIC, precise timing is closely linked to particle identification, enabling time-of-flight methods with resolutions around 20--30~ps, and helping to suppress beam-related backgrounds. For FCC-ee, the emphasis is somewhat different: precision timing supports pattern recognition in high-occupancy environments, increases sensitivity to displaced vertices and long-lived particle signatures, and may provide further discrimination in jet substructure analyses. In both cases, the development of timing layers based on Low-Gain Avalanche Diodes (LGADs) is considered highly promising.

Finally, both projects depend on advanced simulation and reconstruction software. The EIC faces the challenge of reconstructing events in a strongly asymmetric kinematic environment with a large number of subsystems, while FCC-ee must analyze unprecedentedly large datasets with extreme statistical precision. Nevertheless, the two communities share a reliance on particle-flow algorithms, advanced track-based flavor tagging, and flexible detector description frameworks. 

\subsubsection{Shared Technology R\&D}

Several specific technology areas stand out as mutually relevant to both projects. Monolithic Active Pixel Sensors (MAPS) are perhaps the most important, offering ultra-thin silicon detectors with high spatial resolution, ideally suited for vertexing and tracking in both collider environments. LGADs are another example: their ability to deliver timing resolutions below 30~ps makes them attractive for both the time-of-flight systems of the EIC and the precision timing layers foreseen at FCC-ee. Silicon Photomultipliers (SiPMs) are equally versatile, serving as photon detectors in Cherenkov systems and as readout devices in finely segmented calorimeters. Digital SiPM are becoming increasingly important at the FCC-ee and might be beneficial for EIC detectors as well.

The development of high-granularity calorimetry is itself a major area of overlap. Concepts such as silicon--tungsten electromagnetic calorimeters and imaging hadronic calorimeters are already under active investigation in both communities, and both stand to benefit from the advances in electronics, cooling, and integration that are being pursued. Finally, Cherenkov detectors such as DIRC and RICH remain indispensable at the EIC and are also of potential interest at FCC-ee. In both cases, the use of  focusing optics and advanced photon sensors provides the prospect of extended momentum coverage and improved performance.

\subsubsection{Comparison and Outlook}

Table~\ref{tab:overlap} summarizes the areas of overlap between EIC and FCC-ee requirements, alongside the technologies that could be shared. While the precise performance targets differ, the table illustrates how both projects rely on a similar technological foundation.

\begin{table}[h]
\centering
\caption{Comparison of common detector requirements for EIC and FCC-ee, with shared technology areas.}
\label{tab:overlap}
\begin{tabular}{|p{2cm}|p{4cm}|p{4cm}|p{4cm}|}
\hline
\textbf{Area} & \textbf{EIC Requirement} & \textbf{FCC-ee Requirement} & \textbf{Shared Technology} \\
\hline
Tracking \& Vertexing & Wide acceptance; few-$\mu$m vertex precision for heavy flavor; low material budget & Exceptional momentum and impact-parameter resolution; vertexing beyond LHC benchmarks & MAPS, lightweight supports, advanced cooling \\
\hline
PID & $\pi/K/p$ separation from $\sim$0.3--10~\gevc in central and backward rapidities and up to 50 GeV at forward rapidities & Efficient lepton/hadron ID; flavor tagging in clean $e^+e^-$ environment & DIRC/RICH with SiPM readout; d$E$/d$x$; TOF with LGADs \\
\hline
Calorimetry & Hermetic EM+HAD calorimetry; backward--forward coverage; $e/\gamma$ ID & PF calorimetry with fine granularity; excellent jet and $\gamma/\pi^0$ resolution & SiW-ECAL, imaging HCAL, SiPM-based readout \\
\hline
Timing & 10--30~ps for TOF PID and background rejection & Sub-30~ps for pattern recognition, LLP searches & LGAD-based timing layers \\
\hline
\end{tabular}
\end{table}

Looking ahead, it is clear that the EIC and FCC-ee communities face different challenges: the EIC prioritizes wide kinematic acceptance and forward instrumentation, while FCC-ee is primarily concerned with extreme precision in central measurements. Yet the overlap in enabling technologies is striking. Both experiments demand ultra-thin tracking, highly granular calorimetry, precise timing, and sophisticated reconstruction software. Joint research programs and the sharing of technical developments can therefore accelerate progress, reduce costs, and strengthen the scientific reach of both projects. In this sense, the overlap in detector requirements is not only a coincidence but also an opportunity to foster collaboration between the nuclear and high-energy physics communities.

\section{Summary}  
This report documents the closeout results of BNL LDRD 23-050, ``A Second EIC Detector: Physics Case and Conceptual Design.'' The project was motivated by the strong physics case for a second general-purpose EIC detector and interaction region, and by the opportunity to realize genuine complementarity to the first detector, ePIC, through differences in detector technologies, interaction-region (IR) implementation, and analysis systematics. Over the course of the project we (i) refined and expanded the physics opportunities uniquely strengthened by a second detector, (ii) translated these opportunities into detector-level performance requirements, (iii) investigated conceptual detector layouts and subsystem technology options that could meet those requirements within realistic IR and integration constraints, and (iv) identified a set of R\&D directions that would mature key technologies on the timescale appropriate for a later detector.

A recurring theme is that two detectors enable robustness and breadth of the EIC program through complementary strengths, cross-checks, and different systematic limitations. We highlighted new physics opportunities that become particularly compelling when the second detector is designed around alternative technologies or an IR with additional features. Examples include physics enabled by an IR incorporating a secondary focus, which can extend sensitivity to very small transverse momentum fragments and strengthen the ability to veto nuclear breakup activity. In diffractive vector-meson measurements, this capability supports the extremely high veto performance needed to isolate coherent processes over a wide kinematic range. We also emphasized opportunities tied to species and configuration flexibility (including isotopes) and to searches that benefit from detector capabilities not emphasized in the baseline, such as enhanced muon identification, improved far backward coverage, and $\tau$-sensitive signatures, which can open additional avenues for electroweak and beyond-the-Standard-Model studies.

Lessons learned from the evolution of ePIC were used as direct design input. In particular, the silicon-focused tracking approach delivers excellent intrinsic precision but faces limitations associated with hit redundancy, long integration windows, and robustness against machine-related backgrounds. We also identified practical integration drivers—such as the complexity and cost risk of highly segmented PID systems, acceptance and instrumentation needs between the central detector and far-forward Roman-pot stations, demands for faster timing to mitigate backgrounds and improve event association, and the importance of realistic service routing and maintainability. Finally, we stressed that software choices and reconstruction flexibility (including modern track-reconstruction toolkits and geometry/conditions handling) are central to achieving and validating performance in complex environments.

Guided by these requirements and lessons, we evaluated a suite of second-detector concepts and subsystem options. For the magnetic system, we discussed solenoidal designs spanning conventional solutions and more novel concepts (including compact, low-material designs based on higher-temperature superconductors) that can change the available radial envelope and material distribution seen by calorimetry and PID. For tracking, we explored mixed-technology layouts combining precise inner silicon with a large-volume gaseous tracker to provide high hit redundancy and low material, and we quantified performance with full-simulation studies. The mixed-technology design meets the Yellow Report $p_T$-resolution requirements in the central region, while also clearly indicating where additional outer tracking layers are required at large $|\eta|$ to recover performance near and beyond the gaseous-detector acceptance edge. These studies were accompanied by material-budget evaluations to identify the dominant contributors and to guide mechanical and services optimization.

For particle identification, we investigated complementary strategies spanning next-generation DIRC concepts, forward RICH options, precision time-of-flight based on advanced timing sensors, and the use of gaseous-tracker ionization information (including cluster counting) to extend low-momentum PID reach. In calorimetry, we assessed approaches aimed at substantially improved hadronic energy resolution—motivating dual-readout techniques to separate scintillation and Cherenkov components—as well as options for high-performance electromagnetic calorimetry using homogeneous media (e.g.\ advanced glass or crystal technologies) where space and cost allow. We additionally studied the impact of incorporating dedicated muon detection, including barrel concepts inspired by existing KLM-style systems, as a way to enhance quarkonium measurements and broaden sensitivity to muon-rich final states; these capabilities are particularly relevant for a second detector pursuing complementary BSM and electroweak measurements.

A second detector realized on a later timescale is also an opportunity to re-establish a strong generic detector R\&D pipeline that can deliver mature, validated technologies beyond the baseline. We therefore conclude by outlining priority R\&D directions emerging from this project, including high-granularity gaseous readout concepts (e.g.\ GridPix-class approaches), specialized PID instrumentation for far-forward and secondary-focus tagging, precision timing development and integration, and continued work on dual-readout and homogeneous calorimetry with realistic performance validation. Collectively, the studies in this report demonstrate a coherent and technically motivated path toward a second EIC detector concept that is complementary to ePIC, expands the facility physics reach, and provides a roadmap for the R\&D needed to enable that concept.

\section{Acknowledgment}
We would like to thank Dan Cacace for the discussions regarding space for services. We also thank Shujie Li for her assistance with the simulations for the second EIC detector tracking. This work is supported by the Laboratory Directed Research and Development (LDRD) project “A Second EIC Detector: Physics Case and Conceptual Design” (LDRD-23-050) at Brookhaven National Laboratory.
\bibliographystyle{unsrt}
\bibliography{ref}

\end{document}